\documentclass[aps,nofootinbib,groupedaddress]{revtex4}

\usepackage{url}
\usepackage{graphicx}
\usepackage{amsmath,amssymb,amstext,amssymb,amsfonts,amsthm}
\usepackage{hyperref}
\usepackage{appendix}
\usepackage[margin=1.0in,papersize={8.5in,11in}]{geometry}
\usepackage[utf8]{inputenc}
\usepackage{soul}
\usepackage{color}

\newcommand{\cen}{\mathrm{cen}}
\newcommand{\sat}{\mathrm{sat}}
\newcommand{\taud}{\dot{\tau}}

\usepackage{tabularx}
\newcolumntype{C}{>{\centering\arraybackslash}X}
\newcolumntype{R}{>{\raggedleft\arraybackslash}X}

\begin{document}

\title{Velocity reconstruction with the cosmic microwave background and galaxy surveys}

\author{Juan Cayuso${}^{1,2}$}
\author{Richard Bloch${}^{3}$}
\author{Selim C. Hotinli${}^{4}$}
\author{Matthew C. Johnson${}^{1,3}$}
\author{Fiona McCarthy${}^{1,2,5}$}
\affiliation{${}^1$Perimeter Institute for Theoretical Physics, Waterloo, Ontario N2L 2Y5, Canada}
\affiliation{${}^2$Department of Physics and Astronomy, University of Waterloo, Waterloo, Ontario, N2L 3G1, Canada}
\affiliation{${}^3$Department of Physics and Astronomy, York University, Toronto, Ontario, M3J 1P3, Canada}
\affiliation{${}^4$Department of Physics \& Astronomy, Johns Hopkins University, Baltimore, MD 21218 USA}
\affiliation{${}^5$Center for Computational Astrophysics, Flatiron Institute, 162 5th Avenue, New York, NY 10010 USA}

\date{\today}

\begin{abstract}
The kinetic Sunyaev Zel'dovich (kSZ) and moving lens effects, secondary contributions to the cosmic microwave background (CMB), carry significant cosmological information due to their dependence on the large-scale peculiar velocity field. Previous work identified a promising means of extracting this cosmological information using a set of quadratic estimators for the radial and transverse components of the velocity field. These estimators are based on the statistically anisotropic components of the cross-correlation between the CMB and a tracer of large scale structure, such as a galaxy redshift survey. In this work, we assess the challenges to the program of velocity reconstruction posed by various foregrounds and systematics in the CMB and galaxy surveys, as well as biases in the quadratic estimators. To do so, we further develop the quadratic estimator formalism and implement a numerical code for computing properly correlated spectra for all the components of the CMB (primary/secondary blackbody components and foregrounds) and a photometric redshift survey, with associated redshift errors, to allow for accurate forecasting. We create a simulation framework for generating realizations of properly correlated CMB maps and redshift binned galaxy number counts, assuming the underlying fields are Gaussian, and use this to validate a velocity reconstruction pipeline and assess map-based systematics such as masking. We highlight the most significant challenges for velocity reconstruction, which include biases associated with: modelling errors, characterization of redshift errors, and coarse graining of cosmological fields on our past light cone. Despite these challenges, the outlook for velocity reconstruction is quite optimistic, and we use our reconstruction pipeline to confirm that these techniques will be feasible with near-term CMB experiments and photometric galaxy redshift surveys.
\end{abstract}

\maketitle

\section{Introduction}

Measurements of the Cosmic Microwave Background (CMB) radiation are entering an unprecedented era of high-resolution and low noise. Existing experiments such as the Atacama Cosmology Telescope~\cite{Kosowsky:2003smi} (ACT) and South Pole Telescope~\cite{SPT:2004qip} (SPT), near-term experiments such as Simons Observatory~\cite{Ade:2018sbj} (SO), and future experiments such as CMB-S4~\cite{1610.02743} and CMB-HD~\cite{Sehgal:2019ewc} will map the small-angular scale anisotropies in CMB temperature and polarization with steadily increasing precision. Many of the new opportunities on this frontier arise from CMB secondaries: temperature and polarization anisotropies associated with the gravitational or electromagnetic scattering of CMB photons from structure at low redshift. For example, existing data from ACT has been used to reconstruct the lensing potential at high fidelity over a significant fraction of the sky~\cite{Darwish:2020fwf}, and future experiments will enable precision cosmological constraints on e.g. neutrino masses using information from these reconstructions. Simultaneously, galaxy surveys will map ever larger volumes with increasing numbers of spectroscopic redshifts and photometric redshifts of improving precision; near-term surveys include The Rubin Observatory Legacy Survey of Space and Time~\cite{0912.0201} (LSST), Dark Energy Spectroscopic Instrument~\cite{DESI:2019jxc} (DESI), and Euclid~\cite{Laureijs2011}. Existing galaxy surveys have been combined with CMB datasets to make a (statistically) significant detection of the kinetic Sunyaev Zel'dovich (kSZ) effect~\cite{1203.4219,Planck:2015ywj,2016PhRvD..93h2002S,DES:2016umt,Hill:2016dta,1607.02139,AtacamaCosmologyTelescope:2020wtv}, temperature anisotropies induced by the scattering of CMB photons from free electrons in bulk motion. Again, future experiments will allow for precision science using the kSZ effect (as described in more detail below). Beyond lensing and kSZ, there are other CMB secondaries in temperature, including the moving lens (ML) effect,~\cite{1983Natur.302..315B,1986Natur.324..349G}, other non-linear components of the integrated Sachs Wolfe effect~\cite{1968Natur.217..511R,2002PhRvD..65h3518C,2010MNRAS.407..201C}, patchy reionization~\cite{0812.1566}, and rotational kSZ~\cite{Cooray:2001vy}; additional effects exist in the polarized component of the CMB (e.g.~\cite{Kamionkowski1997,Yasini:2016pby}). Assessing the detectability of CMB secondaries and their utility in improving our understanding of cosmology is an active area of investigation, to which the present paper aims to contribute.
 
In this paper, we focus on the kSZ and the ML effects. Schematically, the kSZ temperature anisotropies are given by a line-of-sight integral over the product of the radial component of the peculiar velocity field~\footnote{More precisely, the kSZ temperature anisotropies are sourced by the remote dipole field: the radially-projected CMB dipole observed along our past light cone~\cite{Terrana2016}.} and the electron density; the ML temperature anisotropies are a line-of-sight integral of the transverse gradient of the gravitational potential and the transverse components of the peculiar velocity field. These secondaries contain significant cosmological information owing to their dependence on large-scale bulk velocities. 

One way of accessing this information is through the technique of kSZ tomography~\cite{Zhang:2015uta,Terrana2016,Deutsch:2017ybc,Smith:2018bpn}: tomographic reconstruction of the radial velocity using the CMB and a tracer of large scale structure. A quadratic estimator for the radial velocity field was developed in Ref.~\cite{Deutsch:2017ybc}, which relies on the anisotropic cross-power between a galaxy redshift survey and the kSZ component of the CMB. Such a reconstruction can be used as a general-purpose cosmological observable, and is in principle a powerful probe of primordial non-Gaussianity~\cite{Munchmeyer:2018eey}, relativistic effects in galaxy surveys~\cite{Contreras_2019}, modified gravity~\cite{Pan:2019dax}, CMB anomalies~\cite{Cayuso:2019hen}, and isocurvature perturbations~\cite{Hotinli:2019wdp,Sato-Polito:2020cil}. 

A quadratic estimator for the transverse velocity based on the correlation between a galaxy redshift survey and the ML component of the CMB temperature anisotropies was presented in Ref.~\cite{Hotinli:2018yyc}. For consistency of terminology we will refer to the tomographic reconstruction transverse velocities using the ML effect as `ML tomography'. The ML effect is roughly an order of magnitude smaller than kSZ on small angular scales ($\sim$ arcmin), so while kSZ has been detected at the greater than 5$\sigma$ level, current datasets are not yet sufficient to have made a detection of  the ML effect. Even so, the ML effect will be measured with high signal-to-noise with upcoming CMB experiments~\citep{Hotinli:2018yyc,Yasini:2018rrl,Hotinli:2020ntd}.  As discussed in Ref.~\citep{Hotinli:2021hih}, transverse velocities from ML tomography, when combined with galaxy measurements, can provide competitive constraints on the combination of the linear-theory growth rate $f$ and the amplitude of matter fluctuations $\sigma_8$ on the scale of $8h^{-1}$Mpc, comparable to the scenario where the redshift-space distortions (RSDs) are modelled with high accuracy. This parameter is useful for studying a large range of physics, including dark energy~\citep{Linder:2005in}, modified gravity~\citep{Linder:2007hg} and the effects of neutrino mass~\citep{1983ApJ...274..443B}. Precision measurement of $f\sigma_8$ also allows one to use the kSZ effect to learn about astrophysics, such as the characteristics of the electron density profiles around halos, by breaking degeneracies in kSZ between the electron-scattering optical depth and the growth rate~\citep{Smith:2018bpn}.

The formalism for velocity reconstruction using the kSZ effect is reasonably mature, and several approaches have been developed to analyze simulations and forecast the capabilities of kSZ tomography in future experiments. In the `Box Picture' of~\cite{Smith:2018bpn}, the kSZ effect is estimated from the momentum field along one direction in a 3D box at the median redshift of a galaxy survey. This formulation is convenient since it sidesteps spherical projection effects and allows one to work in the familiar Fourier domain. This approach is a good approximation for reconstruction on relatively small sky areas and over limited ranges in redshift. The study of various systematics is straightforward in the Box Picture, and Ref.~\cite{Smith:2018bpn} estimated the impact of redshift space distortions and photometric redshift errors on velocity reconstruction. Subsequent work explored important contributions to the variance of the quadratic estimator ~\cite{Giri:2020pkk} and explored the effect of the optical depth bias~\cite{Smith:2018bpn,Madhavacheril:2019buy,Giri:2020pkk}, the uncertainty in the reconstruction induced by imperfect modelling of the correlation between the distribution of electrons and galaxies. Ref.~\cite{Giri:2020pkk} validated the quadratic estimator in the Box Picture using a suite of N-body simulations, and demonstrated that the constraints on primordial non-Gaussianity forecasted in Ref.~\cite{Munchmeyer:2018eey} could be realized in practice. It is, however, cumbersome to accurately incorporate redshift evolution, relativistic contributions to the kSZ effect, and large sky area in the Box Picture. 

In the `Light Cone Picture' introduced in Refs.~\cite{Terrana2016,Deutsch:2017ybc}, kSZ tomography is formulated in terms of observables on our past light cone. In the Light Cone Picture, it is straightforward to incorporate redshift evolution and relativistic contributions to the radial velocity field (promoting it to the remote dipole field). However, it is less straightforward to incorporate photometric redshift errors and other systematics. Another drawback of the Light Cone Picture are the cumbersome projection integrals, which make computing observables more expensive. The quadratic estimator in the Light Cone Picture was validated using N-body simulations in Ref.~\cite{Cayuso:2018lhv}, where the impact of gravitational lensing, redshift space distortions, and the non-linear evolution of structure were taken into account. In an application to future datasets, the Light Cone Picture has a number of advantages. Perhaps most importantly, it is formulated in terms of direct observables (e.g. fields on the sphere), making contact between theory and observation precise. The signal to noise of the reconstruction is largest on the largest scales, where the redshift evolution and projection effects captured by the Light Cone Picture are most important. 

A primary goal of this paper is to further develop the formalism for the Light Cone Picture for both kSZ and ML tomography. We begin by developing a self-consistent theoretical framework based on the Halo Model~\cite{Cooray:2002dia} for predicting the auto and cross spectra of fields on the light cone, including dark matter density, electron density, velocities, galaxy number counts, the Newtonian potential and its time derivative, as well as the frequency-dependent contribution to the CMB from the thermal Sunyaev Zel'dovich (tSZ) effect and the Cosmic Infrared Background (CIB). We employ a coarse-graining scheme in radial distance along the light cone based on Haar wavelets and use this complete basis to perform line of sight integrals for the kSZ, ML, ISW, lensing, tSZ, CIB, and binned galaxy number counts. The contributions we include for the CMB represent the most important (in terms of the amplitude of power spectra) blackbody and frequency dependent components. A challenging aspect of kSZ and ML tomography is that large scale fields are reconstructed from small angular scale anisotropies, implying one must model both small and large scales well. We develop tools to accurately compute spectra from the dipole down to sub-arcminute angular scales. We quantify the level of coarse graining given a fiducial CMB experiment and galaxy survey that will be necessary to capture the relevant cosmological information accessible using kSZ and ML tomography. 

Another goal of this paper is to assess the impact of various foregrounds and systematics on kSZ and ML tomography. Previous work has largely neglected these effects in forecasts. We assess the impact of extragalactic foregrounds by forecasting the level of residuals in auto and cross-spectra given a fiducial CMB experiment and the resulting effect on the variance of the quadratic estimators. We develop a formalism, analogous to bias hardening in CMB lensing~\cite{Namikawa_2013}, to remove biases associated with photometric redshift errors, and compute the variance of the resulting unbiased estimators in both a redshift-binned and principal component basis. The kSZ and ML quadratic estimators are biased by other sources of statistical anisotropy in the CMB-galaxy cross-power such as CMB lensing. We confirm that these biases are small enough to be neglected in near-term experiments. A related systematic arises from redshift calibration errors on large angular scales, or any other effect that modulates the amplitude of the underlying statistically isotropic CMB anisotropies or galaxy number counts. This leads to a statistically anisotropic modulation of the CMB-galaxy cross-power that can bias the kSZ and ML quadratic estimators. While this is the dominant source of estimator bias, we determine that it is below the estimator variance for the fiducial CMB experiment we consider. 

To assess the impact of partial sky coverage for the CMB experiment and galaxy survey, we work with simulations in map space. We develop a numerical framework to produce sets of properly correlated CMB maps and redshift binned galaxy number counts, assuming the underlying fields are Gaussian. We derive and implement a set of real space quadratic estimators and an associated pipeline to reconstruct the radial and transverse velocity fields from an ensemble of simulated CMB maps (including both blackbody components and foreground residuals) and correlated binned galaxy number count maps. We find that no significant bias is introduced by masking, and confirm that the reconstructed power spectra can be approximated by simply scaling by the fraction of the sky that remains unmasked. For ML tomography, we highlight a challenge posed by numerical errors associated with spherical harmonic transforms in the low signal to noise regime that near-term experiments will operate in. 

Our results suggest that the main limitations on kSZ and ML tomography will be various modelling errors that give rise to a biased reconstruction of the velocity field. The largest among these are biases introduced by photometric redshift errors and mis-modelling of the galaxy-electron cross-power spectra, highlighting areas for future work. With a fixed experimental setup, improvements in the fidelity of the reconstruction can be made through better foreground removal techniques on small angular scales. However, our investigations have not found any effect that seriously impacts the performance of kSZ and ML tomography as presented in previous literature, which is good news for these techniques. As a companion to this paper, we have released a publicly available code: (Re)construction (C)ode for (C)osmological (O)bservables, \texttt{ReCCO}~\footnote{\href{https://github.com/jcayuso/ReCCO}{\url{https://github.com/jcayuso/ReCCO}} .}. This code can be used to compute various power spectra, generate properly correlated Gaussian mock maps, and perform radial and transverse velocity reconstruction. We hope that this is a useful tool for future forecasts and data analysis.

The plan of the paper is as follows. In Sec.~\ref{form}, we outline the formalism for kSZ and ML tomography and in Sec.~\ref{sec:observables} we outline our construction of various contributions to the CMB and galaxy density field. In Sec.~\ref{sec:recon_analysis} we perform a detailed forecast for a set of fiducial datasets including various biases to the reconstruction. We present and validate a simulation and reconstruction pipeline for the radial and transverse velocity fields in Sec.~\ref{sec:recon_pipeline}, and conclude in Sec.~\ref{sec:conclusion}. We summarize various technical details in several appendices.

\section{Formalism}\label{form}

In this section, we describe the formalism for kSZ and ML tomography (velocity reconstruction) in the Light Cone Picture. We begin by reviewing the projection of cosmological fields onto our past light cone. Coarse grained cosmological fields on the light cone constitute the inputs to our estimator formalism, and we outline a coarse graining scheme as well as the statistics of the coarse grained fields. We outline the quadratic estimator formalism, proceeding from the simplest to the most realistic scenario. 

\subsection{Continuous fields on the light cone}

Continuous fields defined on our past light cone constitute the basic building blocks of our formalism. A simple way of constructing a field on the light cone is to take the projection of an underlying 4-dimensional space time field $\mathrm{U}(\eta,\mathbf{x})$, where $\eta$ is the conformal time and $\mathbf{x}$ are the comoving spatial coordinates. If one parametrizes the light cone with a unit direction vector $\hat{n}$ (the line-of-sight) and a comoving distance $\chi$, then the projected field is defined by:
\begin{equation}\label{eq:projection}
\mathrm{F}(\hat{n}, \chi) \equiv \mathrm{U}(\eta(\chi),\mathbf{x} = \chi\hat{n}).
\end{equation}
In many cases, it is convenient to express $\mathrm{F}(\hat{n}, \chi)$ in terms of the spatial Fourier moments of the field $\mathrm{U}$, defined by:
\begin{equation}\label{eq:kspace_lcmoments}
\mathrm{U}(\eta,\mathbf{x}) = \int \frac{d^{3} \mathbf{k}}{(2 \pi)^{3}}\, \tilde{\mathrm{U}} (\eta, \mathbf{k})\ e^{i \mathbf{k} \cdot \mathbf{x}},
\end{equation}
which gives
\begin{equation}
\mathrm{F}(\hat{n}, \chi) = \int \frac{d^{3} \mathbf{k}}{(2 \pi)^{3}}\, \tilde{\mathrm{U}} (\eta(\chi), \mathbf{k})\ e^{i \mathbf{k} \cdot \chi\hat{n}}.
\end{equation}
The line-of-sight dependence of the light cone field can be expanded in terms of spherical harmonics:
\begin{equation}\label{eq:lcmoments}
\mathrm{F}(\hat{n}, \chi) =\sum_{\ell m} \mathrm{F}_{\ell m}(\chi)Y_{\ell m}(\hat{n})\,,
\end{equation}
and we will refer to the coefficients $\mathrm{F}_{\ell m}(\chi)$ as the \textit{light cone moments} (LC moments for short) of the field $\mathrm{F}$. The LC moments can be expressed as:
\begin{equation}\label{eq:lcmoments1}
\mathrm{F}_{\ell m}(\chi) =\int \frac{d^2\hat{n}}{(2 \pi)^{3}}Y^{*}_{\ell m}(\hat{n})\int \frac{d^{3} \mathbf{k}}{(2 \pi)^{3}}\,
\tilde{\mathrm{U}} (\eta(\chi), \mathbf{k})\ e^{i \mathbf{k} \cdot \chi\hat{n}}, 
\end{equation}
which we can further simplify to:
\begin{equation}\label{eq:lcmoments2}
\mathrm{F}_{\ell m}(\chi) =\int \frac{d^{3} \mathbf{k}}{(2 \pi)^{3}}\,
4\pi(i)^{\ell}\, j_{\ell}(k\chi)\,
\tilde{\mathrm{U}} (\eta(\chi), \mathbf{k}) \,Y_{\ell m}^{*}(\hat{k}), 
\end{equation}
where $ j_{\ell}(k\chi)$ is a spherical Bessel function. It is possible to define 
fields on the light cone using a more complex projection of the underlying field $\mathrm{U}$ than the one used in Eq.~\ref{eq:projection}. For example, the projection could depend on the direction $\hat{n}$ or introduce weights depending on the conformal time $\eta$. A more general expression for the LC moments is then:
\begin{equation}\label{eq:lcmoments3}
\mathrm{F}_{\ell m}(\chi) =\int \frac{d^{3} \mathbf{k}}{(2 \pi)^{3}}\,
\mathcal{K}_{\ell}(\chi,k)\,
\tilde{\mathrm{U}} (\eta(\chi), \mathbf{k}) \,Y_{\ell m}^{*}(\hat{k}), 
\end{equation}
where $\mathcal{K}_{\ell}(\chi,k)$ is an integral kernel determined by the particular observable and typically containing linear combinations of spherical Bessel functions.

\subsection{Integrated and coarse grained fields on the light cone}\label{sec:int_coarse_fields}

A second type of building block of our formalism are line-of-sight integrals of continuous fields on the light cone. Given a generic window function $W(\chi)$ we define the windowed $\mathrm{F}$ field:
\begin{equation}\label{eq:int_field}
\mathrm{F}^{W}(\hat{n}) \equiv \int d\chi \,W(\chi)\, \mathrm{F}(\hat{n}, \chi),
\end{equation}
and its spherical harmonic moments: 
\begin{equation}\label{eq:int_moment}
\mathrm{F}^{W}_{\ell m} \equiv \int d\chi \ W (\chi)\,  \mathrm{F}_{\ell m}(\chi)\,,
\end{equation}
where $\mathrm{F}_{\ell m}(\chi)$ are the LC moments defined in the previous section. Given a finite portion of the light cone determined by an interval $[\chi_{min},\chi_{max}]$, we can consider a complete set of normalized functions $\mu^{i}(\chi)$ and expand the LC moments of the field F:
\begin{equation}
\mathrm{F}_{\ell m}(\chi) =\sum_{i}\,\mathrm{F}^{i}_{\ell m}\,\mu^{i}(\chi)\,,
\end{equation}
where the coefficients $\mathrm{F}^{i}_{\ell m}$ are obtained using Eq.~\ref{eq:int_moment} with $W(\chi)= \mu^{i}(\chi)$. From now on, we refer to these coefficients as the \emph{$\mu$-binned LC moments} of the field $\mathrm{F}$. In this paper, we choose to expand the LC moments in the radial direction using the Haar basis. Haar wavelets are defined on the interval $\chi_{\rm min} \leq \chi \leq \chi_{\rm max}$ by:
\begin{equation}\label{eq:Haar_modes}
h^{s} (\chi ) = \frac{1}{\sqrt{\chi_{\rm max}-\chi_{\rm min}}}
\begin{cases}
2^{p/2}, & (q-1) \ 2^{-p} \leq (\frac{\chi}{\chi_{\rm max}-\chi_{\rm min}}) <  (q-1/2) \ 2^{-p} \\
-2^{p/2}, & (q-1/2) \ 2^{-p} \leq (\frac{\chi}{\chi_{\rm max}-\chi_{\rm min}}) <  q \ 2^{-p}  \\
0, & {\rm otherwise}
\end{cases}
\end{equation}
with $s = 2^p + q -1$ for integer $p,q$ for $s>0$; for $s=0$, the Haar wavelet is $h^{0} (\chi ) = \frac{1}{\sqrt{\chi_{\rm max}-\chi_{\rm min}}}$. The scale is determined by $p$ and the location is determined by $q$; for each value of $s$ there is a unique choice of $p,q$. The Haar basis functions are orthonormal over the interval $\chi_{\rm min} \leq \chi \leq \chi_{\rm max}$:
\begin{equation}
\int_{\chi_{\rm min}}^{\chi_{\rm max}} d\chi \ h^s (\chi) h^{s'} (\chi) = \delta_{s s'}\,.
\end{equation}
We choose the Haar basis to expand the LC moments because conveniently the truncated Haar expansion up to $s = N-1$ is equivalent to representing the LC moments by their average values in comoving bins of equal size $\Delta\chi = \frac{\chi_{\rm max}-\chi_{\rm min}}{N}$:
\begin{equation}\label{eq:binavgcorr}
\sum_{s=0}^{N-1} \mathrm{F}^{s}_{\ell m} \ h^s (\chi) = \sum_{\alpha=0}^{N-1}  \mathrm{F}^{\alpha}_{\ell m} \ \Pi^{\alpha}(\chi)\,,
\end{equation}
where
\begin{equation}\label{eq:tophat}
\Pi^{\alpha}(\chi) = 
\begin{cases}
\frac{1}{\Delta \chi}, & \chi_{\rm min} + j \Delta \chi \leq \chi <  \chi_{\rm min} + (j+1) \Delta \chi, \\
0, & {\rm otherwise}
\end{cases}
\end{equation}
and $\mathrm{F}^{\alpha}_{\ell m}$ are the $\Pi$-binned LC moments (we reserve Greek letters to index the $\Pi$-binned LC moments and Latin letters to index the Haar-binned LC moments). This property allows us to express the LC moments in a way that coarse-graining in the radial direction is clear:
\begin{equation}
\mathrm{F}_{\ell m}(\chi) =\sum_{\alpha =0}^{N-1}\,\mathrm{F}^{\alpha}_{\ell m}\,\Pi^{\alpha} (\chi)+
\sum_{s=N}^{\infty}\,\mathrm{F}^s_{\ell m}\,h^s(\chi)\,,
\end{equation}
where the first sum represents the `coarse' or `bulk' radial modes and the second sum, orthogonal to the first, represents the `fine' modes that don't contribute to the bulk averages. We note that the spherical Fourier-Bessel decomposition (see e.g.  \cite{Fisher:1994xm}) could have been chosen instead of the Haar basis used here. In the context of galaxy redshift surveys, a comparison between the spherical Fourier-Bessel decomposition and the redshift-binned approach employed here can be found in Ref.~\cite{Lanusse:2014gpa}. Exploring the advantages of various choices of basis is deferred to future work.

\subsection{Statistically isotropic correlations}\label{sec:isotropic_correlations}

The statistically isotropic correlations between $\mu$-binned and integrated light cone moments can conveniently be expressed in terms of a set of angular auto- and cross-spectra that depend on $\ell$ and the window labels only. Let's consider two fields $\mathrm{F}$ and $\mathrm{G}$ on the light cone, constructed from underlying 4-dimensional fields $\mathrm{U}^\mathrm{F}$ and $\mathrm{U}^\mathrm{G}$ as described in above, and integrated on the line of sight with windows $W$ and $W'$ respectively. The cross-spectra is:
\begin{eqnarray}\label{Cl_int1}
C_{\ell}^{F^WG^{W'}} &\equiv& \left\langle \mathrm{F}^{W}_{\ell m} \mathrm{G}^{W'}_{\ell m}  \right\rangle \nonumber \\
&=&  \int d \chi_{1} d \chi_{2} \ W (\chi_{1}) W' (\chi_{2}) \Big{\langle} \mathrm{F}_{\ell m} (\chi_{1})  \mathrm{G}_{\ell m} (\chi_{2})\Big{\rangle} \nonumber \\
&=& \int d \chi_{1} d \chi_{2} \ W \left(\chi_{1}\right) W' \left(\chi_{2}\right) \int \frac{k^{2} d k}{(2\pi)^3} \ \mathcal{K}^{F}_{\ell}(\chi_1,k)\, \mathcal{K}^{G}_{\ell}(\chi_2,k) \ P_{FG}(\chi_1,\chi_2,k), 
\end{eqnarray}
where we have assumed a statistically isotropic cross-correlation power spectrum between the underlying fields:
\begin{equation}
\left\langle  \tilde{\mathrm{U}}^{F} (\eta(\chi_1), \mathbf{k})^*  \tilde{\mathrm{U}}^{G} (\eta(\chi_2), \mathbf{k}')\right\rangle = (2\pi)^3 P_{FG}(\chi_1,\chi_2,k) \delta^{(3)}(\mathbf{k}-\mathbf{k'}).
\end{equation}

Although a brute force computation of the integrals in Eq.~\ref{Cl_int1} is feasible for certain values of $\ell$ and certain $\chi$ ranges, the oscillatory behaviour of the integral kernels make such an approach cumbersome if accuracy across a wide range of multipole moments and redshifts is desired. This is exactly our case, as we aim to have consistent modelling of large-angle and small-angle observables across a large redshift range. The Limber approximation (see e.g. \cite{LoVerde:2008re}) can be used to simplify the oscillatory integrals and provide accurate spectra under certain circumstances. For our purposes, an implementation of the Limber approximation is challenged by several factors: first, part of our calculations require narrow window functions, which can drive the Limber approximation beyond its regime of validity if the multipole $\ell$ is not high enough. Second, the Limber approximation only picks up the equal-time contribution to the cross-correlation power spectra ($\chi_1 = \chi_2$), and does not capture non-negligible contributions from unequal-time correlations~\cite{Kitching:2016xcl}. Here we adopt the `Beyond Limber approximation' method from \cite{Fang:2019xat}, which separates Eq.~\ref{Cl_int1} into a term suitable for the Limber approximation and a term with separable structure that allows for fast Bessel integrations. We briefly summarize this method in Appendix \ref{sec:beyond_limber}.

\subsection{Statistically anisotropic cross-correlations}

We now discuss our modelling for anisotropic cross correlations between the temperature field and a windowed density tracer on the light cone. We write the observed temperature field as the sum of two contributions:
\begin{equation}\label{temp}
\Theta (\hat{n}) = I(\hat{n}) +  \int d\chi \ M (\chi,\hat{n}) B(\chi, \hat{n}) \,,
\end{equation}
where the first term $I(\hat{n})$, analogous e.g. to the primary CMB, represents all the contributions to the temperature coming from integrated light cone fields:
\begin{equation}
I(\hat{n}) = \sum_a \int d\chi \,W_a(\chi)\, \mathrm{F}_a(\hat{n}, \chi),
\end{equation}
and the second term, analogous to various CMB secondaries such as kSZ, consists of the line of sight integration of the product of two light cone fields. Consider as well a tracer of large scale structure obtained as a line of sight integration of a density field $\delta(\chi, \hat{n})$ on the light cone:
\begin{equation}
 \delta^{W}(\hat{n}) = \int d\chi \,W(\chi)\, \delta(\hat{n}, \chi).
\end{equation}
We assume that $I(\hat{n})$, $M (\chi,\hat{n})$, $G(\chi,\hat{n})$ and $\delta^{W}(\hat{n})$ are isotropically correlated among each other as described in Sec.~\ref{sec:isotropic_correlations}. 

The second term in Eq.~\ref{temp} leads to a statistical anisotropy when the temperature is correlated with the density field:
\begin{equation}\label{anis0}
\Big{\langle}\Theta_{\ell m}\;\delta^{W}_{\ell'm'}\Big{\rangle} = 
(-1)^m C^{I\delta^{W}}_{\ell}\delta_{\ell\ell'}\delta_{mm'}+
\sum_{\ell_1 m_1}
\sum_{\ell_2 m_2} 
(-1)^m
W_{m_{1}, m_{2},-m}^{\ell_{1}, \ell_{2}, \ell}
\int d\chi \ \Big{\langle}{M}_{\ell_1 m_1}(\chi)\;{B}_{\ell_2 m_2}(\chi)\delta^{W}_{\ell'm'}\Big{\rangle}\,,
\end{equation}
where $W_{m_{a}, m_{b},m_{c}}^{\ell_{a}, \ell_{b}, \ell_{c}}$ is an angular mode coupling matrix containing Wigner 3-j symbols; $M_{\ell_1 m_1}(\chi)$ and $B_{\ell_1 m_1}(\chi)$ are the LC moments of the fields $M(\chi, \hat{n})$ and $B(\chi, \hat{n})$. The type of mode coupling matrices that we consider in this paper all have the following structure:
\begin{equation}
W_{m_{a}, m_{b},m_{c}}^{\ell_{a}, \ell_{b}, \ell_{c}}=\omega(\ell_a,\ell_b,\ell_c)\sqrt{\frac{\left(2 \ell_{a}+1\right)\left(2 \ell_{b}+1\right)\left(2 \ell_{c}+1\right)}{4 \pi}}\left(\begin{array}{ccc}
\ell_{a} & \ell_{b} & \ell_{c} \\
0 & 0 & 0
\end{array}\right)\left(\begin{array}{ccc}
\ell_{a} & \ell_{b} & \ell_{c} \\
m_{a} & m_{b} & m_{c}
\end{array}\right)\,.
\end{equation}
The three-point function in Eq.~\ref{anis0} can be approximated in the limit that two of the $(\ell_1,\ell_2,\ell)$ multipoles are far greater than the remaining multipole (squeezed limit). In the squeezed limit where $\ell_1\ll \ell, \ell_2$, we consider the field ${M}_{\ell_1 m_1}(\chi)$ to be deterministic, and therefore approximate:
\begin{equation}\label{eq:3point}
\Big{\langle}{M}_{\ell_1 m_1}(\chi)\,{B}_{\ell_2 m_2}(\chi)\,\delta^{W}_{\ell'm'}\Big{\rangle}
\approx\;
{M}_{\ell_1 m_1}(\chi)\;\Big{\langle}{B}_{\ell_2 m_2}(\chi)\,\delta^{W}_{\ell'm'}\Big{\rangle}.
\end{equation}
Inserting the expression above into Eq.~\ref{anis0} and expanding  $
{M}_{\ell_1 m_1}(\chi)$ and ${B}_{\ell_2 m_2}(\chi)$ using the Haar basis gives:
\begin{equation}\label{anis1}
\Big{\langle}\Theta_{\ell m}\;\delta^{W}_{\ell'm'}\Big{\rangle} = 
(-1)^m C^{I\delta^{W}}_{\ell}\delta_{\ell\ell'}\delta_{mm'}+
\sum_{s=0}^{\infty} \sum_{\ell_1 m_1}
(-1)^{m_1}
\begin{pmatrix}
\ell & \ell' & \ell_1 \\
m & m' & - m_1
\end{pmatrix}
f^{M^sW}_{\ell\ell_1\ell'} \,M^s_{\ell_1 m_1}
\end{equation}
where we have the defined the coupling:
\begin{equation}
f^{M^sW}_{\ell \ell_1 \ell'} \equiv \omega(\ell,\ell_1,\ell')\sqrt{\frac{(2\ell+1)(2\ell_1+1)(2\ell'+1)}{4\pi}} 
\begin{pmatrix}
\ell & \ell' & \ell_1 \\
0 & 0 & 0
\end{pmatrix} \;C^{B^s\delta^{W}}_{\ell'} .
\end{equation}
Eq.~\ref{anis1} tells us that the statistical anisotropy in the small angular scale temperature-density cross-correlation $\ell, \ell '\gg 1$ is modulated by the large angular scale field $M$. More precisely, each separate Haar-binned LC moment $M^s_{\ell_1 m_1}$ sources an independent statistically anisotropic term. 

As discussed in Sec.~\ref{sec:int_coarse_fields}, it is possible to use a mixed $\Pi$-binned and Haar-binned expansion by choosing a truncation value $s = N-1$:
\begin{eqnarray}\label{anis2}
\Big{\langle}\Theta_{\ell m}\;\delta^{W}_{\ell'm'}\Big{\rangle} =
(-1)^m C^{I\delta^{W}}_{\ell}\delta_{\ell\ell'}\delta_{mm'}
&+&
\sum_{\alpha=0}^{N-1}\sum_{\ell_1 m_1}
(-1)^{m_1}
\begin{pmatrix}
\ell & \ell' & \ell_1 \\
m & m' & - m_1
\end{pmatrix}
f^{M^\alpha W}_{\ell \ell_1 \ell'}\,M^{\alpha}_{\ell_1 m_1} \nonumber \\
&+&
\sum_{s = N}^{\infty}\sum_{\ell_1 m_1}
(-1)^{m_1}
\begin{pmatrix}
\ell & \ell' & \ell_1 \\
m & m' & - m_1
\end{pmatrix}
f^{M^sW}_{\ell \ell_1 \ell'}\,M^s_{\ell_1 m_1}  \\ \nonumber
\end{eqnarray}
where 
\begin{equation}\label{fcoupling}
f^{M^\alpha W}_{\ell \ell_1 \ell'} \equiv \omega(\ell, \ell_1, \ell')\sqrt{\frac{(2\ell+1)(2\ell_1+1)(2\ell'+1)}{4\pi}} 
\begin{pmatrix}
\ell & \ell' & \ell_1 \\
0 & 0 & 0
\end{pmatrix} \;C^{B^{\alpha}\delta^{W}}_{\ell'}\Delta\chi.
\end{equation}
Eq.~\ref{anis2} tells us that some of the statistical anisotropy comes from the 'coarse', 'bulk' radial modes $M^{\alpha}_{\ell_1 m_1}$ and the rest of the statistical anisotropy comes from the `fine', small-scale modes $M^{s}_{\ell_1 m_1}$ that don't contribute to the bulk averages. Our formalism builds upon previous literature~\cite{Deutsch:2017ybc} in which the contribution from fine modes on the light cone has not been considered and the statistical anisotropy is approximated as only sourced by the bulk modes. This can be a good approximation if $N$ is high enough (how high depends on the radial profile of $M_{\ell_1 m_1}(\chi)$); in this paper we will keep these terms and quantify their relevance in the modelling of the temperature-density statistical anisotropy. 

We can add more generality to our modelling of the temperature-density cross-correlation by adding additional contributions to the temperature signal:
\begin{equation}\label{crosstemp2}
\Theta (\hat{n}) = I(\hat{n}) +  \int d\chi \ M (\chi,\hat{n}) B(\chi, \hat{n}) 
+  \int d\chi \ Q (\chi,\hat{n}) D(\chi, \hat{n}) +\dots
\end{equation}
which translate to:
\begin{eqnarray}\label{anis3}
\Big{\langle}\Theta_{\ell m}\;\delta^{W}_{\ell'm'}\Big{\rangle} =
(-1)^m C^{I\delta^{W}}_{\ell}\delta_{\ell\ell'}\delta_{mm'}
&+&
\sum_{\alpha=0}^{N-1}\sum_{\ell_1 m_1}
(-1)^{m_1}
\begin{pmatrix}
\ell & \ell' & \ell_1 \\
m & m' & - m_1
\end{pmatrix}
f^{M^\alpha W}_{\ell \ell_1 \ell'}\,M^{\alpha}_{\ell_1 m_1} \nonumber \\
&+&
\sum_{s = N}^{\infty}\sum_{\ell_1 m_1}
(-1)^{m_1}
\begin{pmatrix}
\ell & \ell' & \ell_1 \\
m & m' & - m_1
\end{pmatrix}
f^{M^s W}_{\ell \ell_1 \ell'}\,M^s_{\ell_1 m_1} \nonumber  \\ 
&+&
\sum_{\alpha=0}^{N-1}\sum_{\ell_1 m_1}
(-1)^{m_1}
\begin{pmatrix}
\ell & \ell' & \ell_1 \\
m & m' & - m_1
\end{pmatrix}
f^{Q^\alpha W}_{\ell \ell_1 \ell'}\,Q^{\alpha}_{\ell_1 m_1} \nonumber \\
&+&
\sum_{s = N}^{\infty}\sum_{\ell_1 m_1}
(-1)^{m_1}
\begin{pmatrix}
\ell & \ell' & \ell_1 \\
m & m' & - m_1
\end{pmatrix}
f^{Q^s W}_{\ell \ell_1 \ell'}\,Q^s_{\ell_1 m_1} \nonumber \\ 
&+&
\dots\,. \\ \nonumber
\end{eqnarray}

\subsection{Quadratic estimator}\label{sec:estimator}

In this subsection we discuss how the $\Pi$-binned modes appearing in the statistical anisotropy Eq.~\ref{anis3} can be estimated by constructing appropriately weighted sums of products of temperature and density multipoles. The most general case discussed in the previous subsection included statistical anisotropies sourced by the $\Pi$-binned and Haar-binned LC moments of a series of modulating fields $(\,M(\hat{n},\chi)\,, \,Q(\hat{n},\chi)\,,\, \dots\,)$. We will first consider the simplified case in which there is only one modulating field $M(\hat{n},\chi)$ and a single $\Pi$ mode $\alpha$ and derive the quadratic estimator. After that, we progressively add layers of complexity until we reach the most general case. 

\subsubsection{Case 1: Single modulating field and single $\Pi$ mode}

Let's start by considering the simple case in which there is only one modulating bulk mode $M^{\alpha}_{\ell_1 m_1} $ sourcing the statistical anisotropy in the temperature-density cross-correlation. Starting from Eq.~\ref{temp}, we write the temperature multipoles as:
\begin{equation}\label{eq:tempcase1}
\Theta_{\ell m} = I_{\ell m}+
\sum_{\ell_1 m_1}
\sum_{\ell_2 m_2} 
(-1)^m
W_{m_{1}, m_{2},-m}^{\ell_{1}, \ell_{2}, \ell}
{M}^{\alpha}_{\ell_1 m_1}\;{B}^{\alpha}_{\ell_2 m_2}\Delta\chi\,,
\end{equation}
and write the temperature cross-correlation as:
\begin{equation}\label{anisotropic2}
\Big{\langle}\Theta_{\ell m}\;\delta^{W}_{\ell'm'}\Big{\rangle} = 
(-1)^m C^{I\delta^{W}}_{\ell}\delta_{\ell\ell'}\delta_{mm'} +
\sum_{\ell_1 m_1} (-1)^{m_1} 
\begin{pmatrix}
\ell & \ell' & \ell_1 \\
m & m' & - m_1
\end{pmatrix}
f^{M^\alpha W}_{\ell\ell_1\ell'} \, M^{\alpha}_{\ell_1 m_1} \,.
\end{equation}

Let's construct a quadratic sum of temperature and density multipoles with the following structure 
\begin{equation}\label{estimator0}
\hat{M}^{\alpha}_{LM} = A^{M^\alpha}_L \sum_{\ell m; \ell' m'} (-1)^M 
\begin{pmatrix}
\ell & \ell' & L \\
m & m' & -M
\end{pmatrix}
G^{M^\alpha W}_{\ell \ell' L}\ \Theta_{\ell m} \delta^{W}_{\ell'm'}\,,
\end{equation}
and choose weights $G^{M^\alpha W}_{\ell \ell' L}$ such that the estimator is unbiased:
\begin{equation}
\Big{\langle}\hat{M}^{\alpha}_{LM}\Big{\rangle} = M^{\alpha}_{LM}\,,
\end{equation}
and has minimum variance. The first condition translates to
\begin{equation}\label{constraint1}
A^{M^\alpha}_L=(2 L+1)\left(\sum_{\ell ; \ell^{\prime}} G^{M^ \alpha W}_{\ell \ell' L}f^{M^\alpha W}_{\ell L\ell'}\right)^{-1}.
\end{equation}
The minimum variance estimator can be found using the Lagrange multiplier method subject to the constraint Eq.~\ref{constraint1}, which gives:
\begin{equation}\label{gminvar}
G^{M^\alpha W}_{\ell \ell' L} \equiv \frac{C_{\ell' }^{\Theta\Theta} {C}_{\ell}^{\delta^{W}\delta^{W}} f^{M^\alpha W}_{\ell L\ell'} - (-1)^{\ell+\ell'+L} {C}_{\ell}^{I\delta^{W}} {C}_{\ell'}^{I\delta^{W}}f^{M^\alpha W}_{\ell' L\ell}}
{C_{\ell }^{\Theta\Theta}C_{\ell' }^{\Theta\Theta} {C}_{\ell}^{\delta^{W}\delta^{W}} {C}_{\ell'}^{\delta^{W}\delta^{W}} - {\Big{(} {C}_{\ell}^{I\delta^{W}} \Big{)}}^2 {\Big{(} {C}_{\ell'}^{I\delta^{W}} \Big{)}^2}}\,,
\end{equation}
where $C_{\ell }^{\Theta \Theta}$ is the full temperature power spectrum. A complete derivation can be found in Appendix~\ref{sec:quad_est_deriv}. In the computation of the estimator variance used to derive $G^{M^\alpha W}_{\ell \ell' L}$ we have only included the disconnected part of the temperature-galaxy-temperature-galaxy four-point function. Under this approximation, the estimator variance is given by:
\begin{equation}\label{eq:est_var_sig_noise}
\left\langle \hat{M}^{\alpha}_{L'M'} {\hat{M}^{\alpha *}}_{LM}\right\rangle
= {C}_{L}^{{M}^{\alpha}{M}^{\alpha}}+A^{M^\alpha}_L\,.
\end{equation}
However, looking in more detail at the estimator variance:
\begin{eqnarray}\nonumber
\left\langle \hat{M}^{\alpha}_{LM} {\hat{M}^{\alpha *}}_{L'M'}\right\rangle &=& 
A^{M^\alpha}_L A^{M^\alpha}_{L'} \sum_{\ell_a m_a; \ell_b m_b} 
\;\sum_{\ell_c m_c; \ell_d m_d} (-1)^{M+M'} 
\begin{pmatrix}
\ell_a & \ell_b & L \\
m_a & m_b & -M
\end{pmatrix}
\begin{pmatrix}
\ell_c & \ell_d & L' \\
m_c & m_d & -M'
\end{pmatrix} \\ \nonumber
&\;& G^{M^\alpha W}_{\ell_a \ell_b L}\;G^{M^\alpha W}_{\ell_c \ell_d L'}\left\langle  \Theta_{\ell_a m_a} \delta^{W}_{\ell_b m_b}
\Theta^*_{\ell_c m_c} \delta^{W*}_{\ell_d m_d}\right\rangle \\ \nonumber
&=&A^{M^\alpha}_L A^{M^\alpha}_{L'} \sum_{\ell_a m_a; \ell_b m_b} 
\;\sum_{\ell_c m_c; \ell_d m_d} (-1)^{M+M'} 
\begin{pmatrix}
\ell_a & \ell_b & L \\
m_a & m_b & -M
\end{pmatrix}
\begin{pmatrix}
\ell_c & \ell_d & L' \\
m_c & m_d & -M'
\end{pmatrix} \\ \nonumber
&\;& G^{M^\alpha W}_{\ell_a \ell_b L}\;G^{M^\alpha W}_{\ell_c \ell_d L'}\Bigg{[}\left\langle I_{\ell_a m_a} \delta^{W}_{\ell_b m_b}
 I^{*}_{\ell_c m_c} \delta^{W*}_{\ell_d m_d}\right\rangle
 +
 (\Delta\chi)^2\sum_{\ell_1 m_1}
\sum_{\ell_2 m_2} 
 \sum_{\ell_1' m_1'}
\sum_{\ell_2' m_2'} \\ 
&\,&
(-1)^{m_a+m_c}
W_{m_{1}, m_{2},-m_a}^{\ell_{1}, \ell_{2}, \ell_a}
W_{m_{1}', m_{2}',-m_c}^{\ell_{1}', \ell_{2}', \ell_c}
\left\langle{M}^{\alpha}_{\ell_1 m_1}\;{B}^{\alpha}_{\ell_2 m_2}\delta^{W}_{\ell_b m_b}
{M}^{\alpha*}_{\ell_1' m_1'}\;{B}^{\alpha*}_{\ell_2' m_2'}\delta^{W*}_{\ell_d m_d}\right\rangle\Bigg{]}\,,\\ \nonumber
\end{eqnarray}
we see that there is a dependence on a six-point function of the underlying fields. Therefore, even if all the fields are Gaussian, the disconnected four-point function is not a complete description of the estimator variance -- one must in principle include the 15 terms that contribute to the disconnected six-point function. Fortunately, as we describe in Appendix~\ref{sec:variance_appendix}, for the observables in this paper the relevant components of the six-point function do not yield any significant additional variance beyond the terms in Eq.~\ref{eq:est_var_sig_noise}. This additional contribution to the variance was computed for kSZ tomography in the box formalism in Ref.~\cite{Giri:2020pkk}, where in analogy with lensing reconstruction, it was referred to as the $N^{(1)}$ bias and was found to be negigably small. For non-Gaussian fields, one must additionally compute the connected part of the six-point function. This was also computed in Ref.~\cite{Giri:2020pkk}, where it was shown that this `$N^{(3/2)}$ bias' is far larger than the $N^{(1)}$ bias, and can even become comparable to $A_L^{M^\alpha}$ at sufficiently high SNR. A full computation of these additional contributions to the variance within the light cone picture will appear in future work. Moving forward, we will only consider the contribution from ${C}_{L}^{{M}^{\alpha}{M}^{\alpha}}$ and $A_L^{M^\alpha}$ in our estimator variance.

\subsubsection{Case 2: Single modulating field and multiple $\Pi$ modes}

The next step to add more realism is to have multiple $\Pi$-binned LC moments from a single modulating field. Recall that the number of $\Pi$ bins corresponds to the degree of coarse graining in the line of sight integral for the temperature: 
\begin{equation}\label{eq:tempcase2}
\Theta_{\ell m} = I_{\ell m}+
\sum_{\alpha = 0}^{N-1}
\sum_{\ell_1 m_1}
\sum_{\ell_2 m_2} 
(-1)^m
W_{m_{1}, m_{2},-m}^{\ell_{1}, \ell_{2}, \ell}
{M}^{\alpha}_{\ell_1 m_1}\;{B}^{\alpha}_{\ell_2 m_2}\Delta\chi\,,
\end{equation}
and
\begin{eqnarray}\label{anisotropic2.2}
\Big{\langle}\Theta_{\ell m}\;\delta^{W}_{\ell'm'}\Big{\rangle} = 
(-1)^m C^{I\delta^{W}}_{\ell}\delta_{\ell\ell'}\delta_{mm'}  &+&
\sum_{\alpha = 0}^{N-1}
\sum_{\ell_1 m_1} (-1)^{m_1} 
\begin{pmatrix}
\ell & \ell' & \ell_1 \\
m & m' & - m_1
\end{pmatrix}
f^{M^\alpha W}_{\ell\ell_1\ell'} \, M^{\alpha}_{\ell_1 m_1}\,. \\ \nonumber
\end{eqnarray}
We want to construct $N$ unbiased quadratic estimators, one for each 
modulating source $M^{\alpha}_{\ell_1 m_1}$. The strategy we choose is to first construct $N$ \emph{biased} estimators, by taking $N$ versions of the single field estimator Eq.~\ref{estimator0} described in Case 1:
\begin{eqnarray}
\hat{M}^{0}_{LM} &=& A^{M^{0}}_L \sum_{\ell m; \ell' m'} (-1)^M 
\begin{pmatrix}
\ell & \ell' & L \\
m & m' & -M
\end{pmatrix}
G^{M^0 W}_{\ell \ell' L}\ \Theta_{\ell m} \delta^{W}_{\ell'm'}\,, \nonumber \\
\vdots&\,& \nonumber \\
\hat{M}^{N-1}_{LM} &=& A^{M^{N-1}}_L \sum_{\ell m; \ell' m'} (-1)^M 
\begin{pmatrix}
\ell & \ell' & L \\
m & m' & -M
\end{pmatrix}
G^{M^{N-1} W}_{\ell \ell' L}\ \Theta_{\ell m} \delta^{W}_{\ell'm'}\,, \\ \nonumber
\end{eqnarray}
where the weights $\;A^{M^{\alpha}}_L, G^{M^\alpha,W}_{\ell \ell' L}\,$ are chosen exactly as if $M^{\alpha}_{\ell_1 m_1}$ was the only source of statistical anisotropy. These estimators will be biased:
\begin{eqnarray}\label{biasest1}
\Big{\langle}\hat{M}^{0}_{LM}\Big{\rangle} &=&\;M^{0}_{L M}\;+\sum_{\alpha \neq 0}\; \hat{M}^{\alpha}_{LM}\;\frac{A^{M^{0}}_L}{2L+1} \sum_{\ell\ell'}
G^{M^0 W}_{\ell \ell' L}f^{M^\alpha W}_{\ell L \ell'} \,,\\
\vdots&\,& \nonumber \\
\Big{\langle}\hat{M}^{N-1}_{LM}\Big{\rangle} &=&\;M^{N-1}_{L M}\;+\sum_{\alpha \neq N-1}\; \hat{M}^{\alpha}_{LM}\;\frac{A^{M^{N-1}}_L}{2L+1} \sum_{\ell\ell'}
G^{M^{N-1} W}_{\ell \ell' L}f^{M^\alpha W}_{\ell L \ell'}\,. \nonumber \\ 
\end{eqnarray}
We can define a `rotation matrix'':
\begin{equation}\label{rot}
(\mathrm{\bf{R}}_L)^{XY} \equiv \frac{ \sum_{\ell\ell'}G^{XW}_{\ell \ell' L}f^{YW}_{\ell L \ell'} }{ \sum_{\ell\ell'}G^{XW}_{\ell \ell' L}f^{X W}_{\ell L \ell'}} \,,
\end{equation}
with indices $X,Y$ in $(M^{0},\dots,M^{N-1})$. With this matrix we write the system Eqs.~\ref{biasest1} as:
\begin{equation}
\Big{\langle}\hat{\mathbf{M}}_{LM}\Big{\rangle}  = \mathrm{\bf{R}}_L\cdot\mathbf{M}_{LM},
\end{equation}
where $\hat{\mathbf{M}}_{LM} = (\hat{M}^{0}_{LM},\dots,\hat{M}^{N-1}_{LM})$ and $\mathbf{M}_{LM} = (M^{\alpha}_{L M},\dots,M^{N-1}_{L M})$. If the rotation matrix is invertible, we can now define unbiased estimators $\mathbf{\tilde{M}}_{LM} $ for $\mathbf{M}_{LM}$:
\begin{equation}
\mathbf{\tilde{M}}_{LM}  \equiv \mathrm{\bf{R}}_L^{-1} \cdot \hat{\mathbf{M}}_{LM}\,.
\end{equation}
The procedure above serves as an example of \emph{bias hardening} the quadratic estimators~\cite{Namikawa_2013} in the presence of multiple sources of statistical anisotropy in the temperature-density cross-correlation\footnote{It is important to highlight that these unbiased estimators won't necessarily be the minimum variance estimators for $\mathbf{M}_{LM}$. It is possible to construct an unbiased and minimum variance estimator by correlating linear combinations of $\delta^W$ maps with $\Theta$. However, as we demonstrate below, the simpler approach taken here of deriving separate estimators and then rotating yields very good results in practice. Exploring such new estimators is deferred to future work.}. 

The two point function for the unbiased estimator is:
\begin{equation}
\Big{\langle}\mathbf{\tilde{M}}_{LM}\mathbf{\tilde{M}}^{\dagger}_{LM}\Big{\rangle}  = \mathrm{\bf{R}^{-1}}_L\cdot\Big{\langle}\mathbf{\hat{M}}_{LM}\mathbf{\hat{M}}^{\dagger}_{LM}\Big{\rangle}\cdot(\mathrm{\bf{R}^{-1}}_L)^{\dagger}\,,
\end{equation}
where $\Big{\langle}\mathbf{\hat{M}}_{LM}\mathbf{\hat{M}}^{\dagger}_{LM}\Big{\rangle}$ is the
two point function of the biased estimator; note that this is a matrix containing all auto- and cross-spectra. Similarly to Case 1, the biased two point function can be written in terms of 4-point and 6-point functions and broken down into signal and noise terms:

\begin{eqnarray}\label{eq:general2point}\nonumber
\left\langle (\hat{\mathbf{M}}_{LM})^X ({\hat{\mathbf{M}}^{*}}_{L'M'})^Y \right\rangle \!\!&=& 
A^{X}_L A^{Y}_{L'} \sum_{\ell_a m_a; \ell_b m_b} 
\;\sum_{\ell_c m_c; \ell_d m_d} (-1)^{M+M'} 
\begin{pmatrix}
\ell_a & \ell_b & L \\
m_a & m_b & -M
\end{pmatrix}
\begin{pmatrix}
\ell_c & \ell_d & L' \\
m_c & m_d & -M'
\end{pmatrix} \\ \nonumber
&\;&\times G^{X W}_{\ell_a \ell_b L}\;G^{Y W}_{\ell_c \ell_d L'}\left\langle  \Theta_{\ell_a m_a} \delta^{W}_{\ell_b m_b}
\Theta^*_{\ell_c m_c} \delta^{W*}_{\ell_d m_d}\right\rangle \\ \nonumber
&=&A^{X}_L A^{Y}_{L'} \sum_{\ell_a m_a; \ell_b m_b} 
\;\sum_{\ell_c m_c; \ell_d m_d} (-1)^{M+M'} 
\begin{pmatrix}
\ell_a & \ell_b & L \\
m_a & m_b & -M
\end{pmatrix}
\begin{pmatrix}
\ell_c & \ell_d & L' \\
m_c & m_d & -M'
\end{pmatrix} \\ \nonumber
&\;&\times G^{X W}_{\ell_a \ell_b L}\;G^{Y W}_{\ell_c \ell_d L'}\Bigg{[}\left\langle I_{\ell_a m_a} \delta^{W}_{\ell_b m_b}
 I^{*}_{\ell_c m_c} \delta^{W*}_{\ell_d m_d}\right\rangle
 +
 \sum_{\ell_1 m_1}
\sum_{\ell_2 m_2} 
 \sum_{\ell_1' m_1'}
\sum_{\ell_2' m_2'} \\ \nonumber
&\,&
(-1)^{m_a+m_c}
W_{m_{1}, m_{2},-m_a}^{\ell_{1}, \ell_{2}, \ell_a}
W_{m_{1}', m_{2}',-m_c}^{\ell_{1}', \ell_{2}', \ell_c}
\left\langle \left(\mathbf{X}_{\ell_1 m_1}\cdot \mathbf{Y}_{\ell_2 m_2}\right)\delta^{W}_{\ell_b m_b}
\left(\mathbf{X}^{*}_{\ell_1' m_1'}\cdot \mathbf{Y}^{*}_{\ell_2' m_2'}\right)\delta^{W*}_{\ell_d m_d}\right\rangle\Bigg{]}\,,
\end{eqnarray}
where $X,Y$ are indices in $(M^{0},\dots,M^{N-1})$ and $\mathbf{X}_{\ell m}, \mathbf{Y}_{\ell m}$ are the vectors:
\begin{equation}\label{eq:aux_vector_1}
\mathbf{X}_{\ell m} = (M^{0}_{\ell m},\dots,M^{N-1}_{\ell m})\,,
\end{equation}
\begin{equation}\label{eq:aux_vector_2}
\mathbf{Y}_{\ell m} = (\Delta\chi B^{0}_{\ell m},\dots,\Delta\chi B^{N-1}_{\ell m})\,.
\end{equation}
The dominant terms of the 2-point function of the biased estimator are:
\begin{equation}\label{eq:twopointcase2}
\Big{\langle}\mathbf{\hat{M}}_{LM}\mathbf{\hat{M}}^{\dagger}_{LM}\Big{\rangle} =
\mathrm{\bf{R}}_L\mathrm{\bf{C}}^{MM}_{L}(\mathrm{\bf{R}}_L)^{\dagger}+
\mathrm{\bf{N}}^{0}_{L}+\dots
\end{equation}
where $\mathrm{\bf{C}}^{MM}_{L}$ is the modulating field covariance matrix and the elements of the $\mathrm{\bf{N}}^{0}_{L}$ matrix are given by:
\begin{equation}\label{eq:N0definition}
(\mathrm{\bf{N}}^{0}_{L})^{\alpha\beta} = \frac{A^{M^{\alpha}}_LA^{M^{\beta}}_L}{2L+1}
\sum_{\ell_1\ell_2}G^{M^\alpha W}_{\ell_1 \ell_2 L}\Big{[}\;
G^{M^\beta W}_{\ell_1 \ell_2 L}C_{\ell_1}^{\Theta\Theta}C_{\ell_2}^{\delta^{W}\delta^{W}}
+(-1)^{\ell_1+\ell_2+L}G^{M^\beta W}_{\ell_2 \ell_1 L} C_{\ell_1}^{\Theta\delta^{W}}C_{\ell_2}^{\Theta\delta^{W}}\Big{]}\,.
\end{equation}
It is easy to check that the diagonal elements satisfy $(\mathrm{\bf{N}}^{0}_{L})^{\alpha\alpha} = A^{M^{\alpha}}$. This is to be expected because we constructed the unbiased estimators as collections of the Case 1 estimator. Eq.~\ref{eq:twopointcase2} has further contributions from the 6-point function: some relatively simple terms proportional to the covariances $\mathrm{\bf{C}}^{BB}_{L}$,$\mathrm{\bf{C}}^{\delta^W\delta^W}_{L}$,$\mathrm{\bf{C}}^{\delta^W B}_{L}$, etc., and more complicated terms coming from various contractions of the 6-point function. Again, these contributions are expected to be small enough to be neglected. 

\subsubsection{Case 3: Single modulating field, multiple $\Pi$ and Haar modes}

The estimators constructed in Case 2 ignore the contributions to the temperature multipoles and temperature-density statistical anisotropy coming from the Haar-binned LC moments of the field $M(\hat{n},\chi)$. We have 
\begin{eqnarray}\label{eq:tempcase3}
\Theta_{\ell m} = I_{\ell m}&+&
\sum_{\alpha = 0}^{N-1}
\sum_{\ell_1 m_1}
\sum_{\ell_2 m_2} 
(-1)^m
W_{m_{1}, m_{2},-m}^{\ell_{1}, \ell_{2}, \ell}
{M}^{\alpha}_{\ell_1 m_1}\;{B}^{\alpha}_{\ell_2 m_2}\Delta\chi \\ \nonumber
&+&\sum_{s = 0}^{s_{max}}
\sum_{\ell_1 m_1}
\sum_{\ell_2 m_2} 
(-1)^m
W_{m_{1}, m_{2},-m}^{\ell_{1}, \ell_{2}, \ell}
{M}^{s}_{\ell_1 m_1}\;{B}^{s}_{\ell_2 m_2}\,,
\end{eqnarray}
and
\begin{eqnarray}\label{anisotropic2.3}
\Big{\langle}\Theta_{\ell m}\;\delta^{W}_{\ell'm'}\Big{\rangle} = 
(-1)^m C^{I\delta^{W}}_{\ell}\delta_{\ell\ell'}\delta_{mm'}  &+&
\sum_{\alpha = 0}^{N-1}
\sum_{\ell_1 m_1} (-1)^{m_1} 
\begin{pmatrix}
\ell & \ell' & \ell_1 \\
m & m' & - m_1
\end{pmatrix}
f^{M^\alpha W}_{\ell\ell_1\ell'} \, M^{\alpha}_{\ell_1 m_1}\nonumber  \\
 &+&
\sum_{s = N}^{s_{max}}
\sum_{\ell_1 m_1} (-1)^{m_1} 
\begin{pmatrix}
\ell & \ell' & \ell_1 \\
m & m' & - m_1
\end{pmatrix}
f^{M^s W}_{\ell\ell_1\ell'} \, M^{s}_{\ell_1 m_1} \,, 
\end{eqnarray}
where $k_{max}$ is chosen such that the Haar expansion is mostly converged.
If we follow the same steps as in Case 2 to construct $N$ biased estimators, the Haar modes lead to an additional bias:

\begin{eqnarray}
\Big{\langle}\hat{M}^{0}_{LM}\Big{\rangle}\Big{|}_{Case\;3} &\rightarrow& \Big{\langle}\hat{M}^{0}_{LM}\Big{\rangle}\Big{|}_{Case\;2}
+\sum_{s= N}^{s_{max}}\; {M}^{s}_{LM}\;\frac{A^{M^{0}}_L}{2L+1} \sum_{\ell\ell'}
G^{M^0 W}_{\ell \ell' L}f^{M^s W}_{\ell\ell_2\ell'}\,,
\nonumber\\
\vdots&\,& \nonumber \\
\Big{\langle}\hat{M}^{N-1}_{LM}\Big{\rangle}\Big{|}_{Case\;3} &\rightarrow&\Big{\langle}\hat{M}^{N-1}_{LM}\Big{\rangle}\Big{|}_{Case\;2}
+\sum_{s=N}^{s_{max}}\; {M}^{s}_{LM}\;\frac{A^{M^{N-1}}_L}{2L+1} \sum_{\ell\ell'}
G^{M^{N-1} W}_{\ell \ell' L}f^{M^s W}_{\ell\ell_2\ell'}\,.\\ \nonumber
\end{eqnarray}
The relevance of this bias depends on the truncation number $N$, which in principle can be chosen to be high enough such that the contribution from Haar modes can be ignored. Thus, quantifying the size of these terms as a function of $N$ is a useful way of determining the level of coarse graining that we need to model our observables with. 

The 2-point function of the estimator can be computed using the same expression Eq.~\ref{eq:general2point} as in Case 2 just by expanding the 
vectors $\mathbf{X},\mathbf{Y}$  defined in Eqs.~\ref{eq:aux_vector_1}-\ref{eq:aux_vector_2}:
\begin{equation}
\mathbf{X}_{\ell m} = (M^{0}_{\ell m},\dots,M^{N-1}_{\ell m},
M^{s=N}_{\ell m},\dots,M^{s=s_{max}}_{\ell m})\,,
\end{equation}
\begin{equation}
\mathbf{Y}_{\ell m} = (\Delta\chi B^{0}_{\ell m},\dots,\Delta\chi B^{N-1}_{\ell m},
B^{s=N}_{\ell m},\dots,B^{s=s_{max}}_{\ell m}).
\end{equation}
 The resulting 2-point function contains three terms which we identify as dominant:
\begin{equation}\label{eq:twopointcase3}
\Big{\langle}\mathbf{\hat{M}}_{LM}\mathbf{\hat{M}}^{\dagger}_{LM}\Big{\rangle} =
\mathrm{\bf{R}}_L\mathrm{\bf{C}}^{MM}_{L}(\mathrm{\bf{R}}_L)^{\dagger}+
\mathrm{\bf{N}}^{0}_{L}+\mathrm{\bf{N}}^{MM\;fine}_{L}+\dots
\end{equation}
where $\mathrm{\bf{R}}_L$ is the rotation matrix defined in Eq.~\ref{rot}, $\mathrm{\bf{C}}^{MM}_{L}$ is the modulating field covariance matrix, $\mathrm{\bf{N}}^{0}_{L}$ is computed exactly as in Case 2, and $\mathrm{\bf{N}}^{MM\;fine}_{L}$ is given by:
\begin{equation}\label{eq:finemodenoisedef}
\left(\mathrm{\bf{N}}^{MM\;fine}_{L}\right)^{\alpha\beta} = \sum_{s, s' =N}^{k_{max}} \left\langle{M}^{s}_{LM}{M}^{s'*}_{LM}\right\rangle  \left(\sum_{\ell_1 \ell_2} \frac{A^{M^\alpha}_L}{2L+1} G^{M^\alpha W}_{\ell_1 \ell_2 L} f_{\ell_1 L \ell_2}^{M^s W}  \right) \left(\sum_{\ell_1 \ell_2} \frac{A^{M^\beta}_L}{2L+1} G^{M^\beta W}_{\ell_1 \ell_2 L} f_{\ell_1 L \ell_2}^{M^{s'} W}  \right) \,.
\end{equation}
We call this term the fine mode noise, as it is sourced by the Haar modes of the modulating field above the truncation number $N$. $\mathrm{\bf{N}}^{MM\;fine}_{L}$ can become comparable to $\mathrm{\bf{N}}^{0}_{L}$ for low enough $N$. Conversely, one can find a high enough truncation number $N$ such that the fine mode noise can be neglected. In a realistic scenario, the truncation number is limited by the details of the 3-dimensional large scale structure survey that is being used for the reconstruction. In further sections we will show the size of the fine mode noise in the estimation of the radial velocity and transverse velocity $\Pi$-binned LC moments.

\subsubsection{Case 4: Multiple modulating fields, multiple $\Pi$ and Haar modes}

Generalizing the results from the previous cases to the multiple field case is straightforward. The temperature and the statistical anisotropy are:
\begin{eqnarray}\label{eq:tempcase4}
\Theta_{\ell m} = I_{\ell m}&+&
\sum_{\alpha = 0}^{N-1}
\sum_{\ell_1 m_1}
\sum_{\ell_2 m_2} 
(-1)^m
W_{m_{1}, m_{2},-m}^{\ell_{1}, \ell_{2}, \ell}
{M}^{\alpha}_{\ell_1 m_1}\;{B}^{\alpha}_{\ell_2 m_2}\Delta\chi \\
&+&\sum_{s = N}^{s_{max}}
\sum_{\ell_1 m_1}
\sum_{\ell_2 m_2} 
(-1)^m
W_{m_{1}, m_{2},-m}^{\ell_{1}, \ell_{2}, \ell}
{M}^{s}_{\ell_1 m_1}\;{B}^{s}_{\ell_2 m_2} \nonumber \\ 
&+&\sum_{\alpha = 0}^{N-1}
\sum_{\ell_1 m_1}
\sum_{\ell_2 m_2} 
(-1)^m
W_{m_{1}, m_{2},-m}^{\ell_{1}, \ell_{2}, \ell}
{Q}^{\alpha}_{\ell_1 m_1}\;{D}^{\alpha}_{\ell_2 m_2}\Delta\chi \nonumber\\
&+&\sum_{s = N}^{s_{max}}
\sum_{\ell_1 m_1}
\sum_{\ell_2 m_2} 
(-1)^m
W_{m_{1}, m_{2},-m}^{\ell_{1}, \ell_{2}, \ell}
{Q}^{s}_{\ell_1 m_1}\;{D}^{s}_{\ell_2 m_2} \nonumber\\
&\vdots& \nonumber
\end{eqnarray}
and
\begin{eqnarray}\label{anisotropic2.4}
\Big{\langle}\Theta_{\ell m}\;\delta^{W}_{\ell'm'}\Big{\rangle} =
(-1)^m C^{I\delta^{W}}_{\ell}\delta_{\ell\ell'}\delta_{mm'}
&+&
\sum_{\alpha=0}^{N-1}\sum_{\ell_1 m_1}
(-1)^{m_1}
\begin{pmatrix}
\ell & \ell' & \ell_1 \\
m & m' & - m_1
\end{pmatrix}
f^{M^\alpha W}_{\ell \ell_1 \ell'}\,M^{\alpha}_{\ell_1 m_1}  \\
&+&
\sum_{s=0}^{s_{max}}\sum_{\ell_1 m_1}
(-1)^{m_1}
\begin{pmatrix}
\ell & \ell' & \ell_1 \\
m & m' & - m_1
\end{pmatrix}
f^{M^s W}_{\ell \ell_1 \ell'}\,M^{s}_{\ell_1 m_1} \nonumber \\
&+&
\sum_{\alpha=0}^{N-1}\sum_{\ell_1 m_1}
(-1)^{m_1}
\begin{pmatrix}
\ell & \ell' & \ell_1 \\
m & m' & - m_1
\end{pmatrix}
f^{Q^\alpha W}_{\ell \ell_1 \ell'}\,Q^{\alpha}_{\ell_1 m_1} \nonumber \\
&+&
\sum_{s=0}^{s_{max}}\sum_{\ell_1 m_1}
(-1)^{m_1}
\begin{pmatrix}
\ell & \ell' & \ell_1 \\
m & m' & - m_1
\end{pmatrix}
f^{Q^s W}_{\ell \ell_1 \ell'}\,Q^{s}_{\ell_1 m_1} \nonumber \\
&\vdots& \nonumber
\end{eqnarray}

Similar to the previous cases, we can construct N biased estimators for the $\Pi$-binned LC moments of the $M$ field. The 2-point function is calculated using Eq.~\ref{eq:general2point} as in Case 2 just by expanding the vectors $\mathbf{X}_{\ell m},\mathbf{Y}_{\ell m}$ to:
\begin{equation}
\mathbf{X}_{\ell m} = (M^{0}_{\ell m},\dots,M^{N-1}_{\ell m},
M^{k=N}_{\ell m},\dots,M^{k=k_{max}}_{\ell m}, Q^{0}_{\ell m},\dots,Q^{N-1}_{\ell m},
Q^{k=N}_{\ell m},\dots,Q^{k=k_{max}}_{\ell m}, \dots)\,,
\end{equation}
\begin{equation}
\mathbf{Y}_{\ell m} = (\Delta\chi B^{0}_{\ell m},\dots,\Delta\chi B^{N-1}_{\ell m},
B^{k=N}_{\ell m},\dots,B^{k=k_{max}}_{\ell m}, \Delta\chi D^{0}_{\ell m},\dots,\Delta\chi D^{N-1}_{\ell m},
D^{k=N}_{\ell m},\dots,D^{k=k_{max}}_{\ell m}, \dots)\,.
\end{equation}
The number of terms contributing to the 2-point function will clearly increase with the introduction of new modulating fields. In addition to the terms from Case 3, each new modulating field Q will introduce a term $\mathrm{\bf{N}}^{QQ}_{L}$ given by:
\begin{equation}\label{eq:otherfieldbiasdef}
\left(\mathrm{\bf{N}}^{QQ}_{L}\right)^{\alpha\beta} = \sum_{\alpha', \beta' =0}^{N-1} \left\langle{Q}^{\alpha'}_{LM}{Q}^{\beta'*}_{LM}\right\rangle  \left(\sum_{\ell_1 \ell_2} \frac{A^{M^\alpha}_L}{2L+1} G^{M^\alpha W}_{\ell_1 \ell_2 L} f_{\ell_1 L \ell_2}^{Q^{\alpha'} W}  \right) \left(\sum_{\ell_1 \ell_2} \frac{A^{M^\beta}_L}{2L+1} G^{M^\beta W}_{\ell_1 \ell_2 L} f_{\ell_1 L \ell_2}^{Q^{\beta'} W}  \right),
\end{equation}
as well as other less significant terms with a similar structure. In principle, there is no immediate way to determine if $\mathrm{\bf{N}}^{QQ}_{L}$ are negligible with respect to $\mathrm{\bf{N}}^{0}_{L}$ and $\mathrm{\bf{N}}^{MM\;fine}_{L}$, as this depends on the specific details of the modulating fields giving rise to the temperature signal. We show examples in Sec.~\ref{sec:recon_analysis},
where we compute the reconstruction noise for the radial and transverse velocity.

\subsubsection{Multiple density windows}\label{sec:multdenswindow}

We have thus far discussed the construction of estimators for $\Pi$-binned LC moments of a light cone field $M(\hat{n},\chi)$ using only one window window function $W(\chi)$ for the density tracer. We remind the reader that the window function shows up in the coupling Eq.~\ref{fcoupling} through the cross-spectra $C^{B^{\alpha}\delta^{W}}_{\ell'} \equiv \left\langle \mathrm{B}^{\alpha}_{\ell m} \delta^{W}_{\ell m}  \right\rangle$, where $B(\hat{n},\chi)$ is the field integrated together with $M(\hat{n},\chi)$ to form a temperature secondary. Since the estimator relies on large multipoles ($\ell, \ell'$), where correlations along the light cone are relatively small, the cross-spectra will be non-negligible only if the window function $W(\chi)$ overlaps with $\Pi^{\beta}(\chi)$. A density window function with wide support on the light cone will lead to well defined couplings and estimators for all the $\Pi$-bins, but at the same time leads to an increased mixing of the 3-dimensional information we are trying to reconstruct. In contrast, more localized density window functions will be better at isolating contributions coming from different redshifts, but can lead to ill-defined estimators for $\Pi$-bins with zero overlap with the density window. One can remedy this last issue by constructing estimators with a numer of density window functions such that $\Pi^{\alpha}(\chi)$ and $W^{\alpha}(\chi)$ overlap:
\begin{eqnarray}\label{eq:manybinharmonicestimator}
\hat{M}^{0}_{LM} &=& A^{M^{0}W^0}_L \sum_{\ell m; \ell' m'} (-1)^M 
\begin{pmatrix}
\ell & \ell' & L \\
m & m' & -M
\end{pmatrix}
G^{M^0 W^0}_{\ell \ell' L}\ \Theta_{\ell m} \delta^{W^0}_{\ell'm'} \nonumber \\
\vdots&\,& \nonumber \\
\hat{M}^{N-1}_{LM} &=& A^{M^{N-1}W^{N-1}}_L \sum_{\ell m; \ell' m'} (-1)^M 
\begin{pmatrix}
\ell & \ell' & L \\
m & m' & -M
\end{pmatrix}
G^{M^{N-1} W^{N-1}}_{\ell \ell' L}\ \Theta_{\ell m} \delta^{W^{N-1}}_{\ell'm'}\,, \\ \nonumber
\end{eqnarray}
One intuitive choice is to take $W^{\alpha}(\chi) \equiv \Pi^{\alpha}(\chi)$, which could be a possibility if one has 3 dimensional measurements of the large scale structure that can be separated into custom redshift bins \footnote{Surveys with big redshift errors can make this separation more difficult.}. The rotation matrix is defined similarly to Eq.~\ref{rot}, where the only difference comes from changing the window functions to match that used in the estimators:
\begin{equation}\label{rot2}
(\mathrm{\bf{R}}_L)^{XY} \equiv \frac{ \sum_{\ell\ell'}G^{XW^{X}}_{\ell \ell' L}f^{YW^Y}_{\ell L\ell'} }{ \sum_{\ell\ell'}G^{XW^X}_{\ell \ell' L}f^{X W^X}_{\ell L\ell'}} \,,
\end{equation}
where $W^X$ is the density window function associated to the observable $X$. The 2-point function introduced in Case 2 can be easily generalized to include the varying density window functions:
\begin{eqnarray}\label{eq:general2point_expanded}
\left\langle \hat{\mathbf{M}}^X_{LM} {\hat{\mathbf{M}}^{Y*}}_{L'M'}\right\rangle &=& 
A^{XW^X}_L A^{YW^Y}_{L'} \sum_{\ell_a m_a; \ell_b m_b} 
\;\sum_{\ell_c m_c; \ell_d m_d} (-1)^{M+M'} 
\begin{pmatrix}
\ell_a & \ell_b & L \\
m_a & m_b & -M
\end{pmatrix}
\begin{pmatrix}
\ell_c & \ell_d & L' \\
m_c & m_d & -M'
\end{pmatrix} \\ \nonumber
&\;& G^{X W^X}_{\ell_a \ell_b L}\;G^{Y W^Y}_{\ell_c \ell_d L'}\left\langle  \Theta_{\ell_a m_a} \delta^{W^X}_{\ell_b m_b}
\Theta^*_{\ell_c m_c} \delta^{W^{Y}*}_{\ell_d m_d}\right\rangle \\ \nonumber
&=&A^{XW^X}_L A^{YW^Y}_{L'} \sum_{\ell_a m_a; \ell_b m_b} 
\;\sum_{\ell_c m_c; \ell_d m_d} (-1)^{M+M'} 
\begin{pmatrix}
\ell_a & \ell_b & L \\
m_a & m_b & -M
\end{pmatrix}
\begin{pmatrix}
\ell_c & \ell_d & L' \\
m_c & m_d & -M'
\end{pmatrix} \\ \nonumber
&\;& G^{X W^X}_{\ell_a \ell_b L}\;G^{Y W^Y}_{\ell_c \ell_d L'}\Bigg{[}\left\langle I_{\ell_a m_a} \delta^{W^X}_{\ell_b m_b}
 I^{*}_{\ell_c m_c} \delta^{W^Y*}_{\ell_d m_d}\right\rangle
 +
\sum_{\ell_1 m_1}
\sum_{\ell_2 m_2} 
 \sum_{\ell_1' m_1'}
\sum_{\ell_2' m_2'} \\ \nonumber
&\,&
(-1)^{m_a+m_c}
W_{m_{1}, m_{2},-m_a}^{\ell_{1}, \ell_{2}, \ell_a}
W_{m_{1}', m_{2}',-m_c}^{\ell_{1}', \ell_{2}', \ell_c}
\left\langle\left( \mathbf{X}_{\ell_1 m_1}\cdot \mathbf{Y}_{\ell_2 m_2}\right)\delta^{W^X}_{\ell_b m_b}
\left(\mathbf{X}^{*}_{\ell_1' m_1'}\cdot \mathbf{Y}^{*}_{\ell_2' m_2'}\right)\delta^{W^Y*}_{\ell_d m_d}\right\rangle\Bigg{]}\,. \nonumber
\end{eqnarray}
The advantage of using multiple localized density window functions over fewer, wider ones, is clear: more 3-dimensional information of the light cone fields is retained. In this paper we will explore a case with many window functions and a case with a single broad window function to compare these two scenarios.

\subsubsection{Principal component analysis}\label{subsec:principal}
 
Consider a set of estimators for the $\Pi$-binned LC moments of a field $M(\hat{n},\chi)$, constructed used the methods described above. The variance of the estimator is given by:
\begin{equation}
\Big{\langle}\mathbf{{M}}_{LM}\mathbf{{M}}^{*}_{LM}\Big{\rangle}  = \mathrm{\bf{R}}_L\mathrm{\bf{C}}^{MM}_{L}(\mathrm{\bf{R}}_L)^{\dagger}+\mathrm{\bf{N}}\,,
\end{equation}
where $\mathrm{\bf{N}}$ is the sum of all sources of noise. Although the $\Pi$ basis is useful when it comes to separation of scales and localization on the light cone, it can be less useful when it comes to separating the independent information contained in the 2-point function of the estimator. Using a principal component analysis, we can find the uncorrelated linear combinations of bins that yield the highest signal to noise. We do so by the following set of transformations:
\begin{itemize}
    \item {Transform to a basis in which the noise matrix is diagonal.}
    \item {Perform a second transformation to a basis in which the noise matrix is the Identity.}
    \item{Perform a third transformation to a basis in which the signal matrix is diagonal. The noise matrix, due to being equal to the identity, is unchanged by the third transformation. The resulting signal matrix $\mathrm{\bf{C}}^{pp}_{L}$ is diagonal and contains the signal to noise for the different uncorrelated principal components.}
\end{itemize}
The linear combinations of bins associated to the principal components can be found using the transformation matrices $\mathrm{\bf{T_1}},\mathrm{\bf{T_2}},\mathrm{\bf{T_3}},\mathrm{\bf{R}}_L$:
\begin{equation}
\mathrm{\bf{X^p}} = \mathrm{\bf{T_3}\cdot\bf{T_2}\cdot\bf{T_1}\cdot\mathrm{\bf{R}}_L\cdot\bf{X}}\,,
\end{equation}
where $\bf{X} =$ $(M^{0}_{LM},\dots,M^{N}_{LM})$. The j-th
principal component then is characterized by a set of $N$ coefficients $c_L^{j\beta}$ such that:
\begin{equation}\label{eq:principal_coeff}
(\mathrm{\bf{X^p}})_{LM}^{j} = \sum_{\beta}c_L^{j\beta}M^{\beta}_{LM}\,,
\end{equation}
The signal to noise per mode for the j-th principal component is simply given by the diagonal element $(\mathrm{\bf{C}}^{pp}_{L})^{jj}$. We define a signal to noise per harmonic $LM$ mode as 
\begin{equation}\label{eq:SNRpermode}
SN_{LM} = \sum_{j} (\mathrm{\bf{C}}^{pp})_{L}^{jj}.
\end{equation}
We define the total signal to noise as a sum over all principal components and harmonic modes:
\begin{equation}\label{eq:totSN}
SN_{tot} = \sum_{j}\sum_{L}(2L+1)(\mathrm{\bf{C}}^{pp})_{L}^{jj}
\end{equation}

\subsubsection{Multiplicative bias from theory modelling}\label{sec:mult_bias}

In order to construct the quadratic estimators, one has to assume a model for the couplings Eq.~\ref{fcoupling}, which depend on $C^{B^{\alpha}\delta^W}_{\ell'}(\chi)$. If an incorrect $\tilde{C}^{B^{\alpha}\delta^W}_{\ell'}$ is used instead of the true physical $C^{B^{\alpha}\delta^W}_{\ell'}$, a multiplicative bias will be introduced:
\begin{equation}
\left\langle \hat{\mathbf{M}}^X_{LM}\right\rangle = \sum_{Y}\mathrm{\Gamma^{XY}_L}\left(\mathrm{\bf{\tilde{R}}}_L\right)^{XY}\mathbf{M}^Y_{LM}+\tilde{\Delta}\,,
\end{equation}
where $\mathrm{\bf{\tilde{R}}}_L$ is the rotation Eq.~\ref{rot2}, $\tilde{\Delta}$ is the reconstruction noise, and $\mathrm{\Gamma^{XY}_L}$ is the multiplicative bias:
\begin{equation}\label{eq:gamma}
\mathrm{\Gamma^{XY}_L} =  \frac{ \sum_{\ell\ell'}\tilde{G}^{XW^{X}}_{\ell \ell' L}f^{YW^Y}_{\ell L\ell'} }{ \sum_{\ell\ell'}\tilde{G}^{XW^X}_{\ell \ell' L}\tilde{f}^{Y W^Y}_{\ell L\ell'}} \,,
\end{equation}
where $f^{Y W^Y}_{\ell L\ell'}$ is the true physical coupling. The bias Eq.~\ref{eq:gamma} is not symmetric in the indices $X,Y$ so in principle there are $N^2$ bias parameters at each scale $L$. In the context of kSZ velocity reconstruction, where the $B$-field is the optical depth, this multiplicative factor is commonly referred as the optical depth bias; see e.g. Refs.~\cite{Battaglia:2016xbi,Smith:2018bpn,Giri:2020pkk,Madhavacheril:2019buy}. As discussed in more detail in Sec.~\ref{opticaldepthbiasksz}, the bias does not depend on the scale $L$ over the relevant range for reconstruction, leaving a total of $N^2$ bias parameters to account for. Note that in the absence of off-diagonal terms in the rotation matrix Eq.~\ref{rot2} (or if these terms are very small), there would only be $N$ bias parameters. This is the assumption that has been made in previous literature utilizing the light cone picture to forecast cosmological constraints, e.g. Refs.~\cite{Cayuso:2019hen,Contreras_2019,Pan:2019dax}. We comment on this assumption and the general problem of mitigating the optical depth bias in Sec.~\ref{opticaldepthbiasksz}.

\section{Modeling of observables}\label{sec:observables}

In this paper, we will be interested in statistically anisotropic correlations between various contributions to the observed CMB and a tracer of large scale structure (LSS). Our goal is to use such statistical anisotropies to reconstruct (on large angular scales) a set of modulating fields -- here, our focus is on the radial and transverse velocity fields. Our prototype tracer is a photometric galaxy redshift survey, as considered in e.g. Refs.~\cite{Terrana2016,Deutsch:2017ybc,Smith:2018bpn,Munchmeyer:2018eey}; other tracers such as spectroscopic surveys~\cite{Smith:2018bpn,Munchmeyer:2018eey}, the Cosmic Infrared Background (CIB)~\cite{McCarthy:2019xwk}, or line-intensity maps~\cite{CosmicVisions21cm:2018rfq,Sato-Polito:2020cil} are other interesting candidates. Throughout the paper, we assume a fiducial cosmological model consistent with Planck 2018~\cite{Aghanim:2018eyx}; in particular, we set: $\{10^9 A_s = 2.2, n_s = 0.965, \Omega_m = 0.31, \Omega_b = 0.049, H_0 = 68 \ {\rm km \ s^{-1} \ Mpc^{-1}}, \tau = 0.06 \}$. There is no strong dependence on cosmological parameters for any of our conclusions.

When necessary, we present relations in the Newtonian gauge, where at late-times when we can neglect anisotropic stress; the metric is:
\begin{equation}
ds^2 = a(\eta)^2 \left( -\left[ 1+ 2 \Psi(\eta,\mathbf{x}) \right] d\eta^2 + \left[ 1- 2 \Psi(\eta,\mathbf{x}) \right] d\mathbf{x}^2 \right)\,.
\end{equation}
Because our Halo Model code calculates perturbations in the synchronous gauge, it is sometimes necessary to relate Newtonian gauge quantities to synchronous gauge ones. For late-times, simple relations can be written for the Newtonian gauge gravitational potential $\Psi$ and the peculiar velocity field $\mathbf{v}$ of dark matter in terms of the synchronous gauge dark matter perturbations in Fourier space:
\begin{equation}
\Psi(\eta,\mathbf{k}) = -\frac{3\Omega_m H_0^2}{a^2(\eta)k^2}\delta^{(sync)}_m(\eta,\mathbf{k}) \,,
\end{equation}
\begin{equation}
\mathbf{v}(\eta,\mathbf{k}) = i\frac{\mathbf{k}}{k^2}f(\eta)H(\eta)a(\eta)\delta^{(sync)}_m(\eta,\mathbf{k})\,,
\end{equation}
where $H$ is the Hubble rate and $f$ is the growth rate, defined as  $\frac{a}{D}\frac{dD}{da}$, with $D(a)$ the linear theory growth factor of dark matter perturbations. Given that perturbations of the galaxy and electron fields are only needed on small scales for the reconstruction procedure, we approximate $\delta_g^{(Newt)}\approx\delta_g^{(sync)}$ and  $\delta_e^{(Newt)}\approx\delta_e^{(sync)}$. Unless stated otherwise, we work in natural units with $\hbar=c=G_N=1$.

\subsection{Constructing observables}

All of the observables presented below are constructed from a set of fundamental cosmological fields, which we compute using linear cosmological perturbation theory and the Halo Model for large scale structure. Combining Eqs.~\ref{eq:lcmoments3} and \ref{eq:int_moment}, a field on the light cone can be characterized by specifying a window function $W (\chi)$, an integral kernel $\mathcal{K}^{F}_{\ell}(\chi,k)$, and an underlying perturbation field in Fourier space $\tilde{\mathrm{U}}^{F} (\eta(\chi), \mathbf{k})$ :
\begin{equation}\label{eq:buildingfunctions}
\mathrm{F}^{W}_{\ell m} \equiv \int d\chi \ W (\chi)\,\int \frac{d^{3} \mathbf{k}}{(2 \pi)^{3}}\,
\mathcal{K}_{\ell}(\chi,k)\,
\tilde{\mathrm{U}}^{F} (\eta(\chi), \mathbf{k}) \,Y_{\ell m}^{*}(\hat{k}).
\end{equation}
Below, we refer to these functions as the `building functions' and we specify them for each observable we construct.

\subsection{The CMB}

The de-beamed CMB temperature measured in a frequency band $\nu$ through an instrument with an isotropic beam $(B^{\Theta}_{\ell})^\nu$ and noise $n_{\ell m}^{\nu}$ has contributions from a variety of sources. As a baseline model, we take:
\begin{equation}\label{eq:CMB_contributions}
\Theta^\nu_{\ell m} =  \Theta_{\ell m}^{pCMB} + \Theta_{\ell m}^{ISW,lin}+ \Theta_{\ell m}^{ML}  + \Theta_{\ell m}^{kSZ} +  \Theta_{\ell m}^{ReI} + \Theta_{\ell m}^{L} + (\Theta_{\ell m}^{XG})^\nu + (\Theta_{\ell m}^{G})^\nu + n_{\ell m}^{\nu} / (B^{\Theta}_{\ell})^\nu\,.
\end{equation}
There are blackbody contributions including: $\Theta_{\ell m}^{pCMB}$ which contains the Sachs Wolfe (SW), Doppler, and early Integrated Sachs Wolfe (ISW) contributions to the primary CMB, $\Theta_{\ell m}^{ISW,lin}$ the linear contribution to the late-time ISW component, $\Theta_{\ell m}^{ML}$ the non-linear contribution to the late-time ISW component (here referred to as the moving lens effect), $\Theta_{\ell m}^{kSZ}$ the late-time kSZ, $\Theta_{\ell m}^{ReI}$ the reionization kSZ, and $\Theta_{\ell m}^{L}$ the lensing contribution to the primary CMB. There are frequency-dependent extragalactic contributions $(\Theta_{\ell m}^{XG})^\nu$, whose dominant components for the experimental configurations considered below include the CIB and the thermal Sunayev Zel'dovich effect (tSZ). Finally, there is a frequency-dependent galactic component $(\Theta_{\ell m}^{G})^\nu$. 

Below we describe in detail the components which have significant cross-correlation with late-time tracers of LSS, since such components must be computed in a self-consistent way. The SW, Doppler, and early ISW contributions $\Theta_{\ell m}^{pCMB}$ to the primary CMB do not contribute to the cross correlation with tracers of LSS; we compute their power spectra using CAMB~\cite{Lewis:1999bs}. The reionization kSZ component $\Theta_{\ell m}^{ReI}$ is modelled as a Gaussian field with power spectrum $\ell^2 C_\ell^{rei} / 2\pi = 1 \mu {\rm K}^2$. If higher redshift tracers of LSS are considered, then the reionization kSZ can be used to reconstruct the radial velocity field as described in Ref.~\cite{Hotinli:2020csk}; in this paper, we focus on tracers of LSS that do not have any significant cross-correlation with reionization kSZ. For the fiducial CMB experiments considered below, including reionization kSZ does not affect any of our results, and we therefore neglect this contribution in our analysis. Galactic foregrounds on the small angular scales relevant to velocity reconstruction are generally sub-dominant to extragalactic foregrounds on a line of sight away from the galactic plane (see e.g. Ref.~\cite{ACT:2020frw}). We assume that regions with significant galactic contamination can be masked. Other than considering the effect of a mask, we therefore neglect galactic foregrounds. We model instrumental noise $n_{\ell m}^{\nu}$ as a frequency-dependent constant and the beam $(B^{\Theta}_{\ell})^\nu$ as a Gaussian with a frequency-dependent Full Width at Half-Maximum (FWHM) $\theta^\nu_{\rm FWHM}$. Our fiducial CMB experiment is consistent with the properties of the Simons Observatory Large Area Telescope~\cite{Ade:2018sbj}, with:
\begin{equation}
n_{\ell m}^{\nu} = N_{\rm red} \left( \frac{\ell}{1000} \right)^{-3.5} + N_{\rm white}\,,
\end{equation}
where $N_{\rm red}$ describes the level of $1/f$ atmospheric noise and $N_{\rm white}$ describes the sensitivity of the frequency band. The frequencies, beam, and noise levels we use in our analysis below are collected in Table~\ref{tab:SOnoise}. We assume an observation time of 5 years when computing the level of $1/f$ noise, and choose the `baseline' values for noise found in~\cite{Ade:2018sbj}.

\begin{table}
\begin{center}
\begin{tabular}{ |c||c|c|c|  }
 \hline
Freq. [GHz] & FWHM [arcmin] & $N_{\rm white}$ [$\mu K$-arcmin] & $N_{\rm red}$ [$\mu K^2 \ s$]\\
 \hline
27 & 7.4 & 71 & 100 \\ 
39 & 5.1 & 36 & 39 \\ 
93 & 2.2 & 8 & 230 \\ 
145 & 1.4 & 10 & 1,500 \\ 
225 & 1.0 & 22 & 17,000 \\ 
280 & 0.9 & 54 & 31,000 \\ 
 \hline
\end{tabular}
\end{center}
\caption{CMB experimental noise parameters used for our fiducial CMB experiment, consistent with Simons Observatory Large Area Telescope~\cite{Ade:2018sbj}}
\label{tab:SOnoise}
\end{table}

\begin{figure}[h]
  \includegraphics[scale=0.4]{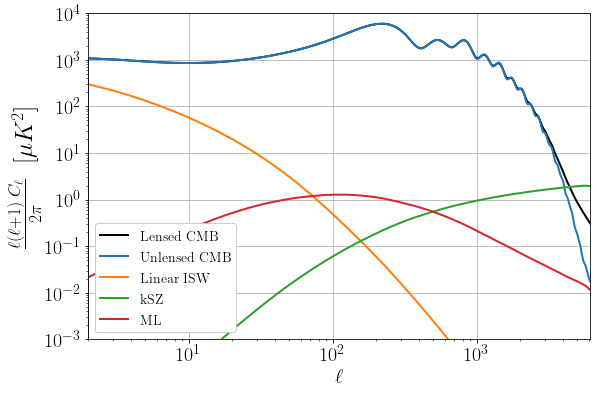}
  \caption{Contributions to blackbody CMB.}
  \label{fig:blackbodyCMB}
\end{figure}

\begin{figure}[h]
  \includegraphics[scale=0.4]{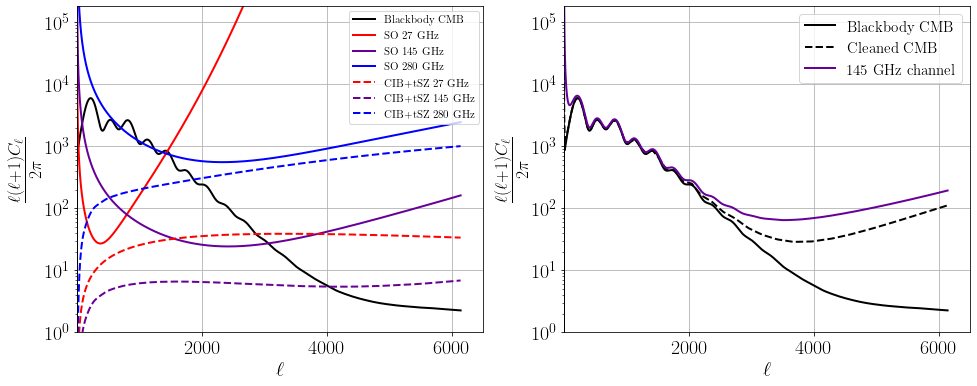}
  \caption{\textbf{Left panel:} Frequency dependent components of CMB compared to the blackbody component. Colored solid lines correspond to the de-beamed instrumental noise and dashed lines correspond to the CIB+tSZ contributions (including their cross-spectra). \textbf{Right panel:} the ILC-cleaned power spectrum compared to the blackbody component and the full 145 GHz channel.}
  \label{fig:cleanCMB}
\end{figure}

In Fig.~\ref{fig:blackbodyCMB}, we summarize the blackbody components of our CMB model: the primary CMB, late-time ISW, lensed CMB, kSZ, and ML. Each of these contributions is discussed in more detail in the following sub-sections. At low-$\ell$, the dominant components are the primary CMB and late-time ISW effects. Crucially, at high-$\ell$ ($\ell \agt 4000$), kSZ is the dominant blackbody component of the CMB. In the left panel of Fig.~\ref{fig:cleanCMB}, we show the frequency-dependent components of our CMB model, including extragalactic foregrounds and instrumental noise. In the right panel of Fig.~\ref{fig:cleanCMB}, we compare the effective noise obtained by using multifrequency information with the blackbody component of the CMB and with the noise and foregrounds in the `cleanest' channel (for velocity reconstruction) of our fiducial experiment, at 145 GHz. Note that the blackbody CMB dominates the noise and foregrounds below $\ell \alt 3000$. We assume that multifrequency information can be used to clean foregrounds using a standard harmonic space internal linear combination (ILC) procedure, described in more detail below. Such a procedure can reduce the level of noise and foregrounds by roughly a factor of 2 at high-$\ell$ when compared to the 145 GHz channel. Unless otherwise specified, in the analyses to follow we will use the ILC-cleaned CMB generated with the specifications in Table~\ref{tab:SOnoise}; we consider a maximum value of $\ell_{\rm max} = 6000$ which roughly corresponds to maps with a HEALPix~\footnote{\texttt{https://healpix.sourceforge.io}} resolution NSIDE of 2048. We now describe in more detail how we model the various CMB components listed above.

\subsubsection{The kSZ effect}

The contribution to the CMB temperature from the late-time kSZ effect is:
\begin{equation}\label{eq:ksztemp}
\Theta^{kSZ} (\hat{n}) = \int_{0}^{\chi_{\rm max}} d\chi \ \dot{\tau} (\hat{n}, \chi) \sum_{m=-1}^{1} \Theta_{1m}(\hat{n},\chi) Y_{1m}(\hat{n})\,,
\end{equation}
where $\dot{\tau} (\hat{n}, \chi)$ is the differential optical depth and $\sum_{m=-1}^{1} \Theta_{1m}(\hat{n},\chi) Y_{1m}(\hat{n})$ is the remote dipole field -- the locally observed CMB dipole at points on our past light cone, projected along the line of sight. The dominant contribution to the remote dipole comes from the radial peculiar velocity of electrons denoted as $v$, and small corrections to the observed dipole come from the intrinsic dipole anisotropy of the CMB. In general, the correction from the intrinsic anisotropies can be safely neglected, only becoming significant when inspecting the correlations of the remote dipole field on ultra-large scales. For simplicity, we will only consider the dominant kinetic term in this paper, and use the terminology radial velocity in lieu of remote dipole. Henceforward, we approximate:
\begin{equation}
\sum_{m=-1}^{1} \Theta_{1m}(\hat{n},\chi) Y_{1m}(\hat{n}) \simeq v (\hat{n},\chi)
\end{equation}
The differential optical depth is
\begin{equation}
\dot{\tau} (\hat{n}, \chi) = - \sigma_T a(\chi) \bar{n}_e(\chi) (1 + \delta_e (\hat{n}, \chi))\,,
\end{equation}
where $\sigma_T$ is the Thompson cross section, $a(\chi)$ the scale factor, $\bar{n}_e(\chi)$ the average electron density, and $\delta_e (\hat{n}, \chi)$ the electron overdensity field. 

We will focus on the late-time kSZ effect here, where the limits of integration extend from the origin out to a radial comoving distance $\chi_{\rm max}$ after reionization ended. We assume a fiducial value of $\chi_{\rm max} = 8.1$ Gpc, which corresponds to a redshift $z_{\rm max} = 5$ in our fiducial cosmology. Computing the multipoles of the kSZ temperature anisotropies Eq.~\ref{eq:ksztemp} in terms of the Haar-binned LC moments, we have:
\begin{eqnarray}
\Theta^{kSZ}_{\ell m} = 
\sum_{\ell_1 m_1 ; \ell_2 m_2} (-1)^m  \sqrt{\frac{(2\ell+1)(2\ell_1+1)(2\ell_2+1)}{4\pi}} 
\begin{pmatrix}
\ell_1 & \ell_2 & \ell \\
0 & 0 & 0
\end{pmatrix}
\begin{pmatrix}
\ell_1 & \ell_2 & \ell \\
m_1 & m_2 & -m
\end{pmatrix}
\sum_{s=1}^{\infty} v_{\ell_1 m_1}^s  \dot{\tau}_{\ell_2 m_2}^s\,.
\end{eqnarray}
For the radial velocity we use the following building functions:
\begin{eqnarray}
W^{v^{s}}(\chi) &=& h^{s} (\chi ) \;\;\;\;\mbox{from  Eq.~}\ref{eq:Haar_modes}\,, \\
\mathcal{K}^{v^{s}}_{\ell}(\chi,k) &=& 
4\pi i^{\ell}\frac{f(\chi)H(\chi)a(\chi)}{(2\ell+1)k}[\ell j_{\ell-1}(k\chi)-(\ell+1)j_{\ell+1}]\,,\\
\tilde{\mathrm{U}}^{v^{s}} (\eta(\chi), \mathbf{k}) &=& \delta_m(\eta(\chi), \mathbf{k})\,,
\end{eqnarray}
where $H(\chi)$ is the Hubble rate and $f(\chi)$ is the growth rate, defined as  $\frac{a}{D}\frac{dD}{da}$, with $D(a(\chi))$ the linear theory growth factor of dark matter perturbations. For the differential optical depth we use the following building functions:
\begin{eqnarray}
W^{\taud^{s}}(\chi) &=& h^{s} (\chi ) \;\;\;\;\mbox{from  Eq.~}\ref{eq:Haar_modes}\,, \\
\mathcal{K}^{\taud^{s}}_{\ell}(\chi,k) &=& -4\pi i^{\ell}j_{\ell}(k\chi) \sigma_T a(\chi) \bar{n}_e(\chi)\,,
\\
\tilde{\mathrm{U}}^{\taud^{s}} (\eta(\chi), \mathbf{k}) &=& \delta_e(\eta(\chi), \mathbf{k})\,.
\end{eqnarray}

The kSZ temperature power spectrum is 
\begin{eqnarray}
C_\ell^{kSZ} &=& \sum_{\ell_1 m_1 ; \ell_2 m_2} \sum_{\ell_1' m_1' ; \ell_2' m_2'}  \sqrt{\frac{(2\ell+1)(2\ell_1+1)(2\ell_2+1)}{4\pi}}  \sqrt{\frac{(2\ell+1)(2\ell_1'+1)(2\ell_2'+1)}{4\pi}} \\
&\times&
\begin{pmatrix}
\ell_1 & \ell_2 & \ell \\
0 & 0 & 0
\end{pmatrix}
\begin{pmatrix}
\ell_1' & \ell_2' & \ell \\
0 & 0 & 0
\end{pmatrix}
\begin{pmatrix}
\ell_1 & \ell_2 & \ell \\
m_1 & m_2 & -m
\end{pmatrix}
\begin{pmatrix}
\ell_1' & \ell_2' & \ell \\
m_1' & m_2' & -m
\end{pmatrix}\nonumber \\
&\times&\sum_{s, s'=1}^{\infty} \langle (v_{\ell_1 m_1}^s)^*  (\dot{\tau}_{\ell_2 m_2}^s)^* v_{\ell_1' m_1'}^{s'} \dot{\tau}_{\ell_2' m_2'}^{s'} \rangle\,.
\end{eqnarray}
Keeping the disconnected parts of the four-point function only, the power spectrum is:
\begin{eqnarray}\label{eq:ksz_power}
C_\ell^{kSZ} &=& \sum_{\ell_1; \ell_2}  \frac{(2\ell_1+1)(2\ell_2+1)}{4\pi} 
\begin{pmatrix}
\ell_1 & \ell_2 & \ell \\
0 & 0 & 0
\end{pmatrix}^2
\sum_{s, s' =1}^{\infty} \left[ (C^{vv})_{\ell_1}^{ss'} (C^{\dot{\tau} \dot{\tau}} )_{\ell_2}^{ss'}+ (C^{v \dot{\tau}})_{\ell_1}^{ss'} (C^{v \dot{\tau}})_{\ell_2}^{ss'} \right]\,,
\end{eqnarray}
where $(C^{vv})_{\ell_1}^{ss'}$, $(C^{\dot{\tau} \dot{\tau}} )_{\ell_2}^{ss'}$, $(C^{v \dot{\tau}})_{\ell_1}^{ss'}$, and$(C^{v \dot{\tau}})_{\ell_2}^{ss'}$are calculated using the building functions and Eq.~\ref{Cl_int1}. Focusing on $\ell \gg 1$, the majority of the power will come from $\ell_1 \ll \ell$ where $\ell_2 \sim \ell$. In this regime, we also expect that there is little bin-bin correlation in the differential optical depth, so we can take $(C^{\dot{\tau} \dot{\tau}} )_{\ell_2}^{ss'} \simeq (C^{\dot{\tau} \dot{\tau}} )_{\ell_2}^{ss} \ \delta_{ss'}$. Finally, the first term in parentheses above will dominate the second on small angular scales. In this limit, the kSZ power can be approximated by:
\begin{eqnarray}\label{eq:asymptotic_ksz}
C_\ell^{kSZ} &\simeq& \sum_{s=1}^{s_{\rm max}}  \left[\sum_{\ell_1}  \frac{(2\ell_1+1)}{4\pi}   (C^{vv})_{\ell_1}^{ss}  \right]  (C^{\dot{\tau} \dot{\tau}} )_{\ell}^{ss} \\
&=& \sum_{s=1}^{s_{\rm max}} \langle \bar{v}^s (0)^2 \rangle (C^{\dot{\tau} \dot{\tau}} )_{\ell}^{ss}\,. 
\end{eqnarray}
Taking $s_{\rm max} \rightarrow \infty$, this is equivalent to the expression:
\begin{equation}\label{eq:ksz_continuum}
C_\ell^{kSZ} = \int d\chi \ \langle v (0,\chi)^2 \rangle \ \dot{\tau}(\chi)^2 \ P^{ee} (\frac{\ell}{\chi},\chi)\,,
\end{equation}
which is consistent with previous literature~\cite{Ma:2001xr}. 

\begin{figure}[h]
  \includegraphics[scale=0.5]{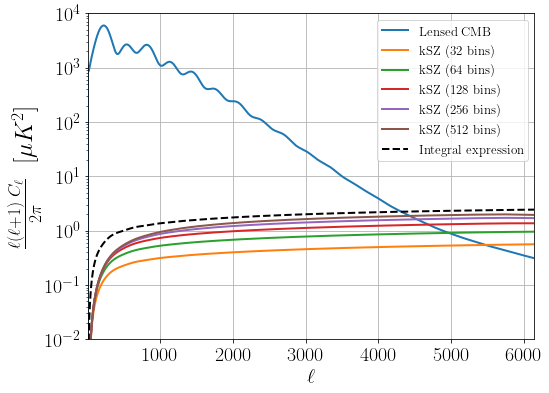}
  \caption{Convergence of kSZ power with number of bins.}
  \label{fig:kszN}
\end{figure}

In Fig.~\ref{fig:kszN} we show the coarse grained kSZ power spectrum for $\chi_{\rm max} = 8.1$ Gpc (corresponding to a redshift $z_{\rm max} = 5$ in our fiducial cosmology) with $s_{\rm max} = 32, 64, 128, 256$, and $512$ bins (corresponding to $\Delta\chi = 263, 131, 66, 33, 17$ Mpc). We compare with the continuum expression Eq.~\ref{eq:ksz_continuum}. It can be seen from this figure that $\sim 512$ bins, corresponding to a coarse graining scale of $\sim 17$ Mpc, is sufficient to capture the majority of the kSZ power. Based on this, we take 512 bins to correspond to the continuum limit below. 

\subsubsection{Late-time ISW (linear)}

Gravitational potentials that evolve in time induce a temperature anisotropy known as the integrated Sachs-Wolfe (ISW) effect. The late-time ISW contribution to the CMB is given by
\begin{equation}
\Theta_{\ell m}^{ISW} = -2 \int_{0}^{\chi_{\rm max}} d\chi \ {\frac{\partial\Psi_{\ell m}}{\partial\chi}}{}(\chi)\,, 
\end{equation}
where $\chi_{\rm max}$ is a fiducial maximum range in comoving distance large enough to capture the majority of the late-time decay of the potential due to the presence of a cosmological constant. The building functions (see Eq.~\ref{eq:buildingfunctions}) for the linear late-time ISW effect are:
\begin{eqnarray}
W^{ISW,lin}(\chi) &=&  \begin{cases}
1, & 0 \leq \chi <  \chi_{\rm max}, \\
0, & {\rm otherwise}
\end{cases}\\
\mathcal{K}^{ISW,lin}_{\ell}(\chi,k) &=& 4\pi i^{\ell}\frac{j_{\ell}(k\chi)}{k^2}\frac{3\Omega_m H_0^2}{a^2(\chi)}
\left(-\frac{da}{d\chi}(\chi)+a(\chi)\frac{d}{d\chi}\right)\,,\\
\tilde{\mathrm{U}}^{ISW,lin} (\eta(\chi), \mathbf{k}) &=& \delta^{(lin)}_m(\eta(\chi), \mathbf{k})\,,
\end{eqnarray}
 where $\delta^{(lin)}_m$ are the Fourier modes of the linear dark matter perturbations. The power spectrum of the linear late-time ISW is calculated using Eq.~\ref{Cl_int1}.

\subsubsection{Moving lens effect (non-linear ISW)}

In the non-linear regime, the ISW effect can be sourced on small-scales by the long-wavelength peculiar velocities of dark matter halos. Taking the limit $k'\ll k$, the non-linear evolution of the gravitational potential due to the coupling of small-wavelength density fluctuations and long wavelength velocity modes can be approximated as:
\begin{equation}
\dot{\Psi}_{\rm NL}(\boldsymbol{k})\sim i \boldsymbol{k}\Psi( \boldsymbol{k})\cdot\int\frac{{\rm d}^3k'}{(2\pi)^3}\mathbf{v}(\boldsymbol{k'})\,,
\end{equation}
which translates in real space to:
\begin{equation}
\dot{\Psi}_{\rm NL}(\eta,\mathbf{x}) \sim \nabla{\Psi}_{\rm NL}(\eta,\mathbf{x}) \cdot\mathbf{v}(\eta,\mathbf{x})\,.
\end{equation}
The ML effect is sourced by motions transverse to the line of sight:
\begin{equation}
\label{eq:ml_def}
\Theta^{\rm ML}(\hat{n}) \approx  2 \int_{0}^{\chi_{ls}} d \chi \ \frac{(\mathbf{\nabla}_\perp \Psi )}{\chi}(\hat{n},\chi) \cdot \mathbf{v}_{\perp}(\hat{n},\chi),
\end{equation}
 where $\mathbf{v}_\perp$ is the peculiar (comoving) transverse velocity and $\mathbf{\nabla}_\perp$ is the gradient on 2-sphere.  In this work, we assume that the large-scale velocity is pure-gradient, and therefore the transverse velocity component can be expressed as  $\mathbf{v}_\perp=\mathbf{\nabla}_\perp\Upsilon(\hat{n},\chi)$. We refer to $\Upsilon$ as the \textit{transverse velocity potential}. In spherical harmonics, the effect on the CMB temperature takes the form
\begin{equation}
\Theta^{\rm ML}_{\ell,m}=\int_{0}^{\chi_{ls}} d \chi\sum\limits_{\ell\ell'\,mm'}\Upsilon_{\ell' m'}(\chi)\psi_{\ell''m''}(\chi)\int{\rm d}^2\hat{n}\,Y_{\ell m}^*\nabla_i Y_{\ell'm'}\nabla^iY_{\ell'' m''}\,,
\end{equation}
where 
\begin{equation}
\psi_{\ell m}(\chi)\equiv 2 \frac{\Psi_{\ell m}(\chi)}{\chi}\,,
\end{equation}
and we will refer to this quantity as the \textit{moving lens potential}. 
We can expand the signal in terms of the Haar-binned LC moments of $\Upsilon$ and $\psi$:
\begin{equation}\label{eq:ml_alms}
\Theta^{\rm ML}_{\ell,m}=\sum_{s=0}^{\infty}\sum\limits_{\ell\ell'\,mm'}\Upsilon_{\ell' m'}^{s}\psi^{s}_{\ell''m''}\int{\rm d}^2\hat{n}\,Y_{\ell m}^*\nabla_i Y_{\ell'm'}\nabla^iY_{\ell'' m''}.
\end{equation}
For the transverse velocity potential we use the following building functions: 
\begin{eqnarray}
W^{\Upsilon^{s}}(\chi) &=& h^{s} (\chi ) \;\;\;\;\mbox{from  Eq.~}\ref{eq:Haar_modes}\,, \\
\mathcal{K}^{\Upsilon^{s}}_{\ell}(\chi,k) &=& 4\pi i^{\ell}\frac{j_{\ell}(k\chi)}{k^2}\frac{f(\chi)H(\chi)a(\chi)}{\chi}\,,\\
\tilde{\mathrm{U}}^{\Upsilon^{s}} (\eta(\chi), \mathbf{k}) &=& \delta_m(\eta(\chi), \mathbf{k})\,.
\end{eqnarray}
For the moving lens potential we use the following building functions:
\begin{eqnarray}
W^{\psi^{s}}(\chi) &=& h^{s} (\chi ) \;\;\;\;\mbox{from  Eq.~}\ref{eq:Haar_modes}\,, \\
\mathcal{K}^{\psi^{s}}_{\ell}(\chi,k) &=& -4\pi i^{\ell}\frac{j_{\ell}(k\chi)}{k^2}\frac{3\Omega_m H_0^2}{a(\chi)\chi}\,,\\
\tilde{\mathrm{U}}^{\psi^{s}} (\eta(\chi), \mathbf{k}) &=& \delta_m(\eta(\chi), \mathbf{k})\,.
\end{eqnarray}
The angular integral in Eq.~\ref{eq:ml_alms} is given by:
\begin{equation}
\begin{split}
\int{\rm d}^2\hat{n}&\,Y_{\ell m}^*\nabla_i Y_{\ell'm'}\nabla^iY_{\ell'' m''}\\
&=\frac{1}{2}[\ell (\ell +1)+\ell'(\ell'+1)-\ell''(\ell''+1)]\sqrt{\frac{(2\ell+1)(2\ell'+1)(2\ell''+1)}{4\pi}}
\begin{pmatrix}
\ell & \ell' & \ell'' \\
m & m' & m''
\end{pmatrix}
\begin{pmatrix}
\ell & \ell' & \ell'' \\
0 & 0 & 0
\end{pmatrix}\,.
\end{split}
\end{equation}

\begin{figure}[t]
  \includegraphics[scale=0.5]{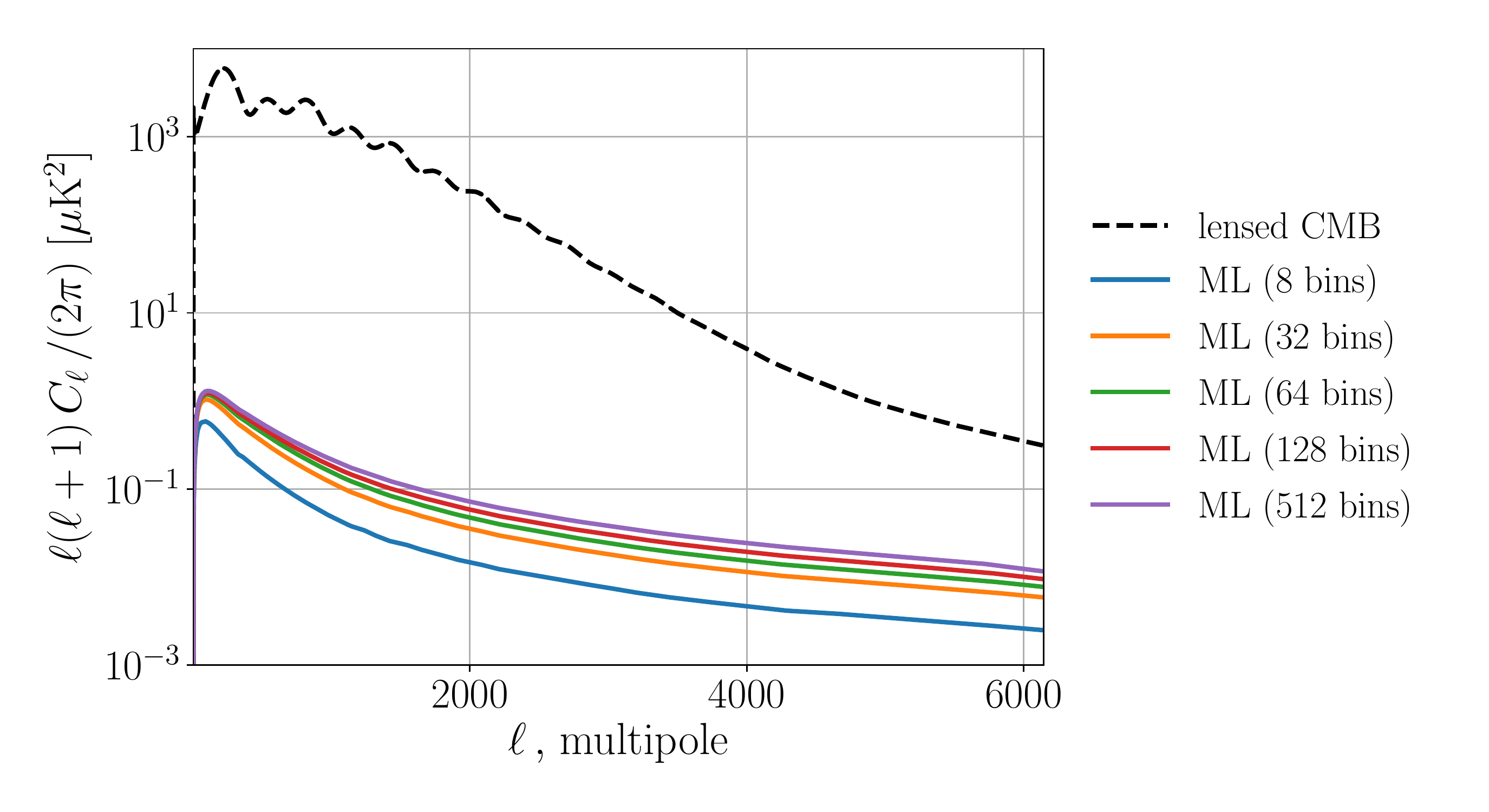}
  \caption{The moving lens power spectrum Eq.~\ref{eq:ML_power_binned} evaluated for $s_{\rm max} = 32, 64, 128, 256$, and $512$ bins.}
  \label{fig:MLN}
\end{figure}

The ML power spectrum can be calculated in terms of the auto- and cross-spectra of the Haar-binned moments of $\Upsilon$ and $\phi$,
\begin{eqnarray}\label{eq:ML_power_binned}
C_{\ell}^{\rm ML}= \sum\limits_{\ell',\ell''}\frac{(2\ell'+1)(2\ell''+1)}{4\pi}&\frac{1}{4}&
[\ell'(\ell'+1)+\ell''(\ell''+1)-\ell(\ell+1)]^2
\begin{pmatrix}
\ell & \ell' & \ell'' \\
0 & 0 & 0
\end{pmatrix}^2\\
&\times& \sum\limits_{s,s'=0}^{\infty}\left[(C^{\Upsilon\Upsilon})_{\ell'}^{ss'}(C^{\psi\psi})_{\ell''}^{ss'}+(C^{\Upsilon\psi})_{\ell'}^{ss'}(C^{\Upsilon\psi})_{\ell''}^{ss'}\right]\,,
\end{eqnarray}
where $(C^{\Upsilon\Upsilon})_{\ell'}^{ss'}$, $(C^{\psi\psi})_{\ell''}^{ss'}$, $(C^{\Upsilon\psi})_{\ell'}^{ss'}$, and $(C^{\Upsilon\psi})_{\ell''}^{ss'}$ are calculated using the building functions and Eq.~\ref{Cl_int1}. To evaluate Eq.~\ref{eq:ML_power_binned}, it is necessary to truncate the sum at some $s_{\rm max}$. In Fig.~\ref{fig:MLN}, we plot the moving lens power spectrum for $\chi_{\rm max} = 8.1$ Gpc (corresponding to a redshift $z_{\rm max} = 5$ in our fiducial cosmology) with $s_{\rm max} = 32, 64, 128, 256$, and $512$ bins (corresponding to $\Delta\chi = 263, 131, 66, 33, 17$ Mpc). For $512$ bins, the moving lens power spectrum is nearly converged for $\ell < 1000$, but still missing some power at large-$\ell$. Unfortunately, 512 bins is already challenging to compute, so we adopt $512$ bins as our model for the continuum limit of ML. As we demonstrate below, the missing fine-grained information is not relevant for velocity reconstruction with ML for near-term experiments, so this is not an important restriction.

Although we do not use it in the analysis below, we note that it is possible to find an approximate formula for the continuum ML power spectrum. In the squeezed limit where $\ell' \ll \ell''$, we can set $\ell \simeq \ell''$ and neglect cross-correlations between the bins $s$ and $s'$ as well as the cross-power between $\Upsilon$ and $\psi$:
\begin{eqnarray}
C_{\ell}^{\rm ML} &=& \frac{1}{4} \sum\limits_{s=1}^{\infty} \left[ \sum\limits_{\ell'}\frac{(2\ell'+1)}{4\pi}
[\ell'(\ell'+1)]^2 (C^{\Upsilon\Upsilon})_{\ell'}^{ss}\right] (C^{\psi\psi})_{\ell}^{ss} .
\end{eqnarray}
Defining $\theta_{\perp} \equiv \frac{1}{2} \nabla_{\perp}^2 \Upsilon$, and taking the continuum limit of the sum, we obtain:
\begin{eqnarray}
C_{\ell}^{\rm ML} &=& \int d\chi \ \langle \theta_{\perp} (0,\chi)^2 \rangle P^{\psi \psi} (\frac{\ell}{\chi},\chi) .
\end{eqnarray}
This expression may be of interest in future analyses of the ML effect.

\subsubsection{Lensing of the primary CMB}

We approximate the lensing of the primary CMB by the first order term:
\begin{equation}
\Theta^{L} (\hat{n}) = (\mathbf{\nabla_{\perp}} \phi) \cdot (\mathbf{\nabla_{\perp}} \Theta^{pCMB} )\,,
\end{equation}
where $\nabla_{\perp}$ is the angular gradient in the unit 2-sphere and $\phi (\hat{n})$ is the lensing potential, defined as:
\begin{equation}
\phi (\hat{n}) = -2 \int_{0}^{\chi_{ls}} d \chi \frac{\chi_{ls} - \chi}{\chi_{ls}\chi}  \Psi (\hat{n}, \chi).
\end{equation}
In terms of the multipole moments of the primary CMB and the lensing potential, the lensed CMB contribution is written as:
\begin{equation}\label{eq:lens_alms}
\Theta^{\rm L}_{\ell m}=\sum\limits_{\ell\ell'\,mm'}\phi_{\ell' m'}\Theta^{pCMB}_{\ell''m''}\int{\rm d}^2\hat{n}\,Y_{\ell m}^*\nabla_i Y_{\ell'm'}\nabla^iY_{\ell'' m''}\,,
\end{equation}
where the primary CMB is computed using CAMB and the lensing potential moments are computed using the building functions:
\begin{eqnarray}
W^{\phi}(\chi) &=& \begin{cases}
1, & 0 \leq \chi <  \chi_{ls}, \\
0, & {\rm otherwise}
\end{cases} \\
\mathcal{K}^{\phi}_{\ell}(\chi,k) &=& 4\pi i^{\ell}\frac{j_{\ell}(k\chi)}{k^2}\frac{3\Omega_m H_0^2}{a(\chi)}\frac{\chi_{ls} - \chi}{\chi_{ls}\chi}\,,\\
\tilde{\mathrm{U}}^{\phi} (\eta(\chi), \mathbf{k}) &=& \delta_m(\eta(\chi), \mathbf{k})\,.
\end{eqnarray}

The power spectrum calculation is similar to the one for the moving lens effect, and yields:
\begin{eqnarray}
C_{\ell}^{\rm L}= \sum\limits_{\ell',\ell''}\frac{(2\ell'+1)(2\ell''+1)}{4\pi}&\frac{1}{4}&
[\ell'(\ell'+1)+\ell''(\ell''+1)-\ell(\ell+1)]^2
\begin{pmatrix}
\ell & \ell' & \ell'' \\
0 & 0 & 0
\end{pmatrix}^2\\
&\times& \left[(C^{\phi\phi})_{\ell'}(C^{\Theta\Theta})_{\ell''}+(C^{\phi\Theta})_{\ell'}(C^{\phi\Theta})_{\ell''}\right]\,.
\end{eqnarray}

\subsubsection{Extragalactic foregrounds}

There are a number of extragalactic foregrounds that contribute to the CMB, whose relative importance depend on the frequency and scale being observed. At low frequencies ($\alt 150 \ {\rm GHz}$) on arcminute scales, the thermal Sunyaev Zel'dovich (tSZ) effect and radio point sources dominate. At high frequencies ($\agt 150 \ {\rm GHz}$) on the same scales, the CIB is the dominant extragalactic foreground. Below, we assume that enough radio point sources can be masked to make tSZ the dominant source at low frequencies. With this assumption, we include the tSZ and CIB only in our extragalactic foreground model.

We model the CIB and tSZ using the Halo Model for large scale structure, combining elements of the models described in Refs.~\cite{Smith:2018bpn,McCarthy:2019xwk,2013PhRvD..88f3526H,2012MNRAS.421.2832S,2014A&A...571A..30P,2020arXiv201016405M}. Our assumptions are outlined in detail in Appendix~\ref{sec:halomodel}. In Fig.~\ref{fig:cleanCMB}, we show angular power spectra of the CIB and tSZ at several frequencies for our fiducial CMB experiment. Since all observables are computed within the same Halo Model, it is possible to capture the correlations between the CIB, tSZ, and galaxy number counts -- e.g. the spectra in Fig.~\ref{fig:cleanCMB} include the CIB-tSZ cross-power. We discuss the detailed properties of the galaxy-foreground cross spectra below in Sec.~\ref{sec:isotropic_cmb_cross}.

\subsection{Foreground cleaning of the CMB}\label{sec:foregroundcleaning}

To access the blackbody components of the CMB necessary for velocity reconstruction, we can estimate 
how well one can use the multifrequency information in the CMB to clean the extragalactic foregrounds in our 
model. Here, we use the harmonic Internal Linear Combination (ILC) algorithm~\cite{Tegmark:2003ve}.

We write the covariance between the de-beamed CMB at different frequencies as a matrix: 
\begin{equation}
{\bf C}_{\ell} = C_{\ell}^{TT} {\bf e}{\bf e}^{\dagger}+{\bf C}_{\ell}^{XG}+({\bf B^{-1}N})_{\ell}\,,
\end{equation}
where $C_{\ell}^{TT}$ contains the blackbody components of the CMB (primary CMB, kSZ, ML, etc.), 
${\bf e}=\{1,1,1,\ldots\}$, ${\bf C}_{\ell}^{XG}$ contains the CIB,
tSZ, and the cross-correlations between these
various components at the measured frequencies, 
and $({\bf B^{-1}N})_{\ell}$ is the de-beamed instrumental noise
covariance (assumed diagonal). Following the ILC method in harmonic
space~\cite{Tegmark:2003ve}, we estimate the blackbody component as:
\begin{equation}
\hat{\Theta}_{\ell m} = {\bf w}_{\ell}^{\dagger}\ {\bf \Theta}_{\ell m}\,,
\end{equation}
where the weights ${\bf w}_{\ell}$ that minimize the variance of the resulting multipole moments $\hat{\Theta}_{\ell m}$ are given by: 
\begin{equation}
{\bf w}_{\ell}=\frac{({\bf C}_{\ell})^{-1}{\bf e}}{{\bf e}^{\dagger}({\bf C}_{\ell})^{-1}{\bf e}}\,.
\end{equation}
The ensemble averaged power spectrum of the cleaned map is: 
\begin{equation}
C_{\ell}^{\Theta \Theta; \text{clean}} =C_{\ell}^{TT} + {\bf w}_{\ell}^{\dagger}\left(  {\bf C}_{\ell}^{XG}+({\bf B^{-1}N})_{\ell} \right){\bf w}_{\ell}\,.
\end{equation}
To the extent that the second term is small, we have successfully isolated the blackbody component of the CMB in the resulting map. 
Note that the residuals represented by the second term include both foreground residuals as well as an effective noise for 
the linear combination of maps. In the right panel of Fig.~\ref{fig:cleanCMB}, we show $C_{\ell}^{\text{clean}}$ for our fiducial CMB experiment. 
For the fiducial experimental parameters we choose, from the left panel of Fig.~\ref{fig:cleanCMB}, we see that the experimental noise 
is somewhat larger than the extragalactic foregrounds. Therefore, much of the improvement of the cleaned CMB over the 145 GHz channel 
comes from a lower effective noise rather than the removal of extragalactic foregrounds.

We can also estimate the cleaned galaxy-Temperature cross power:
\begin{eqnarray}
C_{\ell}^{\Theta \delta^{W^\alpha}\!;\rm clean} =  C_{\ell}^{ISW \ \delta^{W^\alpha}}+{\bf w}_{\ell}^{\dagger}{\bf C}_{\ell}^{XG \ \delta^{W^\alpha}}\,.
\end{eqnarray}
Here, because the CMB noise is uncorrelated with the galaxy field, there is no effective noise term. The ILC algorithm in this case reduces the variance of the cross-power due to the removal of extragalactic foregrounds. 

\subsection{Galaxy number counts}\label{sec:galncounts}

We now consider a tracer of the electron overdensity field, which for the purposes of the present paper we take to be the galaxy overdensity field, measured using a photometric redshift survey. Other tracers such as the redshifted 21cm Hydrogen line (or transitions such as CII) measured by line intensity mapping surveys~\cite{Ansari:2018ury,Sato-Polito:2020cil}, the CIB~\cite{McCarthy:2019xwk}, or the dispersion measure of Fast Radio Bursts~\cite{Madhavacheril:2019buy} have been considered as well. Spectroscopic surveys were considered in Refs.~\cite{Munchmeyer:2018eey,Smith:2018bpn,Giri:2020pkk}, which may be more computationally feasible to analyze in the box picture; we defer a discussion of spectroscopic surveys in the light cone picture to future work.

For the purposes of velocity reconstruction, the three dimensional information in a galaxy redshift survey is used to construct a series of 2-dimensional fields that are later cross-correlated with CMB temperature anisotropies. In harmonic space, these 2-dimensional fields can be expressed as integrals over redshift space:
\begin{equation}
g^{W^{\alpha}}_{\ell m} = \int dz_oW^{\alpha}(z_o)g_{\ell m}(z_o)\,,
\end{equation}
where $z_o$ denotes the observed redshift for the galaxies in the survey, $g_{\ell m}(z_o)$ are spherical harmonic coefficients of the measured 3-dimensional galaxy overdensity field, and $W^{\alpha}(z_o)$ is the window function used to construct the average. The equation above is not immediately related to the comoving space integral Eq.~\ref{eq:int_moment} of a light cone field introduced in Sec.~\ref{form}. First, the observed redshift $z_o$ of a galaxy may be subject to instrumental errors and therefore different from the actual redshift $z'$. Second, due to redshift space distortions (RSD), the redshift $z'$ can be different from the background cosmological redshift $z$ of the galaxy (which is simply related to the comoving distance $\chi$). The second issue can be safely ignored for high enough multipoles, where the RSD correction to the power spectrum is unimportant~\cite{Smith:2018bpn,2011PhRvD..84d3516C}. Since only small angular scale galaxy data is necessary for velocity reconstruction, we don't include RSD in our modelling (for the impact of RSD on correlations between velocity reconstruction and number counts see~\cite{Contreras_2019}) and will treat the actual redshift $z'$ as the cosmological redshift $z$. The issue of measurement errors is discussed below in the context of a redshift galaxy survey subject to photometric redshift errors. In our analysis below, we consider two prototype galaxy surveys: a LSST-like survey with many photometric redshift bins and a WISE-like survey with a single wide photometric redshift bin. For velocity reconstruction, these two surveys will be used as prototypes for the `multiple density window' and `single density window' cases for the quadratic estimators described in Sec.~\ref{sec:estimator}.

\subsubsection{LSST-like survey}\label{sec:Rubin-like}

\begin{figure}[ht]
  \includegraphics[scale=0.35]{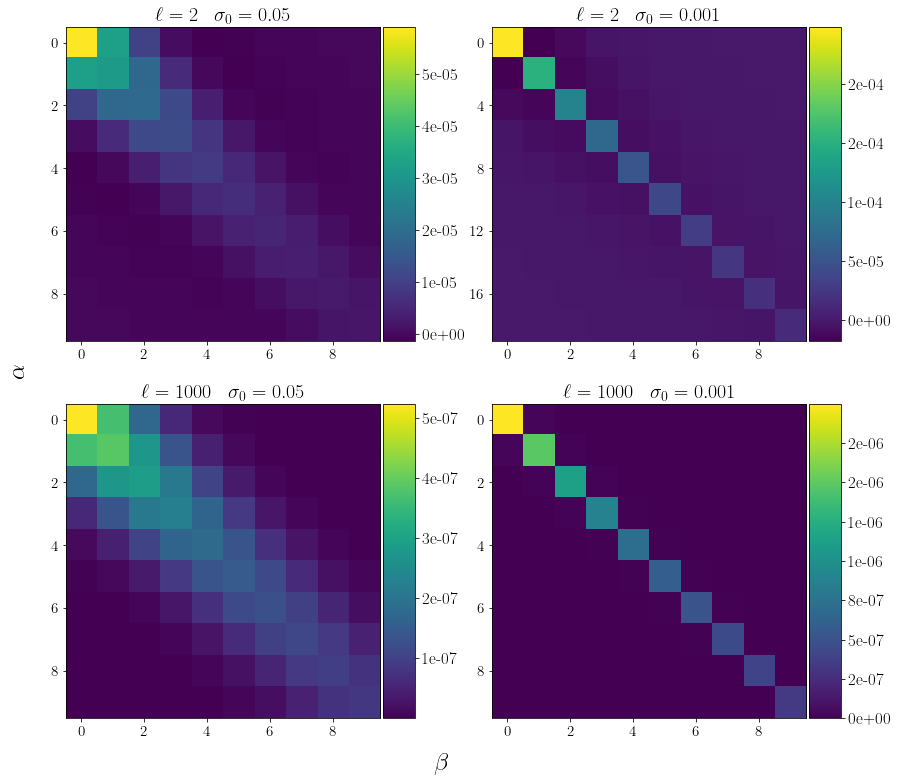}
  \caption{Section of the $32\times32$ redshift bins galaxy-galaxy covariance matrix, for different values of $\sigma_{0}$ and multipole $\ell$. The bin-bin correlations for large redshift errors at both large and small $\ell$ are apparent.}
  \label{fig:galaxy_photosz}
\end{figure}

\begin{figure}[ht]
  \includegraphics[scale=0.4]{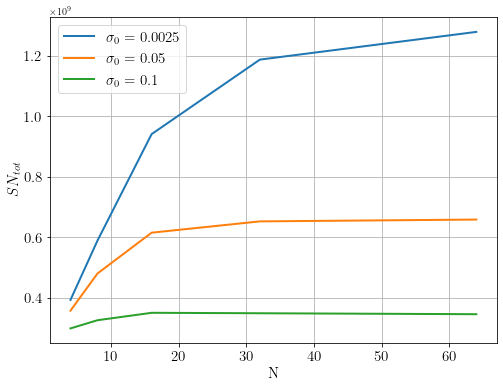}
  \caption{Total signal to noise for a photometric galaxy survey as a function on the number of redshift bins and for different error levels. The number of bins where the signal to noise saturates is a good indicator of the coarse graining scale that should be used in an analysis. For the fiducial LSST-like survey with $\sigma_{0} = 0.05$, the signal to noise saturates around $N \sim 32$ bins.}
  \label{fig:photozcomp}
\end{figure}

For the LSST-like survey, we consider Gaussian errors on photometric redshifts, with the probability of assigning redshift $z_o$ to a galaxy with true redshift $z$ (following Ref.~\cite{0912.0201}) given by: 
\begin{equation}
P(z,z_o) = \frac{\exp \left[-\frac{\left(z_o-z\right)^{2}}{2 \sigma_{z}^{2}}\right]}{\int_{0}^{\infty} d \tilde{z} \exp \left[-\frac{\left(\tilde{z}-z\right)^{2}}{2 \sigma_{z}^{2}}\right]}\,,
\end{equation}
where $\sigma_{z} = \sigma_{0}(1+z)$ with $\sigma_{0}$ parametrizing the size of the photometric errors. We assume a fiducial value of $\sigma_{0} = 0.05$. With this probability distribution, the galaxy average for the window $W^{\alpha}(z_o)$ can be expressed as an integral over the actual redshifts $z$:
\begin{equation}
g^{W^{\alpha}}_{\ell m} = \int dz\;\bigg{[}\int dz_o W^{\alpha}(z_o)P(z,z_o)\bigg{]}\;g_{\ell m}(z),
\end{equation}
and in terms of the comoving distance
\begin{equation}\label{eq:galpharubin}
g^{W^{\alpha}}_{\ell m} = \int d\chi\; {W}_{eff}^{\alpha}(\chi)\;g_{\ell m}(\chi),
\end{equation}
where we have defined the effective window function
\begin{equation}\label{eq:eff_win}
{W}_{eff}^{\alpha}(\chi) = H(z(\chi))\int dz_o W^{\alpha}(z_o)P(z(\chi),z_o),
\end{equation}
and $g_{\ell m}(\chi)$ are the light cone moments of the underlying galaxy overdensity field. The angular power spectrum between two galaxy redshift bins coming from a photometric survey can then be expressed using Eq.~\ref{Cl_int1} plus a shot noise term:
\begin{equation}
C_{\ell}^{g^{W^{\alpha}}g^{W^{\beta}}}  = \int d \chi_{1} d \chi_{2} \ {W}_{eff}^{\alpha} \left(\chi_{1}\right) {W}_{eff}^{\beta} \left(\chi_{2}\right) \int \frac{k^{2} d k}{(2\pi)^3} \ \mathcal{K}^{g}_{\ell}(\chi_1,k)\, \mathcal{K}^{g}_{\ell}(\chi_2,k) \ P_{gg}(\chi_1,\chi_2,k) +\delta_{\alpha\beta}\frac{1}{\bar{n_g}^{\alpha}}\,,
\label{eq:rubinlike_clgg}
\end{equation}
where $P_{gg}(\chi_1,\chi_2,k)$ is the galaxy-galaxy power spectrum computed using the Halo Model (consistent with Refs.~\cite{Munchmeyer:2018eey,Smith:2018bpn}; see Appendix~\ref{sec:halomodel} for a summary), $\mathcal{K}^{g}_{\ell}(\chi,k) = 4\pi i^{\ell}j_{\ell}(k\chi)$ is the galaxy projection kernel from three dimensional Fourier space onto the sky, and $\bar{n_g}^{\alpha}$ is the number of galaxies per steradian in redshift bin $\alpha$. We assume shot noise that is uncorrelated between redshift bins, and compute the number density per bin assuming the galaxy number density $n(z)$ per square arcmin is \cite{0912.0201}:
\begin{equation}
n(z) = \frac{n_{\rm g} }{2z_0} \left( \frac{z}{z_0}\right)^2 \exp\left(\frac{z}{z_0} \right)\ ,
\end{equation}
with $z_0 = 0.3$ and $n_{\rm g}=40/{\rm arcmin}^2$. We construct the effective window functions using:
\begin{equation}
W^{\alpha}(z_o) = \frac{\Pi^{\alpha}(\chi(z_o))}{ H(z_o)}\,,
\end{equation}
with $\Pi^{\alpha}(\chi)$ defined as in Eq.~\ref{eq:tophat}. In the limit of $\sigma_0\rightarrow0$, where photometric redshift errors can be neglected,  these window functions correspond to normalized top-hat windows in comoving space. 
We show the effects of the photometric errors in the galaxy-galaxy covariance matrix in Fig.~\ref{fig:galaxy_photosz}. Bin-bin correlations are enhanced as expected and the auto-power at a particular bin is reduced due to the contamination from distant bins. The principal components of the galaxy survey can be found using the procedure
described in Sec.~\ref{subsec:principal} just by appropriately replacing the signal and noise matrices. Fig.~\ref{fig:photozcomp} compares the effect of different photometric redshift error levels on the total signal to noise Eq.~\ref{eq:totSN} of the galaxy survey as a function of the number of bins $N$. As expected, we observe that the photometric errors put a limit on how much radial resolution our galaxy survey can have. For our fiducial value of $\sigma_{0} = 0.05$, the signal to noise is mostly saturated for more than 32 redshift bins. 

\subsubsection{unWISE-like survey}

For the unWISE-like survey considered in this paper, we model the `blue' sample used in Refs.~\cite{Krolewski:2019yrv,Krolewski:2021yqy,Kusiak:2021hai} from the unWISE catalogue~\cite{Schlafly_2019}, which is based on data from the WISE mission~\cite{Wright_2010}. This sample is characterized by a median redshift of $\bar{z} \sim 0.6$ and is reasonably uniform over a redshift range of $\Delta z \sim 0.3$. The number density of the resulting map is $\bar{n} \simeq 0.95 /{\rm arcmin}^2$. 

Following Ref.~\cite{Krolewski:2019yrv}, we model the unWISE blue sample as a linearly biased tracer of dark matter plus shot noise. In particular, we model the galaxy-galaxy angular power spectrum as 
\begin{equation}
C_{\ell}^{g^{W}g^{W}}  = \int d \chi_{1} d \chi_{2} \ {W}_{eff} \left(\chi_{1}\right) {W}_{eff} \left(\chi_{2}\right) \int \frac{k^{2} d k}{(2\pi)^3} \ \mathcal{K}^{g}_{\ell}(\chi_1,k)\, \mathcal{K}^{g}_{\ell}(\chi_2,k) \ P_{gg}(\chi_1,\chi_2,k) + \frac{1}{\bar{n}_g}\,,
\label{eq:wise_clgg}
\end{equation}
where the power spectrum is:
\begin{equation}
P_{gg} (\chi_1,\chi_2,k) = b(\chi_1) b(\chi_2) P_{mm} (\chi_1, \chi_2, k), \ \ \ b\left[z(\chi) \right] = 1.2 z(\chi) + 0.8\,.
\end{equation}
Here, $P_{mm} (\chi_1, \chi_2, k)$ is computed in the Halo Model as described in Appendix~\ref{sec:halomodel}. The galaxy window function ${W}_{eff}$ is simply the normalized comoving galaxy density
\begin{equation}
    W_{eff} \left(\chi\right) = H\left(z\left(\chi\right)\right) \frac{dN}{dz}\,,
    \label{eq:unwiselike_window}
\end{equation}
where the redshift distribution of galaxies $dN/dz$ is defined to be normalized as $1 = \int dz \ dN/dz$ and is reasonably uniform within a range $\Delta z \sim 0.3$ of the median redshfit $\bar{z} \sim 0.6$; the redshift distribution is shown in Fig.~\ref{fig:pcs}. The total number of galaxies in the survey is $\sim 1.4 \times 10^8$, yielding a shot noise of $1 / \bar{n}_g = 9.2\times10^{-8}$. When performing velocity reconstruction, we must also compute the cross-power with the $\Pi$-binned optical depth and potential. In these cases, it is convenient to expand the observed moments of the galaxy overdensity as:
\begin{eqnarray}
g_{\ell m}^W &=& \int d\chi \ H(\chi) \frac{dN}{dz} g_{\ell m} \\
&=& \sum_\alpha \int d\chi \ \left[ H \frac{dN}{dz} \ \Delta \chi \ \Pi^{\alpha}\right] \  g_{\ell m}\,,
\end{eqnarray}
and define a set of window functions
\begin{equation}
W_{eff}^\alpha \left( \chi \right) = H \frac{dN}{dz} \ \Delta \chi \ \Pi^{\alpha}\,.
    \label{eq:unwiselike_window_binned}
\end{equation}
We then define a set of binned galaxy moments as in Eq.~\ref{eq:galpharubin} using the window functions Eq.~\ref{eq:unwiselike_window_binned}. These binned galaxy moments are used to compute the cross-power with other $\Pi$-binned LC moments.

\subsection{Galaxy survey systematics}\label{sec:galaxy_sys}

Aside from the photometric redshift errors described above, one must consider a wide variety of systematics associated with a galaxy survey, many of which manifest on large angular scales (see e.g. \cite{Leistedt:2013gfa,Elsner_2015,Huterer_2013,PhysRevD.97.043515,PhysRevD.94.043503}). Systematics that modulate the observed number counts of galaxies are the most problematic for velocity reconstruction, as they lead to a statistically anisotropic cross-power between the galaxy overdensity and CMB temperature that mimics the signal of interest. Additive effects that are uncorrelated with extragalactic sources, e.g. mis-identified stars included in the sample, are less problematic, adding only noise to the estimators but not bias. Starting from the observed number counts (following Refs.~\cite{Huterer_2013,PhysRevD.97.043515,PhysRevD.94.043503}), we model systematics effects as:
\begin{equation}
N^{W^\alpha}_{\rm obs} (\hat{n}) = \left( 1+ c(\hat{n}) \right) N^{W^\alpha} (\hat{n})\,,
\end{equation}
where $N^{W^\alpha}_g (\hat{n})$ are the number counts of galaxies in a bin (e.g. sample) defined by the window function $W^\alpha$. The modulating field $c(\hat{n})$ encodes calibration errors which we might expand as a sum of effects associated with the instrument/observation strategy, extinction due to galactic dust, etc. Defining the underlying galaxy overdensity field $g^{W^\alpha} (\hat{n})$ by $N^{W^\alpha} (\hat{n}) = \bar{N}^{W^\alpha} (1 + g^{W^\alpha} (\hat{n}))$, with $\bar{N}^{W^\alpha}$ the mean number of objects on the sky, the moments of the observed galaxy overdensity field are:
\begin{eqnarray}
(g^{W^\alpha}_{\ell m})_{\rm obs} &=& g^{W^\alpha}_{\ell m} + c_{\ell m} \\
&+& \sum_{\ell_1,m_1; \ell_2, m_2} (-1)^m  \sqrt{\frac{(2\ell+1)(2\ell_1+1)(2\ell_2+1)}{4\pi}} 
\begin{pmatrix}
\ell_1 & \ell_2 & \ell \\
0 & 0 & 0
\end{pmatrix}
\begin{pmatrix}
\ell_1 & \ell_2 & \ell \\
m_1 & m_2 & -m
\end{pmatrix} c_{\ell_1 m_1} g^{W^\alpha}_{\ell_2 m_2} \nonumber \\
&+& \mathcal{O}(\epsilon g^{W^\alpha}_{\ell m})\,,\nonumber
\end{eqnarray}
where $g^{W^\alpha}_{\ell m}$ is defined as above in Eq.~\ref{eq:galpharubin} and
\begin{equation}
\epsilon = \frac{1}{\bar{N}^{W^\alpha}} \sum_{\ell m} c_{\ell m} \langle N^{W^\alpha} (\hat{n}) Y_{\ell m} (\hat{n}) \rangle_{\rm sky}\,,
\end{equation}
is the correction to the mean number counts from each moment $c_{\ell m}$. Below, we neglect this correction to the mean. To model the form of the large-angular scale systematics, we assume that the modulating field $c(\hat{n})$ is a Gaussian random field with power spectrum:
\begin{equation}
C_{\ell}^c = A^c e^{-(\ell/10)^2}\,,
\end{equation}
where for a LSST-like experiment we set the fiducial value for the amplitude $A^c$ such that the variance of $c(\hat{n})$ satisfies:
\begin{equation}
    \sum \frac{(2\ell+1)}{4\pi}A^c e^{-(\ell/10)^2} = 10^{-4}\,,
\end{equation}
which corresponds to a level of calibration error somewhere between the best current data sets and futuristic data sets (see Fig.~3 of \cite{Weaverdyck_2018}). For unWISE, we use a value of $10^{-2}$ for the variance of the calibration error field.

\subsection{CMB temperature-galaxy cross power}

As discussed above, there are a number of components of the CMB temperature that are correlated with tracers of large scale structure, such as the galaxy surveys considered above. Some of these contributions, such as the late-time ISW and extragalactic foregrounds, have a statistically isotropic cross-power. On the other hand, secondary components of the CMB  such as lensing, kSZ, and ML will have a statistically anisotropic cross-power with the galaxy survey. Indeed, this statistical anisotropy is the basis for velocity reconstruction. We now consider these two cases in turn.

\subsubsection{Statistically isotropic cross-correlations}\label{sec:isotropic_cmb_cross}

The observed CMB anisotropies have contributions that are isotropically correlated with galaxies, including: extragalactic foregrounds (CIB, tSZ) and the late time linear ISW effect. 
To calculate isotropic cross-correlations we use Eq.~\ref{Cl_int1}:
\begin{equation}
C_{\ell}^{F^WG^{W'}}= \int d \chi_{1} d \chi_{2} \ W \left(\chi_{1}\right) W' \left(\chi_{2}\right) \int \frac{k^{2} d k}{(2\pi)^3} \ \mathcal{K}^{F}_{\ell}(\chi_1,k)\, \mathcal{K}^{G}_{\ell}(\chi_2,k) \ P_{FG}(\chi_1,\chi_2,k),
\end{equation}
and therefore we need to specify the window functions, integral kernels, and underlying power spectra for each of the temperature-galaxy signals. For the galaxies, we use the window functions introduced in Sec.~\ref{sec:galncounts} and the integral kernel $\mathcal{K}^{g}_{\ell}(\chi,k) = 4\pi i^{\ell}j_{\ell}(k\chi)$.

\begin{figure}[h]
\includegraphics[scale=0.4]{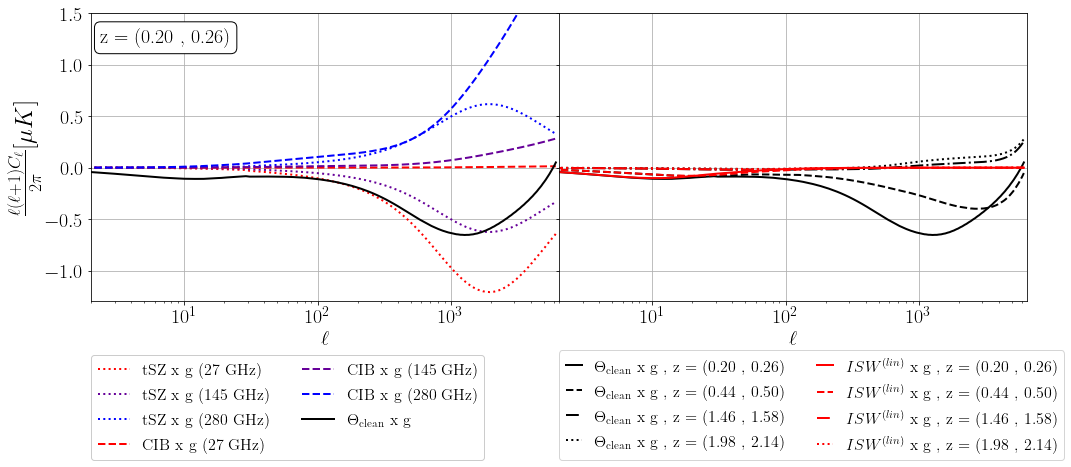}
\caption{\textbf{Left panel:} Extragalactic foregrounds cross galaxies at redshift bin $z=(0.20,0.26)$ and ILC cleaned temperature cross galaxies (solid line). \textbf{Right panel:} ILC cleaned temperature cross galaxies for several redshift bins compared to the linear-ISW cross galaxies. At low-$\ell$, the ISW component is relevant, while at high-$\ell$ it can be safely neglected.}
\label{fig:extragalg}
\end{figure}

Extragalactic foregrounds are themselves tracers of large scale structure, and therefore are well-correlated with binned galaxy density. We assume that extragalactic foregrounds can be described by random Gaussian fields. For these signals we use trivial window functions $W \left(\chi_{1}\right) = 1$, kernels $\mathcal{K}^{extra}_{\ell}(\chi,k) = 4\pi i^{\ell}j_{\ell}(k\chi)$, and underlying spectra $P^{\nu}_{CIB g}(\chi,k)$ and  $P^{\nu}_{tSZ g}(\chi,k)$ computed at each frequency $\nu$ using the Halo Model. In the left panel of Fig.~\ref{fig:extragalg}, we show the cross-correlation between the extragalactic foregrounds at different frequencies and a LSST-like galaxy survey in the redshift bin $z=(0.20,0.26)$. We show as well the cross-power between the ILC cleaned temperature discussed in Sec.~\ref{sec:foregroundcleaning} and the galaxy survey in that same bin. The right panel shows the cross-power between cleaned temperature and galaxies at different redshift bins together with the cross-power between the linear late-time ISW signal and galaxies. The ISW-g correlation is calculated using Eq.~\ref{Cl_int1} with the corresponding building functions.

\subsubsection{Anisotropic cross-correlations}

The main focus of this paper are the statistically anisotropic cross-correlations between the CMB and galaxy surveys, as these are what allow us to perform velocity reconstruction. We work in the basis introduced in Sec.~\ref{sec:estimator}, expanding in terms of $\Pi$ and $\mu$-binned LC moments to define the `bulk' and `fine' modes, respectively. For the kSZ-galaxy cross power we have:
\begin{eqnarray}
\Big{\langle}\Theta^{kSZ}_{\ell m}\;g^{W}_{\ell'm'}\Big{\rangle} &=&
\sum_{\alpha = 0}^{N-1}
\sum_{\ell_1 m_1} (-1)^{m_1} 
\begin{pmatrix}
\ell & \ell' & \ell_1 \\
m & m' & - m_1
\end{pmatrix}
f^{v^\alpha W}_{\ell\ell_1\ell'} \, v^{\alpha}_{\ell_1 m_1}\nonumber  \\
 &+&
\sum_{k = N}^{\infty}
\sum_{\ell_1 m_1} (-1)^{m_1} 
\begin{pmatrix}
\ell & \ell' & \ell_1 \\
m & m' & - m_1
\end{pmatrix}
f^{v^k W}_{\ell\ell_1\ell'} \, v^{k}_{\ell_1 m_1}\,, \nonumber
\end{eqnarray}
where the bulk mode couplings $f^{v^\alpha W}_{\ell\ell_1\ell'}$ and the fine mode couplings 
$f^{v^k W}_{\ell\ell_1\ell'}$ are given by:
\begin{equation}\label{eq:bulkvcoupling}
f^{v^\alpha W}_{\ell \ell_1 \ell'} \equiv \sqrt{\frac{(2\ell+1)(2\ell_1+1)(2\ell'+1)}{4\pi}} 
\begin{pmatrix}
\ell & \ell' & \ell_1 \\
0 & 0 & 0
\end{pmatrix} \;C^{\dot{\tau}^{\alpha}g^{W}}_{\ell'}\Delta \chi\,,
\end{equation}
and
\begin{equation}\label{eq:finevcoupling}
f^{v^k W}_{\ell \ell_1 \ell'} \equiv \sqrt{\frac{(2\ell+1)(2\ell_1+1)(2\ell'+1)}{4\pi}} 
\begin{pmatrix}
\ell & \ell' & \ell_1 \\
0 & 0 & 0
\end{pmatrix} \;C^{\dot{\tau}^{k}g^{W}}_{\ell'}.
\end{equation}
For the Moving Lens-galaxy cross power we have:
\begin{eqnarray}
\Big{\langle}\Theta^{ML}_{\ell m}\;g^{W}_{\ell'm'}\Big{\rangle} &=& 
\sum_{\alpha = 0}^{N-1}
\sum_{\ell_1 m_1} (-1)^{m_1} 
\begin{pmatrix}
\ell & \ell' & \ell_1 \\
m & m' & - m_1
\end{pmatrix}
f^{\Upsilon^\alpha W}_{\ell\ell_1\ell'} \, \Upsilon^{\alpha}_{\ell_1 m_1}\nonumber  \\
 &+&
\sum_{k = N}^{\infty}
\sum_{\ell_1 m_1} (-1)^{m_1} 
\begin{pmatrix}
\ell & \ell' & \ell_1 \\
m & m' & - m_1
\end{pmatrix}
f^{\Upsilon^k W}_{\ell\ell_1\ell'} \, \Upsilon^{k}_{\ell_1 m_1}\,, \nonumber
\end{eqnarray}
where the bulk mode couplings $f^{\Upsilon^\alpha W}_{\ell\ell_1\ell'}$ and the fine mode couplings 
$f^{\Upsilon^k W}_{\ell\ell_1\ell'}$ are given by:
\begin{equation}\label{eq:bulkvtcoupling}
f^{\Upsilon^\alpha W}_{\ell \ell_1 \ell'} \equiv 
\left[ \ell_1(\ell_1+1) + \ell' (\ell' + 1) - \ell (\ell +1) \right]
\sqrt{\frac{(2\ell+1)(2\ell_1+1)(2\ell'+1)}{16\pi}} 
\begin{pmatrix}
\ell & \ell' & \ell_1 \\
0 & 0 & 0
\end{pmatrix} \;C^{\psi^{\alpha}g^{W}}_{\ell'}\Delta \chi
\end{equation}
and
\begin{equation}\label{eq:finevtcoupling}
f^{\Upsilon^k W}_{\ell \ell_1 \ell'} \equiv 
\left[ \ell_1(\ell_1+1) + \ell' (\ell' + 1) - \ell (\ell +1) \right]
\sqrt{\frac{(2\ell+1)(2\ell_1+1)(2\ell'+1)}{16\pi}} 
\begin{pmatrix}
\ell & \ell' & \ell_1 \\
0 & 0 & 0
\end{pmatrix} \;C^{\psi^{k}g^{W}}_{\ell'}.
\end{equation}

The kSZ and ML cross-power with galaxies form the basis of the estimators used for velocity reconstruction. However, there are additional sources of statistical anisotropy in the cross-power that potentially introduce biases on the reconstructed velocity fields. Here, we focus on CMB lensing and large angular scale calibration error in the galaxy survey. For the CMB lensing-galaxy cross power we have:
\begin{eqnarray}
\Big{\langle}\Theta^{L}_{\ell m}\;g^{W}_{\ell'm'}\Big{\rangle} &=& 
\sum_{\alpha = 0}^{N-1}
\sum_{\ell_1 m_1} (-1)^{m_1} 
\begin{pmatrix}
\ell & \ell' & \ell_1 \\
m & m' & - m_1
\end{pmatrix}
f^{\Theta W}_{\ell\ell_1\ell'} \, \Theta^{pCMB}_{\ell_1 m_1}\,,\\ \nonumber
\end{eqnarray}
where the couplings $f^{\Theta W}_{\ell\ell_1\ell'}$ are given by:
\begin{equation}\label{eq:lensingcoupling}
f^{\Theta W}_{\ell\ell_1\ell'} \equiv 
\left[ \ell_1(\ell_1+1) + \ell' (\ell' + 1) - \ell (\ell +1) \right]
\sqrt{\frac{(2\ell+1)(2\ell_1+1)(2\ell'+1)}{16\pi}} 
\begin{pmatrix}
\ell & \ell' & \ell_1 \\
0 & 0 & 0
\end{pmatrix} \;C^{\phi g^{W}}_{\ell'} \,.
\end{equation}
For the calibration error contribution we have:
\begin{eqnarray}
\Big{\langle}\Theta^{\nu}_{\ell m}\;g^{W}_{\ell'm'}\Big{\rangle}\Big{|}_{cal} &=& 
\sum_{\alpha = 0}^{N-1}
\sum_{\ell_1 m_1} (-1)^{m_1} 
\begin{pmatrix}
\ell & \ell' & \ell_1 \\
m & m' & - m_1
\end{pmatrix}
f^{c W}_{\ell\ell_1\ell'} \, c_{\ell_1 m_1}\,
\end{eqnarray}
where the couplings $f^{c W}_{\ell\ell_1\ell'}$ are given by:
\begin{equation}\label{eq:calibrationcoupling}
f^{c W}_{\ell\ell_1\ell'} \equiv 
\left[ \ell_1(\ell_1+1) + \ell' (\ell' + 1) - \ell (\ell +1) \right]
\sqrt{\frac{(2\ell+1)(2\ell_1+1)(2\ell'+1)}{16\pi}} 
\begin{pmatrix}
\ell & \ell' & \ell_1 \\
0 & 0 & 0
\end{pmatrix} \;C^{I g^{W}}_{\ell'}\,. 
\end{equation}

Other effects leading to a statistically anisotropic cross-power which we anticipate will be less important, and which we do not compute here, include: relativistic aberration of the CMB~\cite{PhysRevD.65.103001} (similar effect as calibration error, but with a smaller magnitude), SZ effects at higher order in velocity and temperature (see e.g. Refs.~\cite{Challinor_1998,Challinor_1999,Itoh_1998,Nozawa_1998,Yasini_2016,Coulton_2020}),  anisotropic/ill-characterized beam patterns in the CMB experiment (see e.g. Ref.~\cite{PhysRevD.79.063008} for an assessment of the impact on lensing reconstruction), and perhaps others. In the case of CMB lensing, note that the modulating field is the primary CMB temperature. Although we do not explore it further here, we note that a quadratic estimator for the low-$\ell$ primary CMB can be formulated from the CMB-galaxy cross power using the formalism introduced in Sec.~\ref{sec:estimator}. A similar estimator was introduced in Ref.~\cite{Meerburg:2017xga} as a means to reconstruct the primary CMB dipole, which is not directly measurable due to the contribution from our local peculiar velocity.

\section{Reconstruction analysis}\label{sec:recon_analysis}

In Sec.~\ref{sec:estimator}, we discussed the details involved in constructing quadratic estimators for fields sourcing a statistical anisotropy in the CMB-LSS cross-correlation. We showed that information about these fields can be reconstructed up to a series of noise terms. The purpose of this section is to analyze the relations between signal and noise for the reconstruction of the $\Pi$-binned LC moments of the radial velocity $v(\hat{n},\chi)$ and the transverse velocity potential $\Upsilon(\hat{n},\chi)$, sources for the kSZ-LSS and ML-LSS statistical anisotropies respectively. We estimate the signal and noise for a reconstruction using the modelling for the CMB, LSS, and their correlation presented in Sec.~\ref{sec:observables}. 

\subsection{Radial velocity reconstruction for SO x LSST} \label{sec:recon_rd}

Applying the formalism of Sec.~\ref{sec:estimator} to the reconstruction of $v^{\alpha}_{LM}$ leads to a collection of estimators with the following correlation function:
\begin{equation}\label{eq:rd_twopoint}
\Big{\langle}\hat{v^{\alpha}}_{LM}
\hat{v^{\beta}}^*_{LM}\Big{\rangle} =
\left(\mathrm{\bf{R}}_L\mathrm{\bf{C}}^{vv}_{L}(\mathrm{\bf{R}}_L)^{\dagger}\right)^{\alpha\beta}+
\left(\mathrm{\bf{N}}^{0}_{L}\right)^{\alpha\beta}+\left(\mathrm{\bf{N}}^{fine}_{L}\right)^{\alpha\beta}+
\left(\mathrm{\bf{N}}^{cal}_{L}\right)^{\alpha\beta}+\left(\mathrm{\bf{N}}^{\Upsilon}_{L}\right)^{\alpha\beta}+
\left(\mathrm{\bf{N}}^{lens}_{L}\right)^{\alpha\beta}\,,
\end{equation}
where the various terms are defined by:
\begin{itemize}
\item $\mathrm{\bf{R}}_L \mathrm{\bf{C}}^{vv}_{L}(\mathrm{\bf{R}}_L)^{\dagger}$: the covariance matrix of $\Pi$-binned LC moments of the radial velocity field. The rotation matrix $\mathrm{\bf{R}}_L$ defined in Eq~\ref{rot2} encodes the bin-bin mixing of the signal covariance due to the redshift error in the galaxy survey.
\item $\mathrm{\bf{N}}^{0}_{L}$: the Gaussian reconstruction noise Eq.~\ref{eq:N0definition}, with coupling functions defined by Eq.~\ref{eq:bulkvcoupling}. This term comes from the disconnected contractions in Eq.~\ref{eq:general2point_expanded} (e.g. $\left\langle\Theta\Theta\right\rangle\left\langle\delta\delta\right\rangle$ and $\left\langle\Theta\delta\right\rangle\left\langle\Theta\delta\right\rangle$). Note that we do not include the non-Gaussian contributions to the estimator noise in the present analysis (e.g. the $N^{(3/2)}$ and $N^{(1)}$ noise terms, in the terminology of Ref.~\cite{Giri:2020pkk}); see Appendix~\ref{sec:variance_appendix} for discussion. 
\item $\mathrm{\bf{N}}^{fine}_{L}$: the estimator variance coming from the fine mode bias Eq.~\ref{eq:finemodenoisedef}, with coupling functions for the bulk and fine modes of the radial velocity field defined by Eq.~\ref{eq:bulkvcoupling} and~\ref{eq:finevcoupling}, respectively. The relative importance of this term decreases with an increasing number of bins; we explore this in detail below.
\item $\mathrm{\bf{N}}^{\Upsilon}_{L}$: the estimator variance due to the moving lens effect, defined by Eq.~\ref{eq:otherfieldbiasdef} using the coupling function for the transverse velocity Eq.~\ref{eq:bulkvcoupling}.
\item $\mathrm{\bf{N}}^{cal}_{L}$: the estimator variance due to galaxy survey calibration error systematics, defined by Eq.~\ref{eq:otherfieldbiasdef} using the coupling function for the calibration error Eq.~\ref{eq:calibrationcoupling}.
\item $\mathrm{\bf{N}}^{lens}_{L}$: the estimator variance due the lensing of the primary CMB, defined by Eq.~\ref{eq:otherfieldbiasdef} using the coupling function for the lensing potential Eq.~\ref{eq:lensingcoupling}.
 \end{itemize}
Note that we refer to the contribution $\mathrm{\bf{R}}_L \mathrm{\bf{C}}^{vv}_{L}(\mathrm{\bf{R}}_L)^{\dagger}$ as the `signal' and all other terms as the `noise' in the discussion that follows.
 
\begin{figure}[ht]
  \includegraphics[scale=0.4]{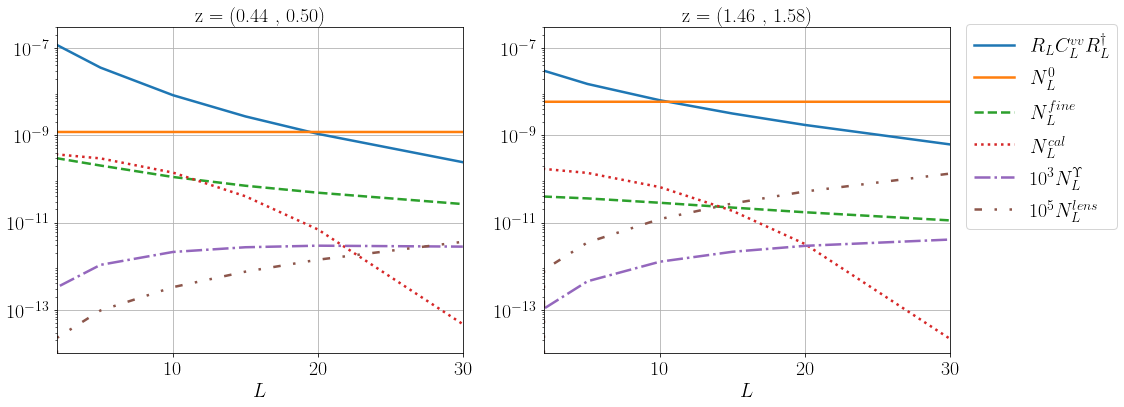}
  \caption{Radial velocity signal and noise sources at redshift bins $z = (0.44\;,\;0.50)$ and $z=(1.46\;,\;1.58)$, corresponding to bins 4 and 16 of 32.}
  \label{fig:diag_rd}
\end{figure}

In Fig.~\ref{fig:diag_rd} we show a few diagonal elements (i.e. $\alpha = \beta$) of Eq.~\ref{eq:rd_twopoint} for a near bin at $z\sim 0.5$ and far bin at $z \sim 1.5$ for SO x LSST with 32 bins. The dominant source of reconstruction noise is the $\mathrm{\bf{N}}^{0}_{L}$ term, followed by the fine mode and calibration error contributions to the variance. The variance arising from the transverse velocity potential and lensing are negligibly small compared to the Gaussian estimator noise; we therefore neglect these terms in our analysis below. 

\begin{figure}[ht]
  \includegraphics[scale=0.45]{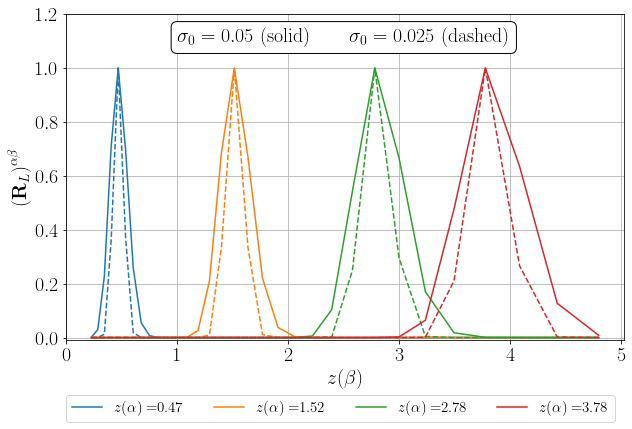}
  \caption{Rows 4, 16 , 24 and 28 of the rotation matrix, corresponding to redshift bins with central redshifts 0.47, 1.52, 2.78 and 3.78. For a given row (fixed $\alpha$), we plot the elements of the row as a function of their redshifts $z(\beta)$. Solid (dashed) lines are obtained using the photometric redshift error model described in Sec.~\ref{sec:Rubin-like} with $\sigma_0 = 0.05$ ($\sigma_0 = 0.025$). As expected, the mixing of radial information is reduced for smaller redshift errors.}
  \label{fig:rotation}
\end{figure}

There are significant bin-bin correlations in the estimator variance both due to the signal and the various noise terms. Photometric redshift errors in the galaxy surveys lead to mixing of radial information that contributes to the bin-bin correlation, and this radial mixing is captured by the rotation matrix $\mathrm{\bf{R}}_L$. In Fig.~\ref{fig:rotation} we show, for fixed $L$, the radial mixing for a set of redshift bins and illustrate how the mixing decreases when the photometric errors are smaller. The rotation matrix is found to be largely independent of the multipole $L$ for $L\lesssim 200$.

\begin{figure}[ht]
  \includegraphics[scale=0.45]{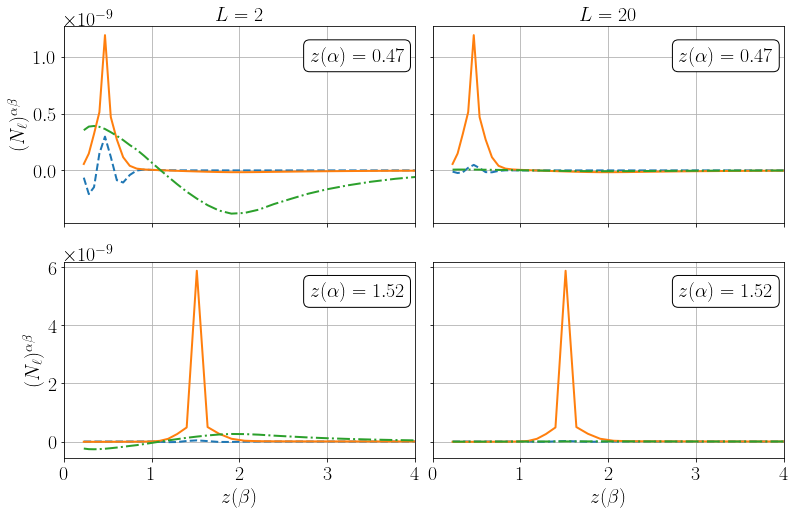}
  \caption{Contributions to the noise covariance matrix from the Gaussian reconstruction noise $\mathrm{\bf{N}}^{0}_{L}$ (solid orange), the fine mode noise $\mathrm{\bf{N}}^{fine}_{L}$ (blue dashed), and the calibration error noise $\mathrm{\bf{N}}^{cal}_{L}$ (green dot-dashed). We show rows 4 (top panels) and 16 (bottom panels) of the noise matrices, corresponding to $z\simeq 0.47$ and $z \simeq 1.52$ respectively, for multipoles $L=2,20$.}
  \label{fig:off_diag_rd}
\end{figure}

In Fig.~\ref{fig:off_diag_rd} we show, for fixed $L$, the contributions to the bin-bin covariance from the various noise terms. The Gaussian reconstruction noise is correlated between bins, mainly due to the the correlation between structures in nearby bins induced by the redshift error in the galaxy survey. This is the largest contribution to the bin-bin covariance in nearby bins, independent of $L$. There is a less significant, but non-negligible, short-range correlation induced by the fine-mode noise which is most important at low-$L$. We observe that the bias from the calibration error induces long-range bin-bin correlations in the estimator variance, as expected due to our assumption that the calibration error is the same for each bin. Had we assumed a different calibration error in each bin, there would be no such correlation.  

\subsubsection{Principal components}
\begin{figure}[ht]
  \includegraphics[scale=0.3]{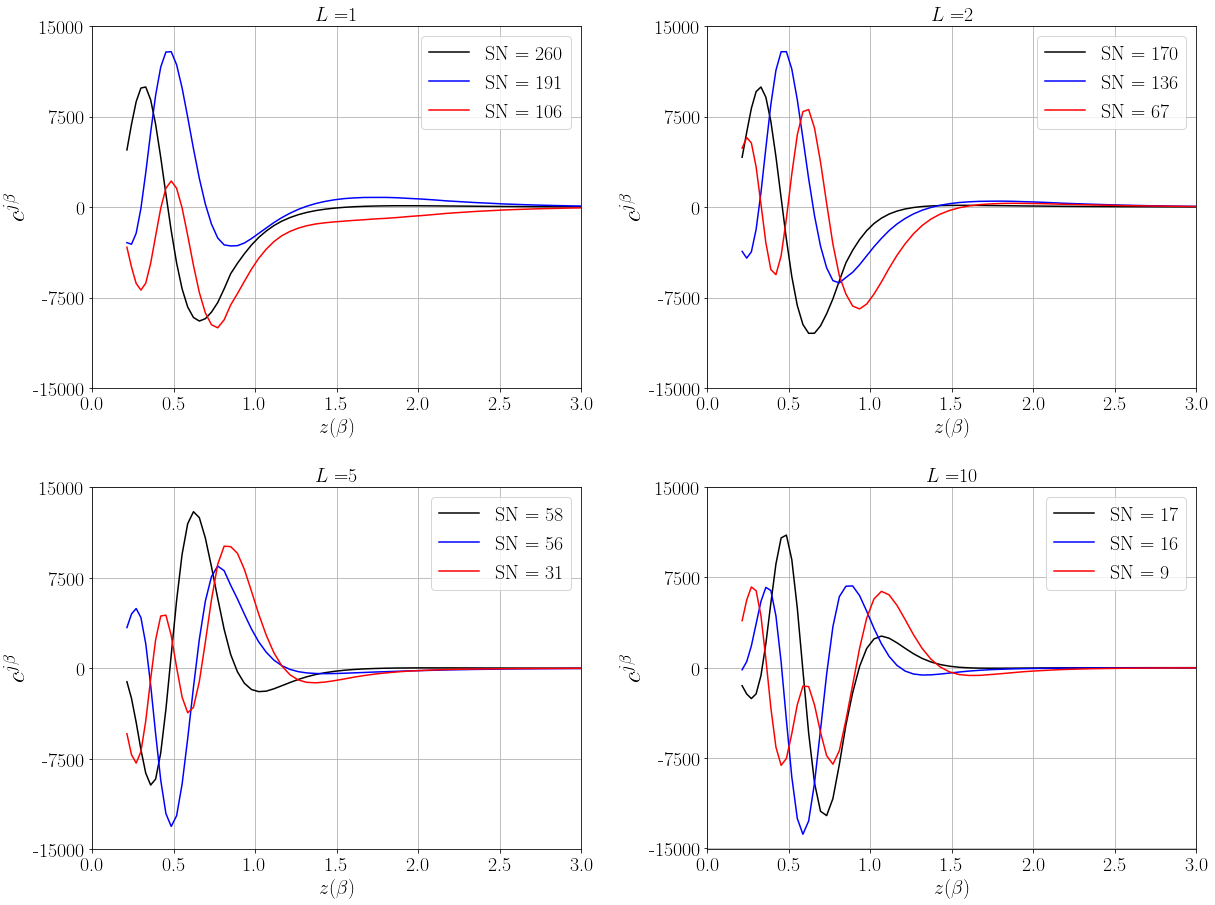}
  \caption{Coefficients $c_{L}^{j \beta}$ for the three largest signal to noise principal components $\sum_\beta c_{L}^{j \beta} \hat{v}^\beta_{LM}$ for a reconstruction with 64 redshift bins, at various $L$ multipoles. }
  \label{fig:pc_64}
\end{figure}

In light of the significant bin-bin covariance present at all scales $L$ in both the signal and the noise terms in the bin basis, it is instructive to consider the principal component basis where there is no covariance. The transformation to the principal component basis was outlined above in Sec.~\ref{subsec:principal} and defined by:
\begin{equation}
(\hat{v}^{P^j})_{LM} = \sum_\beta c_{L}^{j \beta} \hat{v}^\beta_{LM}\,.
\end{equation}
Note that we employ the full signal covariance and all noise terms in Sec.~\ref{eq:rd_twopoint} to define the principal components. In Fig.~\ref{fig:pc_64} we show the $j=1,2,3$ principal component coefficients $c_{L}^{j \beta}$ as a function of bin $\beta$ for $N=64$ bins at $L=1, 2, 5, 10$. Note that at each scale $L$, the weight for the most significant principal components receives support primarily from lowest redshifts. This is where the galaxy density is relatively high (hence the reconstruction noise is minimized) and the amplitude of velocities is relatively large (e.g. due to linear growth). In addition, the number of nodes along the radial direction increases with $L$ and for the lower signal to noise principal components at fixed $L$. 

\begin{figure}[h!]
  \includegraphics[scale=0.35]{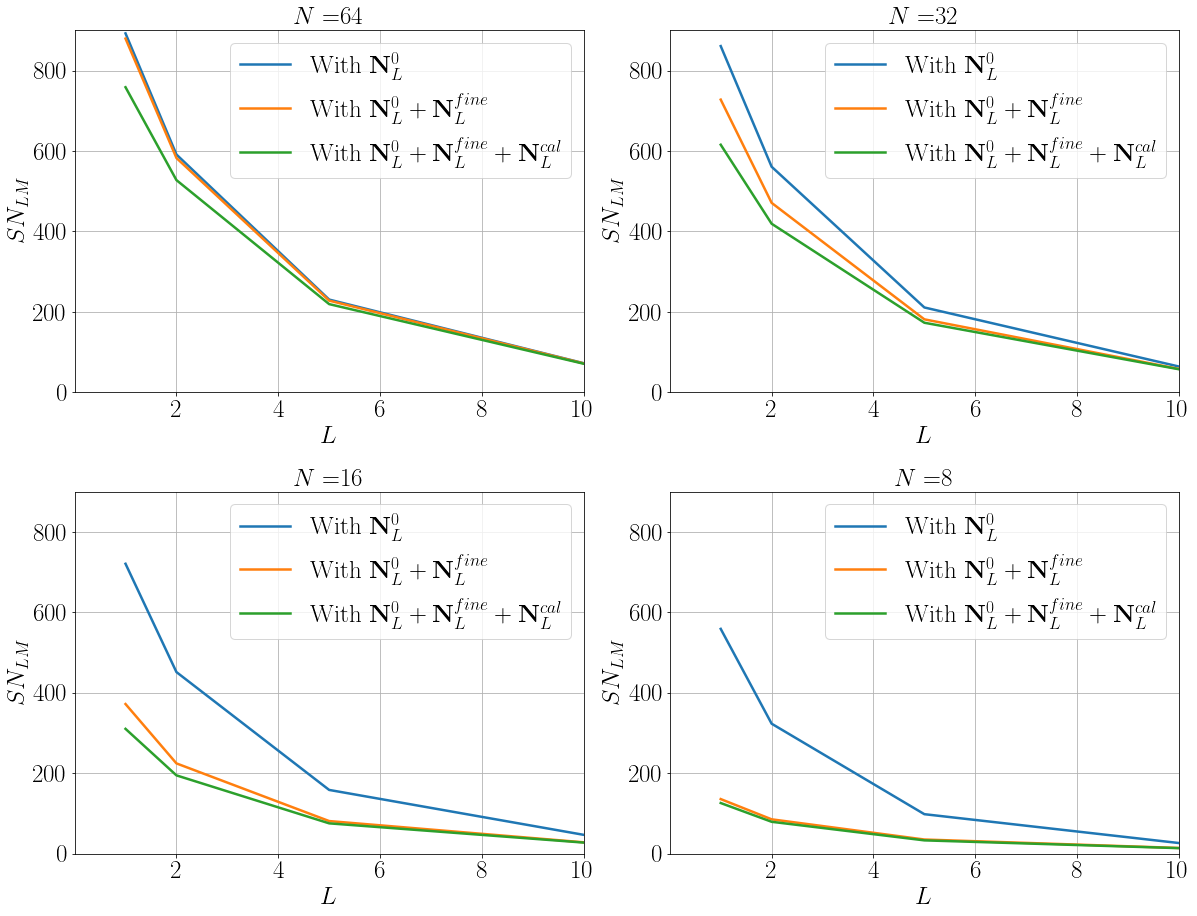}
  \caption{Signal to noise per mode as defined in Eq.~\ref{eq:SNRpermode}, as a function of $L$, for different binning and sources of noise. The loss of signal to noise due to the fine mode noise is accentuated for wider redshift bins.}
  \label{fig:sn_rd}
\end{figure}

In the principal component basis, we can define a measure of the total signal to noise per mode $LM$ by
\begin{equation}
SN_{LM} = \sum_{j=1}^{N} \langle (\hat{v}^{P^j})_{LM}  (\hat{v}^{P^j})_{LM}^\dagger  \rangle \,.
\end{equation}
We evaluate this quantity in Fig.~\ref{fig:sn_rd}. Each panel of that figure compares, for a reconstruction with $N$ redshift bins, the effect of adding the $\mathrm{\bf{N}}^{fine}_{L}$ and $\mathrm{\bf{N}}^{cal}_{L}$ compared with the Gaussian reconstruction noise $\mathrm{\bf{N}}^{0}_{L}$. As expected, with an increasing number redshift bins, the fine mode contribution becomes less important and the calibration error becomes the leading correction to the Gaussian reconstruction noise.  The reconstruction of the $\Pi$-binned LC moments of the radial velocity suffer a considerable loss of signal to noise per mode as we reduce the number of bins even when only including the Gaussian reconstruction noise. For our binning scheme, $N=64$ corresponds to redshift bins of equal comoving size of approximately 110 Mpc. The coherence length of the velocity field is around 70 Mpc, and therefore it makes sense that the fine modes become more relevant for $N=32$ and smaller, as the size of the bins are considerably larger than 70 Mpc. Even with $N=64$ bins, comparing the orange and green curves, we see that calibration error leads to a significant degradation in $SN_{LM}$ of greater than $10 \%$. Efforts to mitigate systematics in galaxy surveys on large angular scales can therefore meaningfully impact the fidelity of the reconstruction. Regardless, we see that velocity reconstruction with SO x LSST will have exceedingly high SNR on large angular scales, with $SN_{LM} > 1$ for $L < 30$ with the most significant principal component.

\subsubsection{Optical depth bias}\label{opticaldepthbiasksz}

As discussed in Sec.~\ref{sec:mult_bias}, incorrect modelling of the correlation between the electron and galaxy density leads to a multiplicative bias on the reconstructed radial velocity, commonly referred to as the optical depth bias. We illustrate how this bias shows up in our formalism by considering a one parameter toy model for the electron-galaxy correlation function, based on the Halo Model. If we fix the model parameters determining how galaxies inhabit dark matter halos, the electron-galaxy cross-power is determined by the model for the electron density profile inside dark matter halos (see Appendix~\ref{sec:halomodel} for more details). Due to physical processes such as AGN feedback, baryonic matter does not trace dark matter inside of halos. In Fourier space, this translates into an electron density profile $u_e(k,M,z)$ that is different from the dark matter density profile $u(k,M,z)$. To explore a one-parameter family of models, we construct the following toy model for the electron density profile:
\begin{equation}\label{eq:opticaldepthmodel}
u^{(toy)}_e(k,M,z) = \frac{u_e(Ak,M,z)}{u(Ak,M,z)}u(k,M,z)\,,
\end{equation}
where $A$ is a continuous parameter that interpolates between the dark matter profile ($A=0$) and our fiducial model ($A=1$) that incorporates feedback. If we take our fiducial model to be the true model for the electron density, we can explore the optical depth bias when `incorrect' $A\neq1$ profiles are used for velocity reconstruction. 

\begin{figure}[h!]
  \includegraphics[scale=0.4]{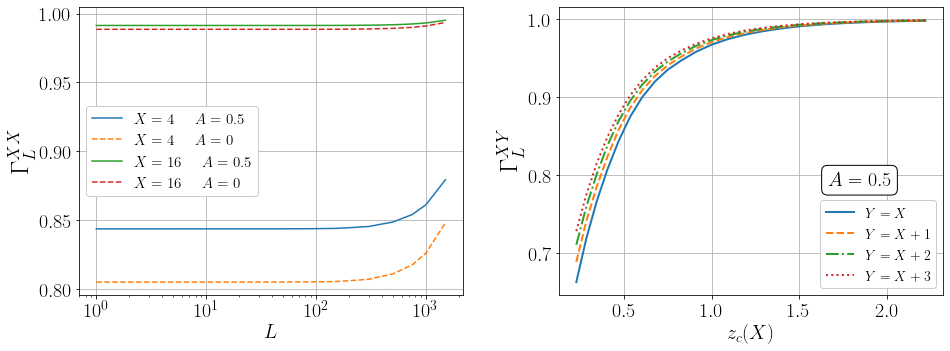}
  \caption{\textbf{Left panel:} Diagonal elements of the optical depth bias matrix as a function of the multipole $L$. Bin 4 and bin 16 correspond to redshift ranges $(0.44,0.50)$ and $(1.46,1.58)$ respectively. Solid (dashed) lines correspond to $A = 0.5\;(A=0)$. \textbf{Right panel:} Diagonal and some off-diagonal elements of the bias matrix as a function of redshift. Bias tends to 1 for higher redshifts as electrons trace dark matter more closely at earlier times.}
  \label{fig:odbias}
\end{figure}

Fig.~\ref{fig:odbias} shows the behaviour of the elements of the bias matrix Eq.~\ref{eq:gamma} as a function of the angular scale (left panel) and redshift (right panel). We find that the bias is practically independent of $L$ for large angular scales $L\lesssim 200$ (in accordance to what was found in \cite{Giri:2020pkk}) and that it is less significant at higher redshifts, where the difference between electron and dark matter perturbations is less pronounced. The off-diagonal elements of the bias matrix are similarly scale-independent and approach 1 at high redshift. The $L$ independence on large angular scales is a feature that we expect to be robust independently of the models under consideration. Note that the bias is always less than 1 for this family of models. This is because feedback in the fiducial model causes the electron halo profiles to be more diffuse than their host dark matter halos, leading to a power suppression at high-$k$, and therefore on small angular scales where the sums in Eq.~\ref{eq:gamma} receive the most weight. Even in the extreme case where baryons are assumed to trace dark matter, the magnitude of the bias over the entire range of redshifts lies within a reasonably small range $0.6 \alt \Gamma_L^{\rm XY} < 1$.

To obtain cosmological constraints from velocity reconstruction using future datasets, it will be necessary to incorporate the optical depth bias into the analysis. For example, if we wish to obtain constraints on a set of cosmological parameters ${\bf m}$ appearing in the radial velocity power spectrum $\mathrm{{\bf C}}^{vv}_{L} ({\bf m})$, it is necessary to compare (via e.g. a likelihood function) the measured velocity spectra to the model:
\begin{equation}
(C^{\rm recon})_L^{\alpha \beta} =
\left[ \left( {\bf \Gamma} \mathrm{{\bf R}} \right) \mathrm{{\bf C}}^{vv}_{L}({\bf \Gamma} \mathrm{{\bf R}})^{\dagger}\right]^{\alpha\beta}+
\left(\mathrm{\bf{N}}_{L}\right)^{\alpha\beta} \,,
\end{equation}
where $\left(\mathrm{\bf{N}}_{L}\right)^{\alpha\beta}$ includes the most relevant noise terms (e.g. $N^{(0)}$, fine-mode, calibration) and $\left( {\bf \Gamma} \mathrm{{\bf R}} \right)^{ij} = \Gamma^{ij} (u_e) \mathrm{R}^{ij} (\sigma_z)$, where we assume the optical depth bias and rotation matrix are independent of $L$ (a good approximation, as shown above) and indicate explicitly the dependence on the electron profile $u_e$ and redshift error $\sigma_z$. To get access to the cosmological information contained in $\mathrm{{\bf C}}^{vv}_{L} ({\bf m})$ it is necessary to encapsulate the redshift errors and electron profile into a set of nuisance parameters that can be marginalized over. In the absence of any modelling, there are $N_{\rm bin}^2$ nuisance parameters. This is the same number of independent entries in $\mathrm{{\bf C}}^{vv}_{L} ({\bf m})$ that determine redshift-redshift correlations, implying that the only residual cosmological information is in the shape of the velocity {\em angular} power spectrum. This impedes one's ability to learn about e.g. the growth function using the reconstructed velocity field. But this scenario is far too pessimistic, as it does not incorporate information from other sources, or physical constraints present in the modelling. 

Optimistically, it may be sufficient to characterize redshift errors and the electron profile by a small number of model parameters. For example, the fiducial model of Gaussian redshift errors considered above contains a single parameter $\sigma_0$. Assuming for the moment that this is an accurate model for LSST redshift errors, there is a single model parameter associated with the rotation matrix. In addition, one can put a prior on the ranges this parameter might take by using other available information: the galaxy-galaxy power spectrum itself, simulations, comparing with a spectroscopic survey, etc. Likewise, if Eq.~\ref{eq:opticaldepthmodel} is a reasonable description of the range of possible electron profiles, then a single model parameter would determine the optical depth bias. Again, one could incorporate additional measurements to provide a prior on $A$, for example by independently measuring the galaxy-electron cross power using Fast Radio Bursts~\cite{Madhavacheril:2019buy} or by correlating the reconstructed velocity field with the galaxy survey~ \cite{sugiyama2017kinematic,Smith:2018bpn,Contreras_2019} or the reconstructed transverse velocity field~\cite{Hotinli:2021hih}. In reality, there are likely more than two model parameters to consider to fully characterize redshift errors and the electron profiles. But by evaluating a range of physical models and finding complementary observations, one can likely put an informative prior on the $N_{\rm bin}^2$ degrees of freedom in the product of the rotation matrix and optical depth bias. 

In a number of previous analyses, e.g. Refs.~\cite{Pan:2019dax,Cayuso:2019hen,Contreras_2019,Hotinli:2019wdp}, it was assumed that the rotation matrix was diagonal, and therefore that the optical depth bias consisted of $N_{\rm bin}$ parameters to be marginalized over. In the presence of photometric redshift errors, we have seen that this assumption does not hold; the off-diagonal nature of the rotation matrix gives rise to greater than $N_{\rm bin}$ parameters. How many additional parameters need to be incorporated depends on how dominant the diagonal terms in $\left(\mathrm{\bf{N}}_{L}\right)^{\alpha\beta}$ are compared with the off-diagonal terms, since small off-diagonal terms can be neglected. This depends primarily on the magnitude of redshift errors, so more accurate photometric redshifts, or spectroscopic redshifts can simplify the analysis of the reconstructed velocity field. In future work, cosmological forecasts and analyses should take into account the off-diagonal terms in the optical depth bias, either through a physical model or by marginalizing over a sufficient number of degrees of freedom.

\subsection{`Double' SO and pre-reconstruction vs post-reconstruction cleaning}

\begin{figure}[h]
  \includegraphics[scale=0.6]{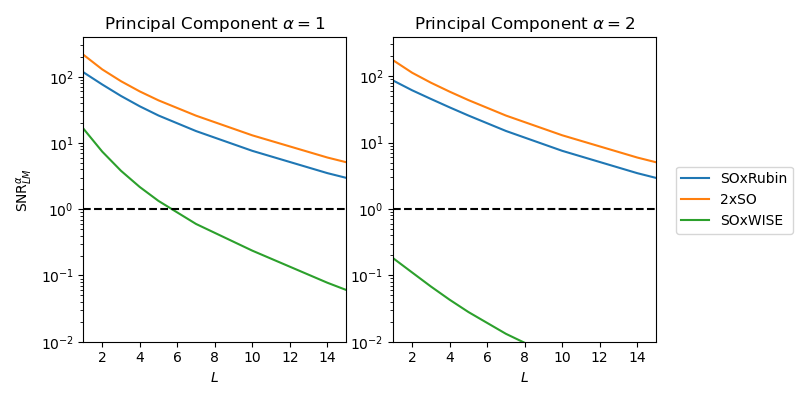}
  \caption{SNR of principal components 1 and 2 of velocity reconstruction.}
  \label{fig:doublesosnr}
\end{figure}

In this section, we investigate two scenarios related to the effect of foregrounds on the reconstruction. First, we investigate whether or not additional frequency channels can help with mitigating the effect of foregrounds on the reconstruction. To do so, we define a hypothetical experiment we refer to as 'Double' SO, which has a set of channels (in GHz) at: 47, 52, 63, 75, 91, 109, 131, 158, 190, 228, 275, 330, 397, 478, 575, 691, 831, 1000. The boundaries and spacing of this selection were chosen to minimize residuals in the cleaned CMB temperature spectrum for our foreground model. Including frequencies below $\sim$50 GHz and above $\sim$1000 GHz provides no improvement for removing extragalactic foregrounds. Our choice of 12 frequency channels in the relevant range is somewhat arbitrary, and is simply meant to be representative of a reasonable number of detectors as compared to SO. To define the noise properties of Double SO, we first take the SO LAT TT noise model \cite{Ade:2018sbj} assumed above and define a linear interpolating function on the three free parameters in the noise model, extrapolating when necessary to higher frequencies that are not in the SO selection. We then analyzed the reconstruction noise for the fiducial $N=64$ bin case assumed for SO x LSST above. In Fig.~\ref{fig:doublesosnr} we show the signal to noise for the first two principal components over a range of scales for SO and Double SO x LSST. It can be seen that the Double SO experiment (true to its name) yields a signal to noise that is about twice as good as SO. This is due to a combination of a lower effective noise in the auto-power as well as a reduction of foreground residuals in the cross-power. This result illustrates the great room for progress in velocity reconstruction with future instruments.

Above, we considered the scenario where a linear combination of CMB maps was used to remove foregrounds before velocity reconstruction. It is also possible to perform velocity reconstruction on each frequency map, and then find the linear combination of reconstructions that minimizes the variance of the level of the reconstruction. To do so, we describe the power spectrum of the reconstruction as 
\begin{equation}
{\bf C}_{\ell}=C_{\ell}^{vv}{\bf e}{\bf e}^{t}+{\bf N}_{\ell}\,.
\end{equation}
We can then apply the same harmonic space ILC method defined above to find a map that minimizes the variance due to reconstruction noise: 
\begin{equation}
C_{\ell}^{{\rm clean}}={\bf w}_{\ell}^{\dagger}{\bf N}_{\ell}{\bf w}_{\ell}+C_{\ell}^{vv}\,,
\end{equation}
where the ILC weights $\bf{w}_\ell$ are defined using the reconstructed spectra:
\begin{equation}
{\bf w}_{\ell}=\frac{{\bf C}_{\ell}^{-1}{\bf e}}{{\bf e}^{t}{\bf C}_{\ell}^{-1}{\bf e}}\,.
\end{equation}
Since we know the signal $C_\ell^{vv}$ we can subtract this from the reconstruction in the pre-reconstruction cleaning scenario to arrive at the residual noise, which we compare directly to ${\bf w}_{\ell}^{\dagger}{\bf N}_{\ell}{\bf w}_{\ell}$ from the post-reconstruction cleaning scenario. We find that the residuals for pre-reconstruction cleaning are smaller than the residuals for post-reconstruction cleaning. Therefore, we focus on the  scenario where foregrounds are mitigated before reconstruction is performed.

\subsection{Radial velocity reconstruction for SO x unWISE}

\begin{figure}[h]
  \includegraphics[scale=0.5]{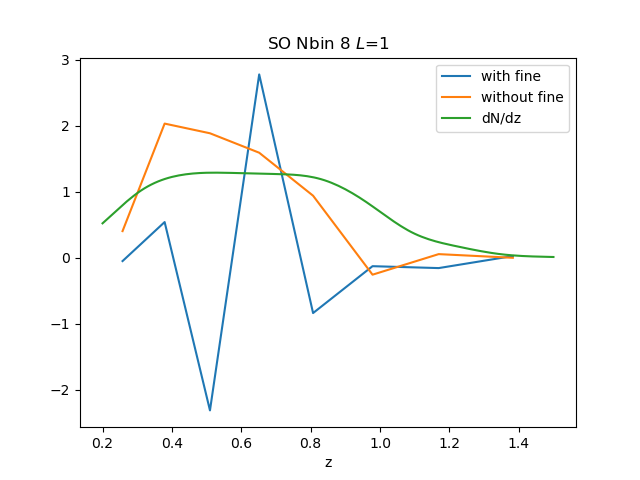}
   \includegraphics[scale=0.5]{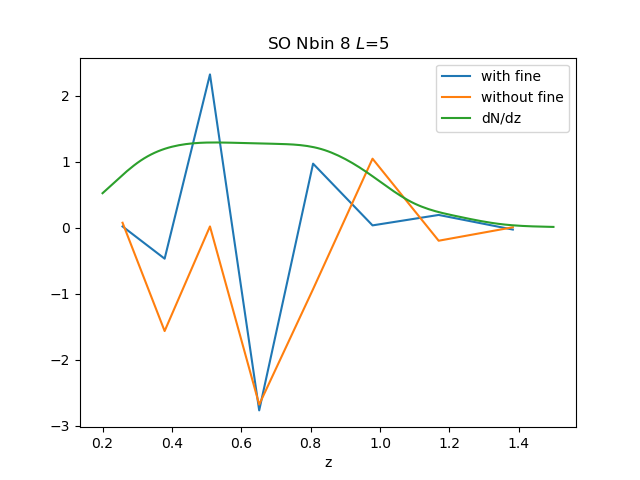}
  \caption{First principal component of the velocity reconstruction for SO x unWISE.}
  \label{fig:pcs}
\end{figure}

We now turn to the second scenario we consider, where velocity reconstruction is performed with SO and data currently available from the unWISE blue sample. Here, there is a single galaxy window function, which is plotted in Fig.~\ref{fig:pcs}. We consider a reconstruction using 8 bins in the redshift range between $0.2<z<1.5$ corresponding to a comoving bin width of $\Delta \chi \simeq 450$ Mpc. To compute the fine mode noise, we use 512 bins in the same redshift range. We increase the calibration error from our LSST framework by a factor of $10^2$ to account for the difference in precision of redshift measurement between the two experiments. Because there is a single galaxy window function, the reconstructed velocity field and reconstruction noise will be highly correlated among the 8 bins in which we perform the reconstruction. Therefore, it is crucial in this case to use the principal component basis. Fig.~\ref{fig:pcs} shows the $\alpha=1$ principal component coefficients $c_{L}^{\alpha \beta}$ both with and without the inclusion of fine-mode noise and calibration errors for $L=1$ and $L=5$. Note that for $L=1$, the first principal component roughly traces out the unWISE window function $dN/dz$ when the fine-mode noise and calibration errors are neglected. However, the first principle component becomes oscillatory once the additional noise is included. This is due to the redshift-redshift correlations of the noise terms obscuring the redshift correlations in the signal. The signal to noise of the first principal component at $L=1$ drops from 25 to 16.4 as the additional noise terms are added. At $L=5$, the first principal component has an oscillatory structure in redshift both with and without the additional sources of noise. The signal to noise of the first principle component at $L=5$ is 1.7 and 1.3 with and without the additional noise terms, respectively. Therefore, most of the signal lies at the lowest $L$. Analyzing the higher principle components, they make an insignificant contribution to the signal to noise at all scales. We therefore can focus on the first principle component only.

\begin{figure}[h]
  \includegraphics[scale=0.6]{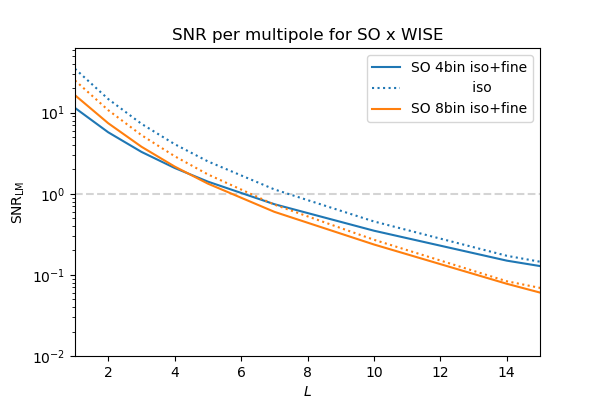}
  \caption{SNR per multipole for SO x unWISE for 4 and 8 coarse bins and 512 fine bins. SNR with and without fine mode contributions are shown.}
  \label{fig:sowisesnr}
\end{figure}

We explore the effect of changing the number of bins used in the analysis by computing the signal to noise $SN_{LM}$ defined in Eq.~\ref{eq:SNRpermode}, summing over principal components at fixed $L$. We find that it is numerically difficult to consider greater than 8 bins. Large bin-bin correlations in the signal covariance and Gaussian reconstruction noise, especially in bins where the redshift distribution is small, lead to poorly conditioned rotation matrices (Eq.~\ref{rot}) that spoil the construction of the principal component basis. Therefore, we consider scenarios with 4 and 8 bins. The result for the signal to noise per mode, summed over principal components, is shown in Fig.~\ref{fig:sowisesnr}. Here, the dependence of the signal to noise on the number of bins is less dramatic than for SO x LSST. This is to be expected, since not much information is gained by finer sampling in redshift due to the fact that there is a single wide galaxy window function. In this figure, we also demonstrate the effect of fine mode noise. For 4 bins, there is a significant correction beyond the Gaussian reconstruction noise. However, we see that for 8 bins, we are able to improve on the signal to noise in the presence of fine mode noise.

Finally, in Fig.~\ref{fig:doublesosnr} we compare the $SN_{LM}^\alpha$ attainable for SO x unWISE compared with SO x LSST. The signal to noise per mode for the first principle component for SO x unWISE is roughly an order of magnitude lower than for SO x LSST; for the second principle component the difference is three orders of magnitude. Although there is a significant galaxy density in the unWISE sample, yielding a small Gaussian reconstruction noise (at least over some range in $L$ for the first principal component), there is little redshift information. We therefore can only expect to obtain coarse-grained knowledge of the velocity field from such an analysis. Nevertheless, this is in principle important information, and the reconstruction of the first principle component at signal to noise greater than unity can be obtained for $L \alt 10$. This represents a modest, but non-trivial, number of well measured modes.

\subsection{Transverse velocity potential reconstruction for SO x LSST}\label{sec:transv_sorubin_forecast}

In this section, we discuss reconstruction of the transverse velocity potential $\Upsilon$ using the ML effect for the fiducial case of SO x LSST. As for the case of the radial velocity field, we must consider both the signal covariance as well as multiple sources of bias and noise:
\begin{equation}\label{eq:ML_twopoint}
\Big{\langle}\hat{(\Upsilon^{\alpha})}_{LM}
(\hat{\Upsilon_{\perp}^{\beta})}^*_{LM}\Big{\rangle} =
\left(\mathrm{\bf{R}}_L\mathrm{\bf{C}}^{\Upsilon \Upsilon}_{L}(\mathrm{\bf{R}}_L)^{\dagger}\right)^{\alpha\beta} +
\left(\mathrm{\bf{N}}^{0}_{L}\right)^{\alpha\beta} + \left(\mathrm{\bf{N}}^{\rm fine}_{L}\right)^{\alpha\beta} +
\left(\mathrm{\bf{N}}^{\rm cal}_{L}\right)^{\alpha\beta} + \left(\mathrm{\bf{N}}^{v}_{L}\right)^{\alpha\beta} +
\left(\mathrm{\bf{N}}^{\rm lens}_{L}\right)^{\alpha\beta}\,.
\end{equation}
These various terms are defined as for the radial velocity estimator variance Eq.~\ref{eq:rd_twopoint}, using the coupling function for the transverse velocity Eq.~\ref{eq:bulkvcoupling}. The term $\left(\mathrm{\bf{N}}^{v}_{L}\right)^{\alpha\beta}$ is the bias induced by radial velocity in the kSZ effect, defined by Eq.~\ref{eq:otherfieldbiasdef} using the coupling function Eq.~\ref{eq:bulkvcoupling}.

\begin{figure}[h]
  \includegraphics[scale=0.4]{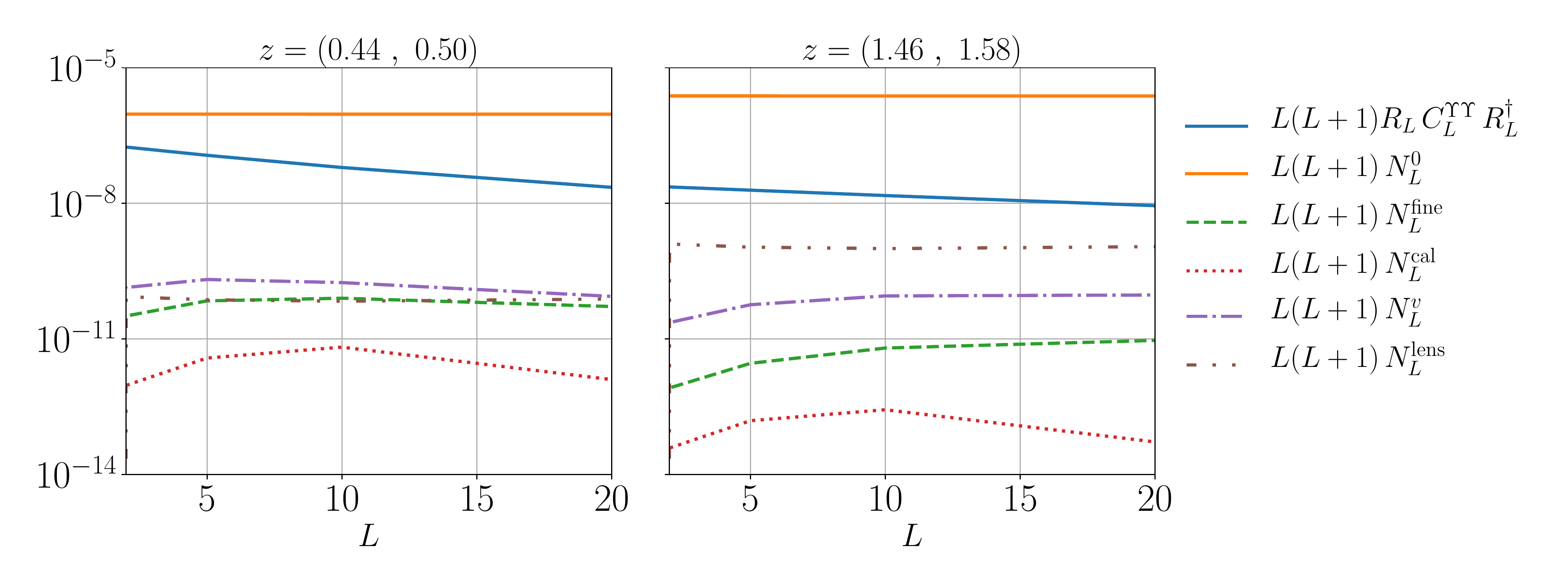}
  \caption{Transverse velocity signal and noise sources at redshift bins $z=(0.44,0.50)$ and $z=(1.46,1.58)$, corresponding to bins 4 and 16 of 32.}
  \label{fig:diagML}
\end{figure}

\begin{figure}[h]
  \includegraphics[scale=0.33]{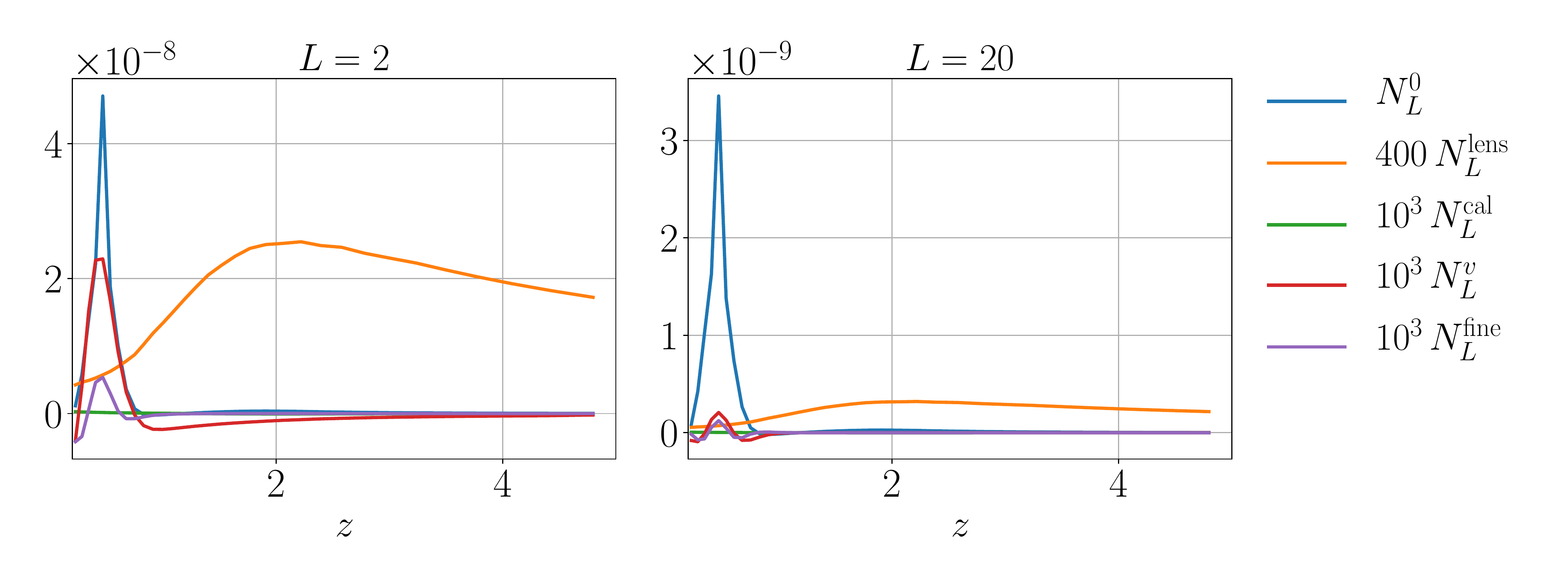}
  \includegraphics[scale=0.33]{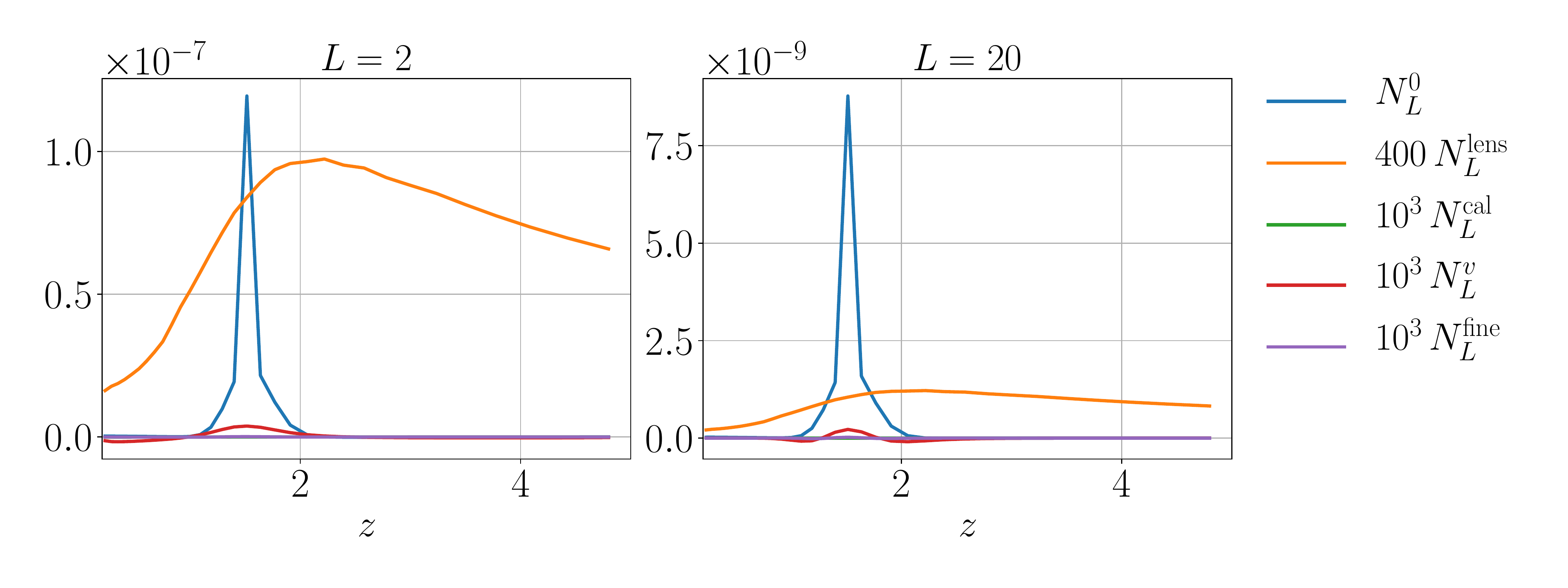}
  \caption{Off-diagonal behaviour of the moving-lens estimator, bin 4 of 32 (top) and bin 16 of 32 (bottom).}
  \label{fig:offdiagML}
\end{figure}

In Fig.~\ref{fig:diagML}, we plot the diagonal components of the various terms in Eq.~\ref{eq:ML_twopoint} for the same two redshift bins as the radial velocity estimator. For SO x LSST with $N=32$ bins, the fidelity of the transverse velocity potential reconstruction in each bin is not high, with the signal covariance below the reconstruction noise for all redshifts and scales $L$. Note that the contribution to the estimator variance from lensing is significant, comprising roughly $10 \%$ of the signal in the high redshift bin shown in the right panel of Fig.~\ref{fig:diagML}. We also explore the off-diagonal correlations fo of the various noise terms, as shown in Fig.~\ref{fig:offdiagML}. The fine-mode and reconstruction noise lead to small and fairly localized contributions to the off-diagonal estimator variance; calibration error in the case of ML is negligible (in contrast to the case of radial velocity reconstruction, where calibration error was significant and led to long-range correlations). Lensing, which is the most significant contribution after the reconstruction noise and signal covariance, leads to long-range correlations between bins.  

\begin{figure}[h]
  \includegraphics[scale=0.3]{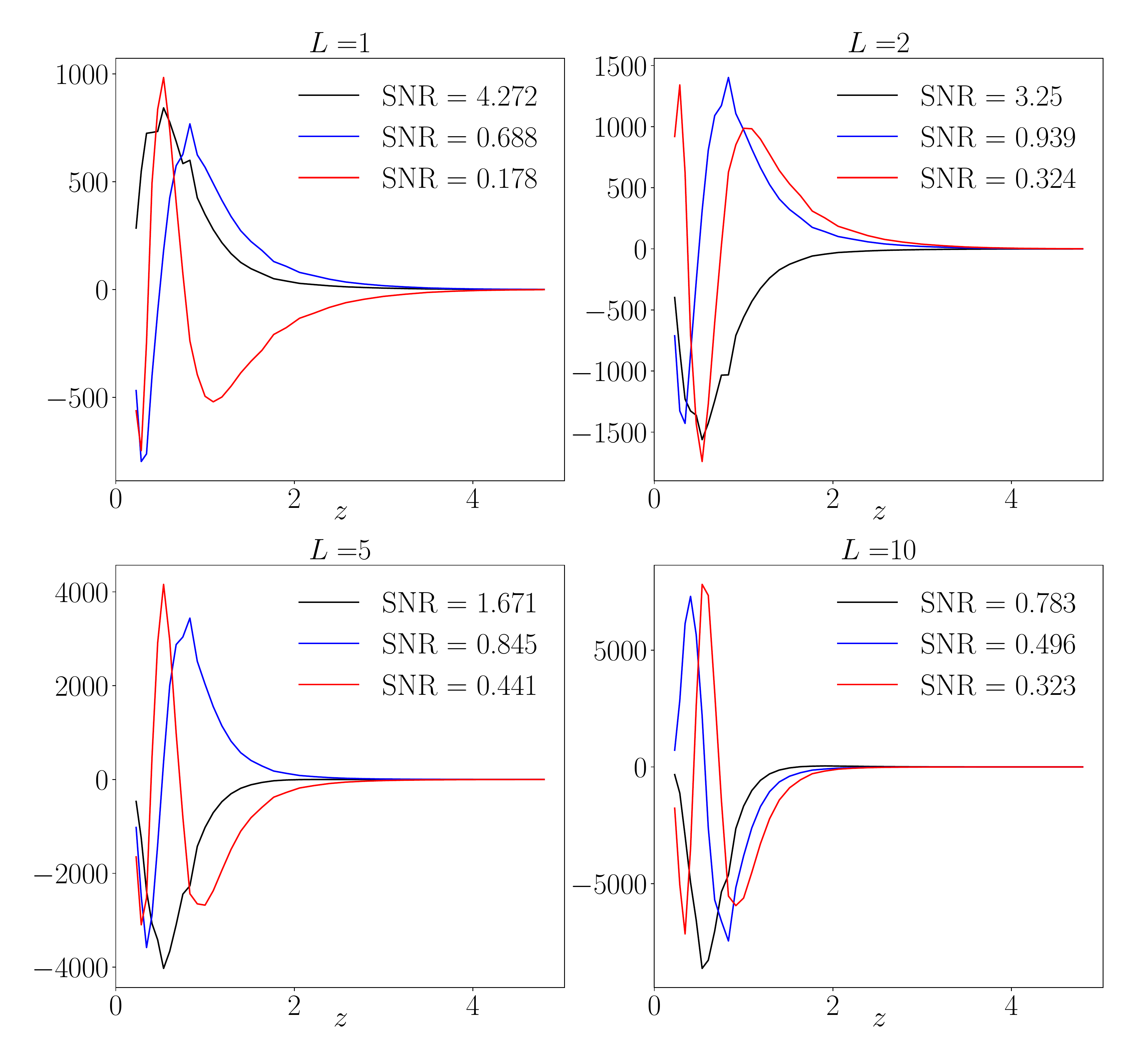}
  \caption{Three largest signal to noise principal components from ML reconstruction for SO x LSST with 32 redshift bins. The $x$-axis is the central redshift corresponding to each bin.}
  \label{fig:ML_PC_coeffs}
\end{figure}

\begin{figure}[h]
  \includegraphics[scale=0.468]{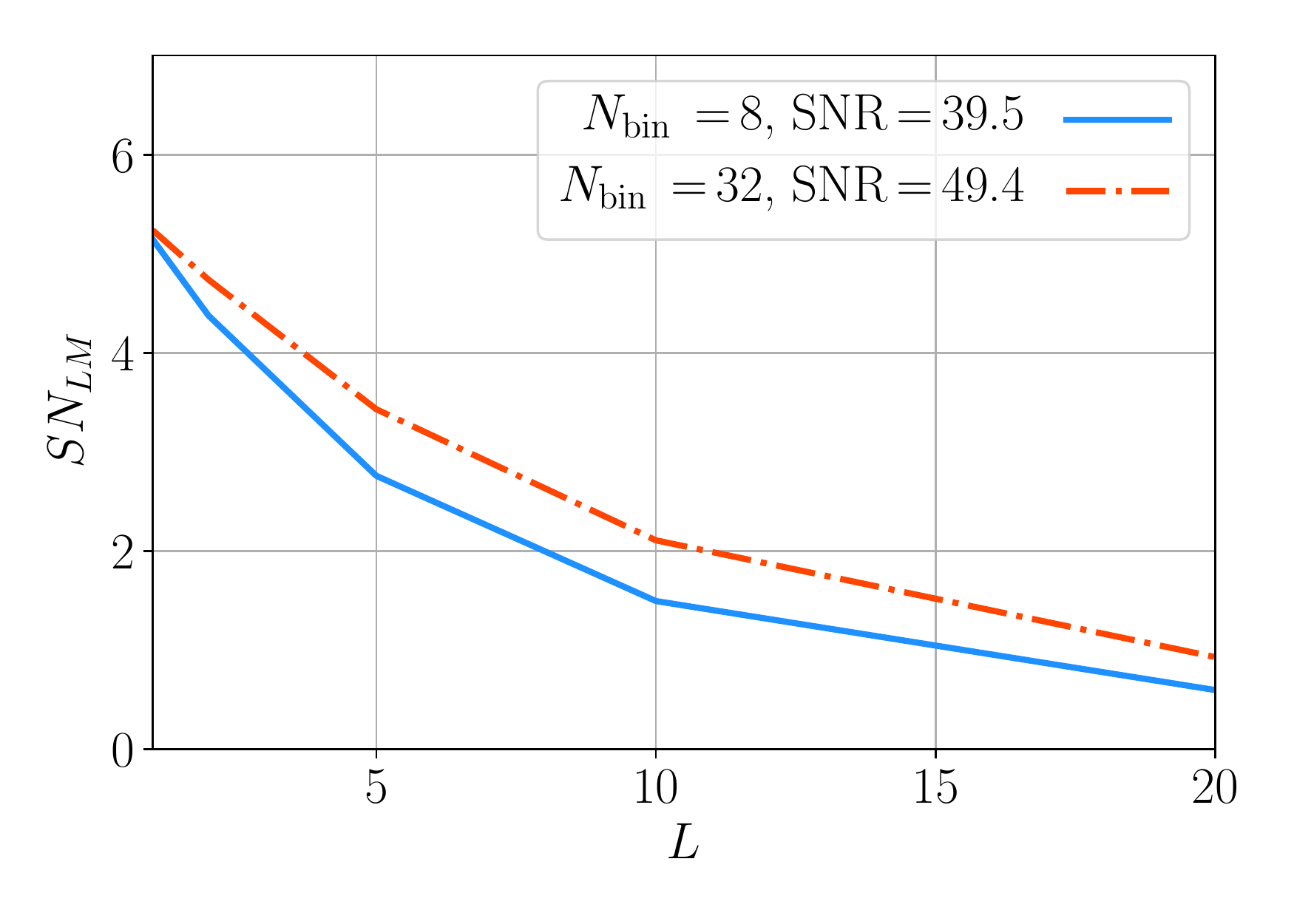}
  \caption{Signal to noise per mode for the ML reconstruction, as a function of $L$, for different binning and sources of noise. Note the fine mode noise and calibration error are too small to affect the SNR.}
  \label{fig:ML_PC_SNR}
\end{figure}

Given the degree of bin-bin correlation, it is useful to define a principal component basis for the transverse velocity potential estimator. The coefficients for the first few principal components are shown in Fig.~\ref{fig:ML_PC_coeffs} over a variety of scales $L$. The first principal component, which is a weighted average of the transverse velocity potential over a reasonable range in redshift, has a significant signal to noise over a reasonable range of scales $L \alt 10$. The next principal components have an increasing number of nodes in the radial direction. There is less structure generally in the radial direction than for the radial velocity estimator, owing mainly to the lower signal to noise. In Fig.~\ref{fig:ML_PC_SNR} we show the total signal to noise per mode $SN_{LM}$ at scales $L$ summed over all principal components. As can be seen in this figure, the signal to noise per mode is most significant at the largest scales, falling below unity at $L \sim 15$. In contrast to radial velocity reconstruction, the increase in signal to noise per mode does not increase dramatically with an increasing number of bins.

\section{Velocity reconstruction pipeline}\label{sec:recon_pipeline}

In this section we assess the performance of velocity reconstruction with future datasets using a suite of simulations and a reconstruction pipeline based on the quadratic estimators described in previous sections. Previous work has demonstrated the effectiveness of the quadratic estimator for reconstruction of the radial velocity field using N-body simulations both on the light cone~\cite{Cayuso:2018lhv} and in the box geometry~\cite{Giri:2020pkk}. There has been no previous work demonstrating the feasibility of transverse velocity potential reconstruction with the moving lens effect, a gap which we fill with the present work. Here, we focus on simulated data that consists of properly correlated random Gaussian fields including: the velocity field, galaxy number counts with photometric redshift errors, the electron density field, the primary CMB, the kSZ and moving lens contributions to the CMB, and extragalactic foreground contributions to the CMB. We develop a reconstruction pipeline for the radial and transverse velocity fields using fast real-space versions of the quadratic estimators described above. Theoretical modelling is an important component of velocity reconstruction, since it appears in the estimator for the $\Pi$-binned moments of the velocity fields and also in the rotation matrices required to de-bias the estimators. This makes a combined pipeline including both the simulation of the maps and the application of the estimators essential.

The benefit of using a Gaussian simulation framework is that the ensemble-average properties of the estimator are well-understood on the full sky using the results of previous sections, which allows us to validate the analysis pipeline. Another benefit is that we can isolate and investigate the effect of map-based systematics such as masking on the reconstruction to compare with results on the full sky. Since the generation of correlated random Gaussian fields is far less computationally intensive than running a suite of N-body simulations, it is possible to explore ensemble averages, and quantify underlying numerical inaccuracies or biases. A disadvantage of this approach is that we miss important non-linear contributions to the reconstruction. As shown in Ref.~\cite{Giri:2020pkk}, for radial velocity reconstruction this includes a contribution to the reconstruction noise analogous to the $N^{(3/2)}$ bias in lensing reconstruction~\cite{Bohm:2016gzt}. At the resolutions considered in Ref.~\cite{Giri:2020pkk}, this was in fact larger than the Gaussian contributions to the reconstruction noise by a factor of $\sim 2$. At the somewhat lower resolution and higher instrumental noise we consider, we expect this contribution to be smaller, and sub-dominant to the Gaussian contributions. Another non-linear effect included in Ref.~\cite{Cayuso:2018lhv} is redshift space distortions, which were found to have minimal impact on the reconstruction at the resolutions simulated. All previous work has relied on dark matter-only N-body simulations, making the approximation that baryons follow the N-body particles. This assumption will fail on the small scales relevant for velocity reconstruction. Under the assumption of statistical isotropy made above, this mis-modelling of baryons shows up as a multiplicative bias (see Sec.~\ref{opticaldepthbiasksz} for a discussion), but at the non-linear level there may be additional effects. Future work with simulations should certainly include baryonic effects to explore the impact on simulations at the non-linear level.

\subsection{Simulations}

In this work, we approximate the primary CMB, galaxy number counts, components of the velocity field, electron density, moving lens potential, and extragalactic foreground contributions to the CMB as correlated random Gaussian fields. Using the complete set of spectra and cross-spectra between these fields, we can construct a multivariate Gaussian distribution from which to drawn properly correlated realizations. These realizations can be used to compute the kSZ and moving lens contributions to the CMB.  Signals constructed this way will show the expected statistical anisotropy when correlated with the galaxy density. For each set of realizations, the quadratic estimator for the underlying radial velocity or transverse velocity potential can be applied, allowing us to validate the statistics of the estimators by averaging over many realizations. Here is the list of steps we take to generate a suite of Gaussian simulations for radial velocity and transverse velocity potential reconstruction:

\begin{enumerate}
    \item \textbf{Determine fields for simulation:} The first step we take is to determine which fields need to be simulated `simultaneously'', that is, from a single multivariate Gaussian distribution capturing all the crucial correlations. Ideally, all of the cosmological fields we consider in this paper should be simulated simultaneously. This can become a difficult computational task if we want to simulate the $\Pi$-binned moments of various fields in many redshift bins, which translates into large covariance matrices with non-vanishing off-diagonal terms and many high-resolution maps. The smaller the covariance matrix, the more likely it is that numerical errors can be avoided, so there is good motivation to be as economical as possible. We can ask ourselves, for example, which fields are necessary for a simulation of the kSZ signal. Certainly, joint simulations of the $\Pi$-binned moments of the radial velocity $v$, the differential optical depth $\dot{\tau}$, and the galaxy fields $g$ are necessary if we want to ensure that the kSZ-g cross-correlation has the correct statistical anisotropy. Having identified these completely necessary fields for velocity reconstruction, we can ask ourselves if the fields that source other forms of temperature-galaxy statistical anisotropy should also be considered in the multivariate Gaussian distribution. The analysis of Sec.~\ref{sec:recon_analysis} answers this question for us: the bias introduced by the moving lens effect and the CMB lensing are negligible. This means that, for simulating radial velocity reconstruction, the moving lens and the CMB lensing signals can be simply treated as `effective' sources of CMB anisotropies with no statistical correlation with the galaxy distribution. Finally, we ask ourselves if the fields that are isotropically correlated with the galaxy distribution need to be simulated together with $v$, $\dot{\tau}$ and $g$. These are the linear late-time ISW signal and the frequency cleaned extragalactic temperature foregrounds. The isotropic correlation between temperature and galaxies appears in the estimator weights Eq.~\ref{gminvar}. A quick inspection shows that, for our fiducial experimental noise levels, the relative difference in these weights when the small-angle $(\ell>200)$ temperature-galaxy cross-power is ignored is at most $3\%$. Thus, we consider it to be safe to ignore all isotropic correlations between temperature and galaxies in the reconstruction pipeline. Summarizing, we only need to generate simultaneous simulations of the $\Pi$-binned moments of $v$, $\dot{\tau}$, and g in order to capture all the important correlations for radial velocity reconstruction. The non-kSZ CMB anisotropies can be simulated separately from a single temperature spectra and later combined with the kSZ map from the joint simulations. The considerations above also apply for the transverse velocity potential reconstruction from moving lens.    
    \item \textbf{Simulate the fields:} Once we determine which fields have to be simulated from a single multivariate Gaussian distribution, we construct a joint covariance matrix $\mathbf{C}_\ell$ for $0 < \ell \leq \ell_{\rm max}$ including all spectra and cross-spectra. Our fiducial resolution is $N_{\rm bin} = 32$ and $\ell_{\rm max} = 6144$ (corresponding to the band-limited multipole for a HEALPix map of $N_{\rm side} = 2048$). The choice of number of bins and angular resolution are fixed by computational resources; it would be desirable to include more bins when possible to incorporate the effects of fine-mode noise. At each $\ell$ we find the Cholesky decomposition $\mathbf{L}_{\ell} \mathbf{L}^\dagger_{\ell} = \mathbf{C}_\ell$ and generate data vectors $\mathbf{a}_{\ell m}$ corresponding to the $2 \ell + 1$ spherical harmonic coefficients at fixed $\ell$ of for all maps using the relation:
    \begin{equation}
    \mathbf{a}_{\ell m} = \mathbf{L}_{\ell} \cdot \mathbf{X}_{\ell m}\,,
    \end{equation}
    where $\mathbf{X}$ is a vector of random Gaussian numbers with zero mean and unit variance.
    In general we find good agreement between the ensemble-average spectra and cross-spectra from simulations and the input spectra. At $\ell \alt 30$ the built-in routine for generating Gaussian maps in HEALPix (which generates realizations using a Cholesky decomposition as above) performs somewhat better than our algorithm. For large $\ell$, the HEALPix algorithm performs worse than ours. We therefore employ a hybrid method, generating low-$\ell$ moments using HEALPix and high-$\ell$ moments using our algorithm. In either case, it is necessary to compute spectra and cross-spectra at sufficiently high accuracy to ensure that $\mathbf{C}_\ell$ is numerically positive definite and therefore the Cholesky decomposition well-defined. In this respect, our code for computing spectra is sufficiently accurate at the resolutions we have explored, but we expect increasingly accurate spectra are necessary for larger numbers of radial bins.    
     
    \item\textbf{Construct the temperature signals:} For radial velocity reconstruction, the kSZ signal is constructed from products of the simulated maps as:
    \begin{equation}
    \Theta^{kSZ} (\hat{n}) = \sum_\alpha \Delta \chi \ \dot{\tau}^\alpha (\hat{n}) v^\alpha (\hat{n})\,.
    \end{equation}
    The part of the CMB temperature that we approximate as uncorrelated with galaxies is simulated as above, using a single temperature spectrum including: primary CMB, lensing contribution, moving lens contribution, and ILC cleaned extragalactic + intrumental noise components. The total temperature map is the sum of the uncorrelated map and the kSZ signal.
    
    For transverse velocity potential reconstruction, the moving lens signal is constructed from products of the simulated maps as:
    \begin{equation}
    \Theta^{ML} (\hat{n}) =\sum_\alpha \Delta \chi \ \nabla_{\perp} \psi^{\alpha}(\hat{n})\cdot\nabla_{\perp}\Upsilon^\alpha (\hat{n})\,.
    \end{equation}
    The part of the CMB temperature that we approximate as uncorrelated with galaxies is simulated as above, using a single temperature spectrum including: primary CMB, lensing contribution, kSZ contribution, and ILC cleaned extragalactic + instrumental noise components. The total temperature map is the sum of the uncorrelated map and the moving lens signal.
    
    For both reconstruction scenarios, we apply a mask corresponding to an SO-like experiment, consisting of a cut between a declination of -70 degrees and +20 degrees and a Galactic mask that removes $\sim 30\%$ of the sky. The total sky fraction covered by the joint mask is $f_{\rm sky} = 0.45$.

    \item {\textbf{Run the estimator pipeline:} The galaxy maps and temperature maps generated using steps 1-3 above are processed using the real space estimators described below and compared with the expected results based on the input fields. }

    \end{enumerate}

\subsection{Real-space estimators}

The harmonic-space quadratic estimators for the radial velocity and transverse velocity potential cannot be implemented efficiently at the resolutions we wish to explore. We therefore derive mathematically equivalent real-space estimators that take advantage of fast forward- and inverse-spherical harmonic transforms, which can be efficiently implemented. A version of the real-space quadratic estimator for radial velocity reconstruction in the absence of foregrounds appeared in Ref.~\cite{Cayuso:2018lhv}. Here, we derive the equivalent estimator including foregrounds, and present a real-space version of the quadratic estimator for the transverse velocity potential.

\subsubsection{Radial velocity estimator}\label{sec:rve}

To derive an efficient real-space estimator for the radial velocity, we start from the harmonic space estimator defined in Eq.~\ref{eq:manybinharmonicestimator}. First, re-write $G_{\ell \ell' L}^\alpha$ as:
\begin{eqnarray}\label{eq:real_space_filt}
G_{\ell \ell' L}^\alpha &=&\left( 1 - \frac{(C_{\ell}^{g^{W^\alpha} \Theta})^2}{C_{\ell}^{\Theta \Theta} C_{\ell}^{g^{W^\alpha}g^{W^\alpha}}} \frac{(C_{\ell'}^{ g^{W^\alpha} \Theta})^2}{C_{\ell'}^{\Theta\Theta} C_{\ell'}^{g^{W^\alpha} g^{W^\alpha}}} \right)^{-1} \left[ \frac{f^{v^\alpha W^\alpha}_{\ell L \ell'} }{C_{\ell}^{\Theta\Theta} C_{\ell' }^{g^{W^\alpha} g^{W^\alpha}}} - \frac{ (-1)^{\ell+\ell'+L} f^{v^\alpha W^\alpha}_{\ell' L \ell} }{C_{\ell'}^{\Theta\Theta} C_{\ell}^{g^{W^\alpha} g^{W^\alpha}}} \frac{C_{\ell}^{ g^{W^\alpha} \Theta} C_{\ell'}^{g^{W^\alpha} \Theta}}{C_{\ell}^{\Theta \Theta} C_{\ell'}^{g^{W^\alpha} g^{W^\alpha}}} \right] \nonumber \\
&=&\left[ \sum_{n=0}^\infty  \left( \frac{(C_{\ell}^{g^{W^\alpha} \Theta})^2}{C_{\ell}^{\Theta \Theta} C_{\ell}^{g^{W^\alpha}g^{W^\alpha}}} \right)^n  \left( \frac{(C_{\ell'}^{ g^{W^\alpha} \Theta})^2}{C_{\ell'}^{\Theta\Theta} C_{\ell'}^{g^{W^\alpha} g^{W^\alpha}}} \right)^n \right] \nonumber \\ 
&& \times \left[ \frac{f^{v^\alpha W^\alpha}_{\ell L \ell'} }{C_{\ell}^{\Theta\Theta} C_{\ell' }^{g^{W^\alpha} g^{W^\alpha}}} - \frac{ (-1)^{\ell+\ell'+L} f^{v^\alpha W^\alpha}_{\ell' L \ell} }{C_{\ell'}^{\Theta\Theta} C_{\ell}^{g^{W^\alpha} g^{W^\alpha}}} \frac{C_{\ell}^{ g^{W^\alpha} \Theta} C_{\ell'}^{g^{W^\alpha} \Theta}}{C_{\ell}^{\Theta \Theta} C_{\ell'}^{g^{W^\alpha} g^{W^\alpha}}} \right]  \,.
\end{eqnarray}
Next, we use the relation
\begin{eqnarray}\label{eq:sht_to_wigner}
\int d^2 \hat{n} \ Y_{\ell m} Y_{\ell' m'} Y_{LM}^{*}  = \sqrt{\frac{(2\ell+1)(2\ell_1+1)(2\ell'+1)}{4\pi}} (-1)^M 
\begin{pmatrix}
\ell & \ell' & L \\
m & m' & -M
\end{pmatrix}
\begin{pmatrix}
\ell & \ell' & L \\
0 & 0 & 0
\end{pmatrix}\,,
\end{eqnarray}
and the definitions:
\begin{eqnarray}\label{eq:kSZxizeta1}
\xi_n^\alpha (\hat{n}) = \sum_{\ell m} \left( \frac{(C_{\ell}^{g^{W^\alpha} \Theta})^2}{C_{\ell}^{\Theta \Theta} C_{\ell}^{g^{W^\alpha}g^{W^\alpha}}}\right)^n  \frac{\Theta_{\ell m}}{C_{\ell}^{\Theta \Theta}} \ Y_{\ell m} (\hat{n}), \ \ \   \zeta_n^\alpha (\hat{n}) = \sum_{\ell m} \left( \frac{(C_{\ell}^{g^{W^\alpha} \Theta})^2}{C_{\ell}^{\Theta \Theta} C_{\ell}^{g^{W^\alpha}g^{W^\alpha}}} \right)^n \frac{C_{\ell}^{\dot{\tau}^\alpha g^{W^\alpha}} g^{W^\alpha}_{\ell m} }{C_{\ell}^{g^{W^\alpha}g^{W^\alpha}}} \ Y_{\ell m} (\hat{n})\,,\nonumber\\
\end{eqnarray}
\begin{eqnarray}\label{eq:kSZxizeta2}
\bar{\xi}_n^\alpha (\hat{n}) = \sum_{\ell m} \left( \frac{(C_{\ell}^{g^{W^\alpha} \Theta})^2}{C_{\ell}^{\Theta \Theta} C_{\ell}^{g^{W^\alpha}g^{W^\alpha}}} \right)^{n+1}  \frac{ C_{\ell}^{\dot{\tau}^\alpha g^{W^\alpha}}  \Theta_{\ell m}}{C_{\ell}^{g^{W^\alpha} \Theta}} \ Y_{\ell m} (\hat{n}), \ \ \ \   \bar{\zeta}_n^\alpha (\hat{n}) = \sum_{\ell m} \left(  \frac{(C_{\ell}^{g^{W^\alpha} \Theta})^2}{C_{\ell}^{\Theta \Theta} C_{\ell}^{g^{W^\alpha}g^{W^\alpha}}} \right)^{n+1} \frac{ g^{W^\alpha}_{\ell m} }{ C_{\ell}^{g^{W^\alpha} \Theta}} \ Y_{\ell m} (\hat{n}).\nonumber\\ 
\end{eqnarray}
Substituting into Eq.~\ref{eq:manybinharmonicestimator}, the real-space estimator is given by:
\begin{eqnarray}
\hat{v}_{LM}^\alpha = N_L^\alpha \sum_{n=0}^\infty \int d^2 \hat{n} \ Y_{LM}^*(\hat{n}) \left[ \xi_n^\alpha (\hat{n}) \zeta_n^\alpha (\hat{n}) - \bar{\xi}_n^\alpha (\hat{n}) \bar{\zeta}_n^\alpha (\hat{n})  \right]\,.
\end{eqnarray}
For the simulations presented below, where we effectively set $C_{\ell}^{g^{W^\alpha} \Theta} = 0$ by not including the statistically isotropic correlations between the galaxy and temperature fields (as argued above, these contributions are insignificant for our fiducial CMB experiment), we can work at the $n=0$ level. For different experimental configurations, it may become necessary to consider higher order terms if $(C_{\ell}^{g^{W^\alpha} \Theta})^2 \sim C_{\ell}^{\Theta \Theta} C_{\ell}^{g^{W^\alpha}g^{W^\alpha}}$.

Note that the fields $\xi_n^\alpha (\hat{n})$, $\zeta_n^\alpha (\hat{n})$, $\bar{\xi}_n^\alpha (\hat{n})$ and $\bar{\zeta}_n^\alpha (\hat{n})$ are convolutions of an azimuthally symmetric function and the moments of the CMB and galaxy density maps. In pixel-space, we can therefore write:
\begin{equation}\label{eq:beamformula}
\xi_n^\alpha (\hat{n}) = \int d^2 \hat{n}' \ B^{\xi_n^\alpha} (|\hat{n}-\hat{n}'|) \Theta (\hat{n}')\,,
\end{equation}
where the `beam' $B^{\xi_n^\alpha} (|\hat{n}-\hat{n}'|)$ for the field $\xi_n^\alpha (\hat{n})$, and the beams for the other filtered fields, are given by:
\begin{eqnarray}
B^{\xi_n^\alpha} (\theta) &=& \sum_{\ell} \sqrt{\frac{2\ell+1}{4\pi}}  \left(  \frac{(C_{\ell}^{g^{W^\alpha} \Theta})^2}{C_{\ell}^{\Theta \Theta} C_{\ell}^{g^{W^\alpha}g^{W^\alpha}}}  \right)^n  \frac{1}{ C_{\ell}^{\Theta \Theta}} P_{\ell} (\cos \theta)\,, \\ 
B^{\zeta_n^\alpha } (\theta) &=& \sum_{\ell}  \sqrt{\frac{2\ell+1}{4\pi}} \left( \frac{(C_{\ell}^{g^{W^\alpha} \Theta})^2}{C_{\ell}^{\Theta \Theta} C_{\ell}^{g^{W^\alpha}g^{W^\alpha}}}  \right)^n \frac{C_{\ell}^{\dot{\tau}^\alpha g^{W^\alpha}}}{ C_{\ell}^{g^{W^\alpha}g^{W^\alpha}}} P_{\ell} (\cos \theta)\,,
\end{eqnarray}
and
\begin{eqnarray}
B^{\bar{\xi}_n^\alpha} (\theta)&=& \sum_{\ell}  \sqrt{\frac{2\ell+1}{4\pi}}  \left( \frac{(C_{\ell}^{g^{W^\alpha} \Theta})^2}{C_{\ell}^{\Theta \Theta} C_{\ell}^{g^{W^\alpha}g^{W^\alpha}}}  \right)^{n+1}  \frac{ C_{\ell}^{\dot{\tau}^\alpha g^{W^\alpha}} }{ C_{\ell}^{g^{W^\alpha} \Theta}}  P_{\ell} (\cos \theta)\,, \\ 
B^{\bar{\zeta}_n^\alpha } (\theta) &=&\sum_{\ell }  \sqrt{\frac{2\ell+1}{4\pi}}  \left( \frac{(C_{\ell}^{g^{W^\alpha} \Theta})^2}{C_{\ell}^{\Theta \Theta} C_{\ell}^{g^{W^\alpha}g^{W^\alpha}}}  \right)^{n+1} \frac{1}{C_{\ell}^{g^{W^\alpha} \Theta}}  P_{\ell} (\cos \theta)\,.
\end{eqnarray}
Some insight into the map-based properties of the estimator can be gained by examining the shape of these functions, which we plot in Fig.~\ref{fig:beams}. The beams receive support only over a scale of {$\sim\!4$} arcmin for the experimental parameters considered here, corresponding to {$\sim\!3$} pixels at HEALPix resolution $N_{\rm side} = 2048$. This implies that the quadratic estimator is highly local, and that systematic errors due mixing information from masked or contaminated regions of the sky will be minimal. Unlike the case of CMB lensing (see e.g.~\cite{astro-ph/0301031}), we therefore expect that there is only a very small bias from the mask on the reconstructed velocity field. 

\subsubsection{Transverse velocity potential estimator}\label{sec:tve}

Following the derivation of the real space radial velocity estimator, we expand $G_{\ell\ell'L}^{\alpha}$, use the coupling function $f^{\Upsilon^\alpha W^\alpha}_{\ell' L \ell}$ defined in Eq.~\ref{eq:bulkvtcoupling}, and use the definition Eq.~\ref{eq:sht_to_wigner} to obtain:
\begin{eqnarray}\label{eq:ML_realspace}
\hat{\Upsilon}_{LM}^{\alpha} &=& L(L+1) N_{L}^{\rm \alpha}\sum\limits_{n=0}^{\infty}\int{\rm d}^2\hat{n}\ Y_{LM}^*(\hat{n}) \ \left( [\xi_{1;n}^{\alpha}(\hat{n})\zeta_{1;n}^{\alpha}(\hat{n})-\bar{\xi}_{1;n}^{\alpha}(\hat{n})\bar{\zeta}_{1;n}^{\alpha}(\hat{n})]\!  \right. \\
&+&\left.  \![\xi_{1;n}^{\alpha}(\hat{n})\zeta_{2;n}^{\alpha}(\hat{n})-\bar{\xi}_{2;n}^{\alpha}(\hat{n})\bar{\zeta}_{1;n}^{\alpha}(\hat{n})]\!-\![\xi_{2;n}^{\alpha}(\hat{n})\zeta_{1;n}^{\alpha}(\hat{n})-\bar{\xi}_{1;n}^{\alpha}(\hat{n})\bar{\zeta}_{2;n}^{\alpha}(\hat{n})] \right) \,,\nonumber
\end{eqnarray}
where the auxiliary functions are given by
\begin{eqnarray}\label{eq:MLxizeta1}
\xi_{1;n}^\alpha (\hat{n}) = \sum_{\ell m} \left( \mathcal{F}^\alpha_\ell \right)^n  \frac{\Theta_{\ell m}}{C_{\ell}^{\Theta \Theta}} \ Y_{\ell m} (\hat{n}), \ \ \  \zeta_{1;n}^\alpha (\hat{n}) = \sum_{\ell m} \left( \mathcal{F}^\alpha_\ell  \right)^n \frac{C_{\ell}^{\psi g^{W^\alpha}} g^{W^\alpha}_{\ell m} }{ C_{\ell}^{g^{W^\alpha} g^{W^\alpha}}} \ Y_{\ell m} (\hat{n}), \,,
\end{eqnarray}
\begin{eqnarray}\label{eq:MLxizeta2}
\bar{\xi}_{1;n}^\alpha (\hat{n}) = \sum_{\ell m} \left( \mathcal{F}^\alpha_\ell  \right)^{n+1}  \frac{ C_{\ell}^{\psi g^{W^\alpha}}  \Theta_{\ell m}}{C_{\ell}^{g^{W^\alpha} \Theta}} \ Y_{\ell m} (\hat{n}), \ \ \  
 \bar{\zeta}_{1;n}^\alpha (\hat{n}) = \sum_{\ell m} \left( \mathcal{F}^\alpha_\ell\right)^{n+1} \frac{g^{W^\alpha}_{\ell m}}{ C_{\ell}^{g^{W^\alpha} \Theta}} \ Y_{\ell m} (\hat{n})\,,\ \ \ \
\end{eqnarray}
\begin{eqnarray}\label{eq:MLxizeta1}
\xi_{2;n}^\alpha (\hat{n}) = \sum_{\ell m}\,\ell(\ell+1)\,\left( \mathcal{F}^\alpha_\ell  \right)^n \frac{\Theta_{\ell m}}{C_{\ell}^{\Theta \Theta}} \ Y_{\ell m} (\hat{n}), 
&&\zeta_{2;n}^\alpha (\hat{n}) = \sum_{\ell m} \,\ell(\ell+1)\,\left( \mathcal{F}^\alpha_\ell \right)^n \frac{C_{\ell}^{\psi g^{W^\alpha}} g^{W^\alpha}_{\ell m} }{ C_{\ell}^{g^{W^\alpha} g^{W^\alpha}}} \ Y_{\ell m} (\hat{n})\,,\ \ 
\end{eqnarray}
\begin{eqnarray}\label{eq:MLxizeta2}
\bar{\xi}_{2;n}^\alpha (\hat{n}) = \sum_{\ell m}\,\ell(\ell+1)\,\left( \mathcal{F}^\alpha_\ell  \right)^{n+1} \frac{ C_{\ell}^{\psi g^{W^\alpha}}  \Theta_{\ell m}}{C_{\ell}^{g^{W^\alpha} \Theta}} \ Y_{\ell m} (\hat{n}), \ \ \ \bar{\zeta}_{2;n}^\alpha (\hat{n}) = \sum_{\ell m}\,\ell(\ell+1) \left( \mathcal{F}^\alpha_\ell \right)^{n+1} \frac{ g^{W^\alpha}_{\ell m} }{C_{\ell}^{g^{W^\alpha} \Theta}} \ Y_{\ell m} (\hat{n})\ \ \ \ 
\end{eqnarray}
where we have defined the factor
\begin{equation}
\mathcal{F}^\alpha_\ell \equiv \frac{( C_{\ell}^{g^{W^\alpha} \Theta} )^2}{C_{\ell}^{\Theta \Theta} C_{\ell}^{g^{W^\alpha} g^{W^\alpha} }}\,,
\end{equation}
for convenience. 

\begin{figure}[h]
  \includegraphics[scale=0.6]{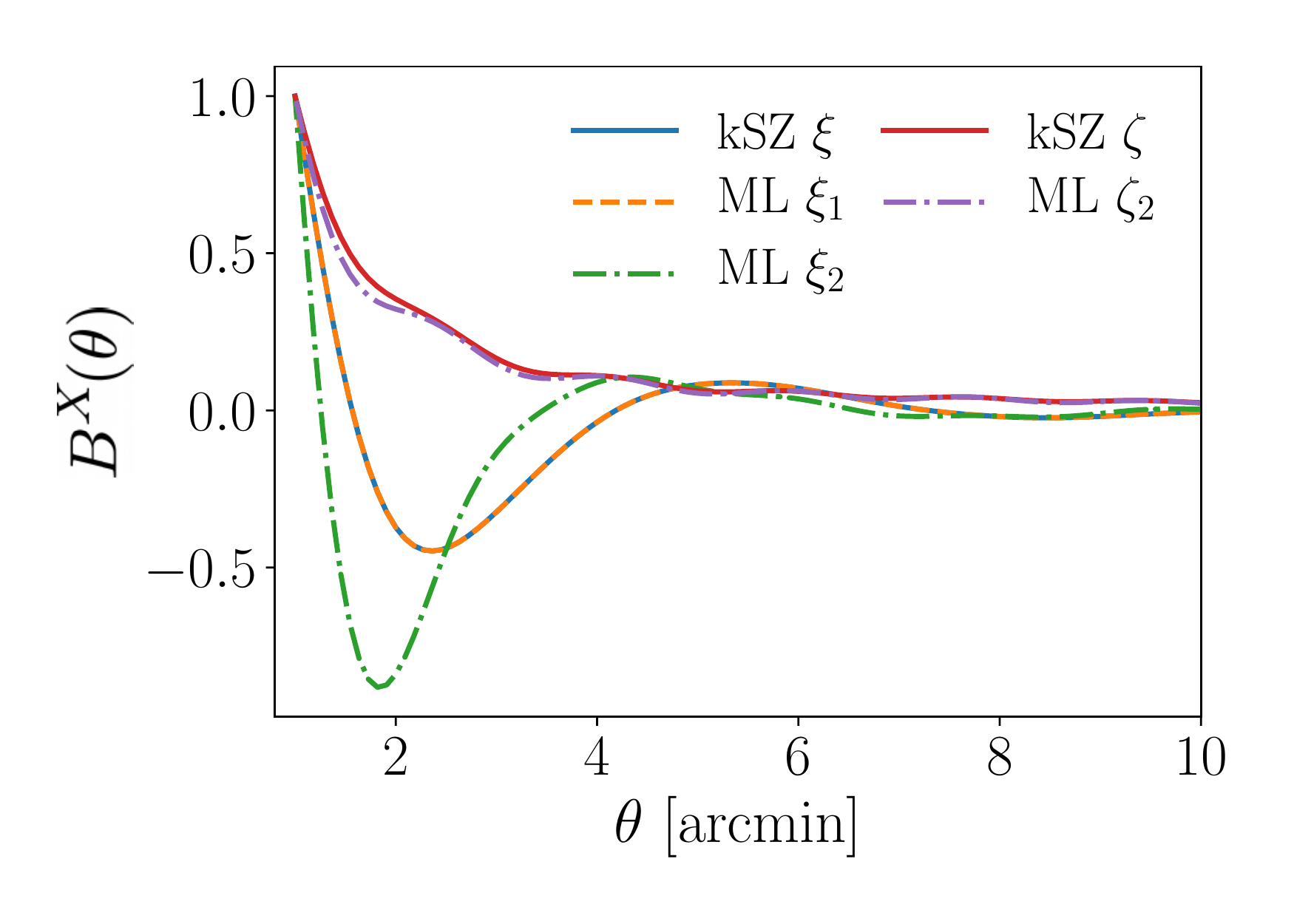}
  \caption{Various `beams' defined in Secs.~\ref{sec:rve}~and~\ref{sec:tve} for the fiducial SO x LSST spectra, normalized to unity at $\theta\ll1$. The ingredients of the real space quadratic estimator are formed by convolving these beams with the temperature or galaxy maps. For comparison, the size of a HEALPix pixel at $N_{\rm side} = 2048$, the default map resolution employed in later sections, is $\sim1.7$~armin. The beams are quite local, implying that the estimators will be rather insensitive to masking or local contaminants. This same property can lead to significant numerical errors, which we discuss further below.}
  \label{fig:beams}
\end{figure}

As described in the previous subsection for the radial velocity estimator, these auxiliary functions can be viewed as a convolution in map space (Eq.~\ref{eq:beamformula}) with a set of azimuthally symmetric 'beams' defined by  
\begin{eqnarray}
B^{\xi_{1; n}^\alpha} = \sum_{\ell}  \sqrt{\frac{2\ell+1}{4\pi}}   \frac{ \left( \mathcal{F}^\alpha_\ell \right)^n}{C_{\ell}^{\Theta \Theta}} P_{\ell} (\cos \theta)\,, \ \ \ 
B^{\zeta_{1; n}^\alpha } = \sum_{\ell}  \sqrt{\frac{2\ell+1}{4\pi}} \frac{  \left( \mathcal{F}^\alpha_\ell \right)^n C_{\ell}^{\psi g^{W^\alpha}} }{ C_{\ell}^{g^{W^\alpha} g^{W^\alpha}}} P_{\ell} (\cos \theta)\,,
\end{eqnarray}
\begin{eqnarray}
B^{\bar{\xi}_{1;n}^\alpha} = \sum_{\ell} \sqrt{\frac{2\ell+1}{4\pi}}  \frac{ \left( \mathcal{F}^\alpha_\ell \right)^{n+1} C_{\ell}^{\psi g^{W^\alpha}} }{C_{\ell}^{g^{W^\alpha} \Theta}} P_{\ell} (\cos \theta)\,, \ \ \ 
B^{\bar{\zeta}_{1;n}^\alpha } = \sum_{\ell m} \sqrt{\frac{2\ell+1}{4\pi}}   \frac{  \left(  \mathcal{F}^\alpha_\ell \right)^{n+1} }{ C_{\ell}^{g^{W^\alpha} \Theta}} P_{\ell} (\cos \theta)\,,
\end{eqnarray}
\begin{eqnarray}
B^{\xi_{2;n}^\alpha} = \sum_{\ell}\,\ell(\ell+1) \sqrt{\frac{2\ell+1}{4\pi}} \frac{\left( \mathcal{F}^\alpha_\ell  \right)^n}{C_{\ell}^{\Theta \Theta}} P_{\ell} (\cos \theta)\,, \ \ \ 
B^{\zeta_{2;n}^\alpha} = \sum_{\ell}\,\ell(\ell+1) \sqrt{\frac{2\ell+1}{4\pi}} \frac{ \left( \mathcal{F}^\alpha_\ell  \right)^n C_{\ell}^{\psi g^{W^\alpha}}  }{ C_{\ell}^{g^{W^\alpha} g^{W^\alpha}}} P_{\ell} (\cos \theta)\,,\ \ \ 
\end{eqnarray}
\begin{eqnarray}
B^{\bar{\xi}_{2;n}^\alpha} = \sum_{\ell}\,\ell(\ell+1) \sqrt{\frac{2\ell+1}{4\pi}} \frac{ \left( \mathcal{F}^\alpha_\ell  \right)^{n+1} C_{\ell}^{\psi g^{W^\alpha}} }{C_{\ell}^{g^{W^\alpha} \Theta}}  P_{\ell} (\cos \theta)\,, \ \ \ 
B^{\bar{\zeta}_{2;n}^\alpha } = \sum_{\ell m}\,\ell(\ell+1)\sqrt{\frac{2\ell+1}{4\pi}}  \frac{ \left( \mathcal{F}^\alpha_\ell \right)^{n+1} }{C_{\ell}^{g^{W^\alpha} \Theta}} P_{\ell} (\cos \theta)\,. \ \ \  
\end{eqnarray}
Note the terms that appear with an additional factor of $\ell(\ell+1)$ in the beams defined above, which cause the transverse velocity potential estimator to be more localized than the radial velocity estimator. This is demonstrated in Fig.~\ref{fig:beams}, where $B^{\xi_{2;n=0}^\alpha} (\theta)$ is seen to have support on smaller angular scales than the corresponding auxiliary field $B^{\xi_{n=0}^\alpha} (\theta)$ for the radial velocity. As we describe further in Sec.~\ref{sec:transverse_velo_from_sims}, this property makes the transverse velocity potential estimator more susceptible to numerical errors.

\subsection{Reconstruction on simulated maps: Radial velocity}

\begin{figure}[ht]
  \includegraphics[scale=0.35]{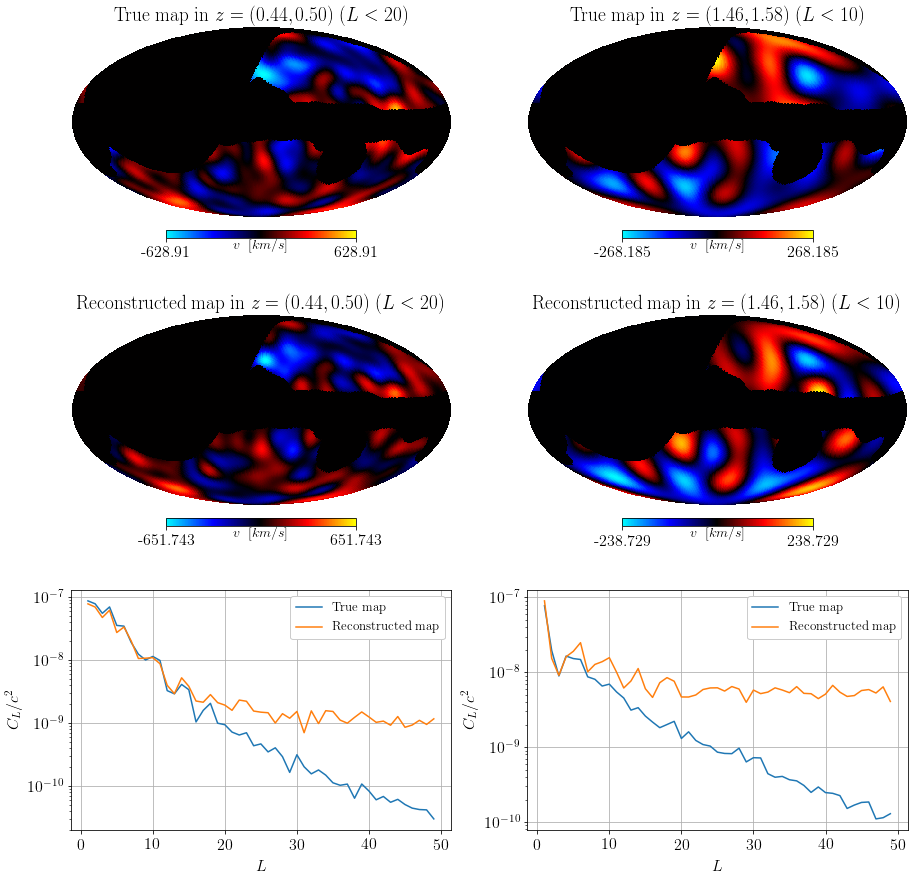}
  \caption{\textbf{Top panels}: Low-pass filtered maps of the true rotated velocities for an example realization in a low-redshift and high-redshift bin. \textbf{Middle panels}: Low-pass filtered reconstructed maps for the same example realization. \textbf{Bottom panels}: power spectra comparison between true and reconstructed maps in the two redshift bins. At the level of the maps and the power spectra, we see that large scales can be reconstructed with SO x LSST.}
  \label{fig:maps}
\end{figure}

We now present the results obtained by applying the real-space estimators derived above to the simulated maps for the fiducial data combination of SO x LSST employed in previous sections. We generated a set of 30 realizations, each with 32 bins in the redshift range $0.2 \leq z \leq 5$ and output resolutions of NSIDE 2048. The resolution and number of simulations were dictated by available computational resources. Note that our reconstructions do not include fine-mode noise, since the simulations are constructed using a limited number of bins. We show examples of the reconstruction in Fig.~\ref{fig:maps}, where we compare the rotated true velocities $\mathrm{\bf{R}}_L\cdot\bf{v}_{LM}$ to the output of the estimator $\bf{\hat{v}}_{LM}$ at two representative redshift bins located at $z \sim 0.5$ and $z \sim 1.5$. These maps have been filtered to show only the largest angular scales, where the reconstruction is signal-dominated. A visual inspection of these maps indicates that a successful reconstruction of the radial velocity has been achieved. A comparison of the angular power spectra for the reconstruction and the masked actual velocity field indicate good quantitative agreement. Before undertaking a full quantitative analysis of the full set of realizations, and comparing to theoretical expectations, we take a brief digression to discuss the effects of the mask on the reconstruction.

\begin{figure}[ht]
  \includegraphics[scale=0.35]{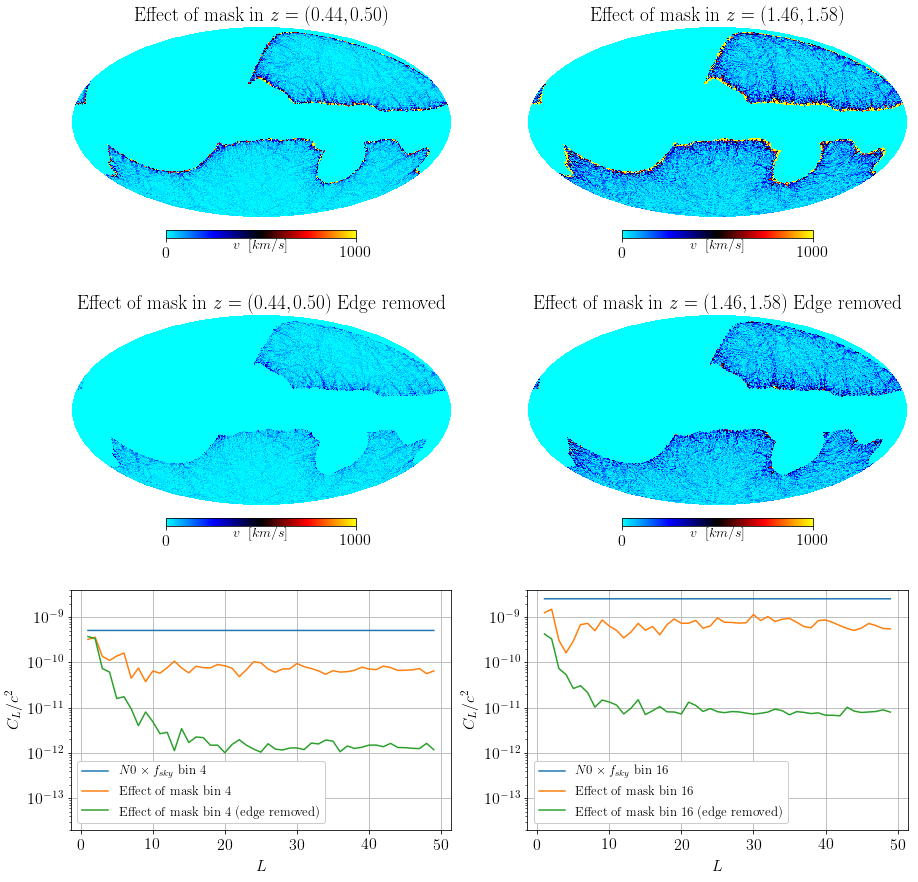}
  \caption{\textbf{Top panels}: Mask bias at Nside 64 when no edge pixels are removed. \textbf{Middle panels}: mask bias after removing a 1-pixel thick border at Nside = 64, which leads to a reduction of $f_{sky}$ from 0.45 to 0.44. \textbf{Bottom panels}: power spectrum of the mask bias with and without an extended mask compared to the $N^0$ noise for radial velocity reconstruction.}
  \label{fig:mask_effect}
\end{figure}

As we discussed in Sec.~\ref{sec:rve}, we expect contamination from masked regions to extend only a few pixels from the mask boundary due to the local nature of the radial velocity estimator. We corroborate this by fixing the realization and subtracting the full-sky reconstruction noise (defined as the reconstruction minus the actual radial velocity) from the masked-sky reconstruction noise and studying the residuals, which we refer to as the mask bias. We can see from Fig.~\ref{fig:mask_effect} that the dominant effect of mask is concentrated on the edge of the unmasked region as expected, and that this contamination can be removed post-reconstruction by extending the mask by a few pixels. Comparing the maps in the top and middle panels of Fig.~\ref{fig:mask_effect}, extending the mask by a single pixel removes the most contaminated regions of the map. In the bottom panel of Fig.~\ref{fig:mask_effect}, we see that the mask bias is always below the reconstruction noise, and that extending the mask by one pixel decreases the mask bias at the level of the power spectrum by orders of magnitude. We conclude that one need not worry about mask bias for reconstruction of the radial velocity. Note that the story presented here will become more complicated for apodized maps, since apodization will introduce a statistical anisotropy that may be picked up by the estimator and which must be accounted for in the reconstruction. In addition, from the perspective of the reconstruction, a smaller bias will be incurred by fitting and subtracting point sources rather than masking them.

Returning to our ensemble of simulations, we now confirm that the statistics of the ensemble are as expected from the analytic estimates presented in Sec.~\ref{sec:recon_rd}. On the full sky this can be thought of as a validation exercise for our simulations and reconstruction pipeline, since in the absence of numerical errors, the agreement should be perfect. On the masked sky, we determine what the effect of the mask is on the reconstructed power spectrum. To mitigate the mask bias, we extend the mask post-reconstruction by one pixel in the results presented below.

The top panel of Fig.~\ref{fig:cl_sims_vr} compares the ensemble average reconstruction signal and noise to the theoretical expectations on the full sky. Comparing the theory signal (blue) to the simulated signal (green dashed), there is excellent agreement in both redshift bins. Comparing the power spectrum of the reconstruction (red dot-dashed) to the theory signal and the theory noise (orange), we see excellent agreement in both the signal-dominated and noise-dominated regimes. To obtain a reconstruction noise, we compute the power spectrum of the reconstruction minus the actual signal (purple dots) and the result of applying the estimator to a temperature map whose kSZ component is uncorrelated with the galaxy density (brown squares). In both cases, the agreement with the theory reconstruction noise is excellent, aside from some excess at low-L, at the level of about one percent of the signal. We attribute this residual to numerical error in the reconstruction. This result is a powerful validation of our simulation and reconstruction pipeline.

The bottom panel of Fig.~\ref{fig:cl_sims_vr} shows the comparison between the theory and ensemble-averaged reconstruction on the masked sky. We make no attempt here to de-project the mask from the power spectra (e.g. using the methods of Ref.~\cite{Alonso:2018jzx}), and simply find the power spectra of the masked maps. The theory curves (signal and noise) have been multiplied by $f_{\rm sky} = 0.44$ to account for the loss of variance from masking. Comparing the theory signal multiplied by $f_{\rm sky}$ (blue) to the simulated masked signal (green dashed) the scale-dependent effect of the mode-coupling with the mask is evident, especially at low redshift. Nevertheless the factor of $f_{\rm sky}$ gives a reasonable estimate of the power spectrum of the velocity field on the masked sky. Comparing the reconstruction on the masked sky to the masked actual velocity and the theory noise reduced by $f_{\rm sky}$, we see good agreement at both high- and low-L. Checking this in more detail by finding the difference between the reconstruction and the actual masked signal (purple dots), we again find good agreement with the expected reconstruction noise aside from a few percent excess at low-L and low-redshift.

\begin{figure}[ht]
  \includegraphics[scale=0.35]{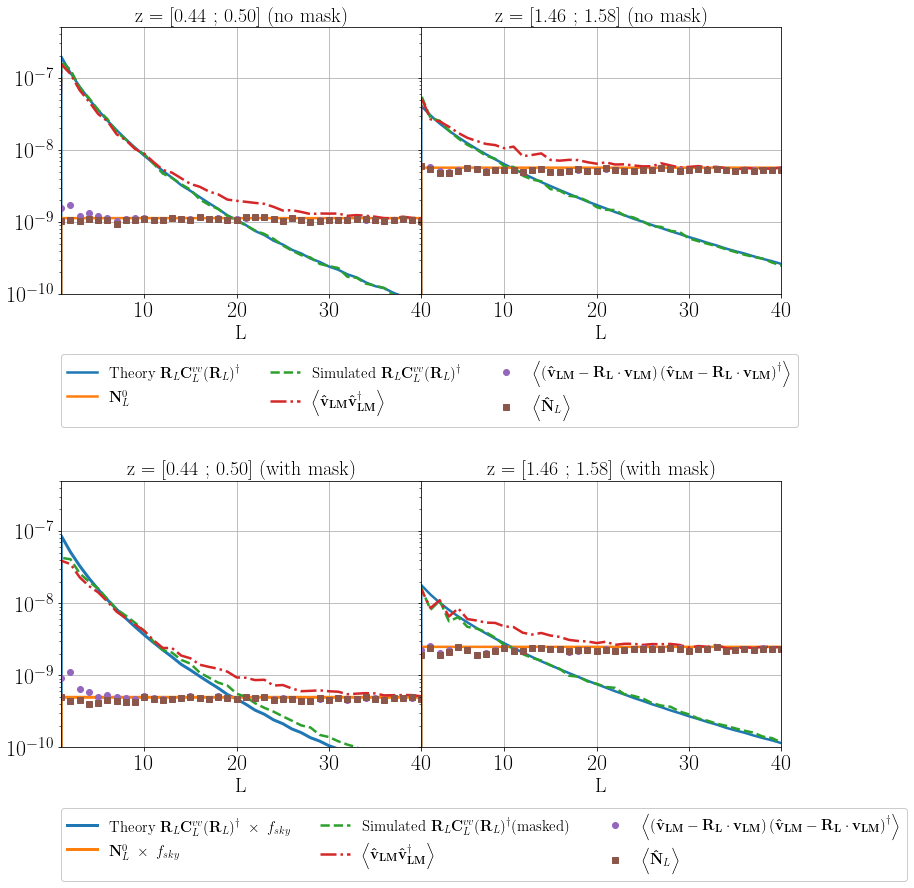}
  \caption{Average power spectrum from 30 simulated reconstructions on the full-sky (top panels) and on masked-sky (bottom panels) for SO x LSST. Solid lines correspond to signal and noise from theory, dashed lines show simulated radial velocities (rotated by $\mathrm{\bf{R}}_L$ ), dot-dashed lines shows the estimator output, circles show the difference between the estimator output and the simulated rotated velocities, and squares are the result of running the estimator on uncorrelated maps (a measure of the reconstruction noise).}
  \label{fig:cl_sims_vr}
\end{figure}

\begin{figure}[ht]
  \includegraphics[scale=0.35]{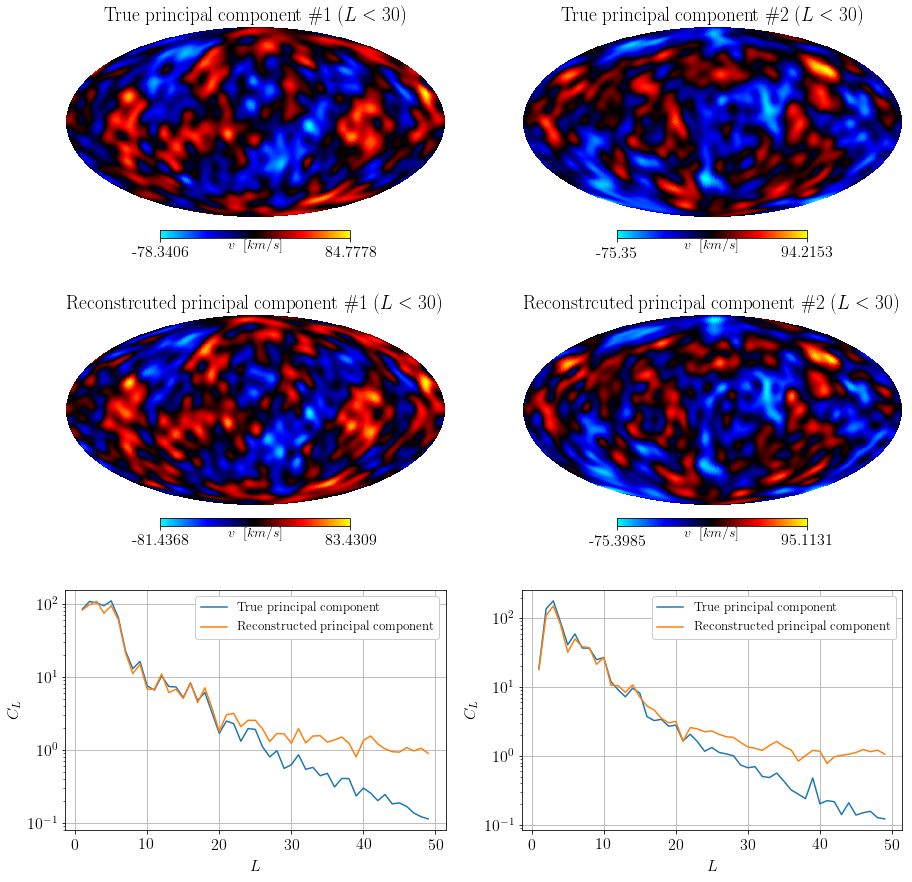}
  \caption{\textbf{Top panels}: Low-pass filtered maps of the 2 highest signal to noise true principal components on the full sky. \textbf{Middle panels}: Low-pass filtered reconstructed principal components. \textbf{Bottom panels}: power spectra comparison between true and reconstructed maps.}
  \label{fig:maps_pc}
\end{figure}

\begin{figure}[ht]
  \includegraphics[scale=0.35]{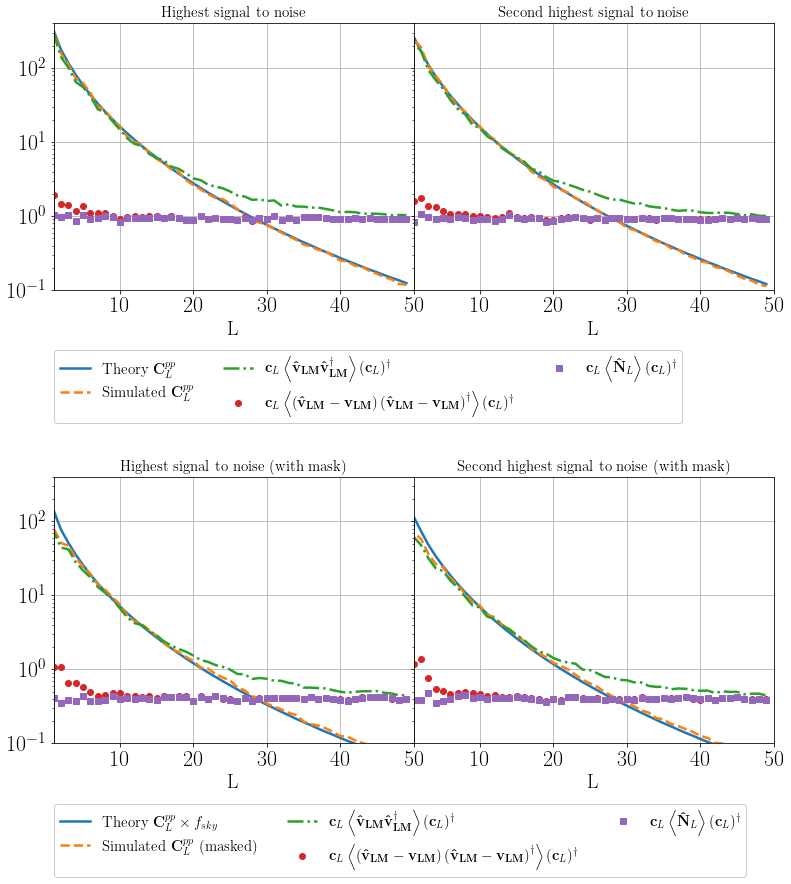}
  \caption{Signal and noise averages for the principal component transformation of reconstructed spectra on the full-sky (top panels) and masked sky (bottom panels).}
  \label{fig:pc_spectra}
\end{figure}

The collection of reconstructed maps are not statistically independent of each other due to velocity correlations along the light cone. On the full-sky, harmonic moments with different multipoles $L,M$ are independent from each other and only correlated in the radial direction. Using the principal component decomposition described in Sec.~\ref{subsec:principal}, we can construct $N_{bin}$ linear combinations of $\Pi$-binned harmonic multipoles $a^{j}_{LM} = \sum_{\alpha}c_L^{j\beta}a^{\beta}_{LM}$ such that $a^{j}_{LM}$ and $a^{i}_{LM}$ are uncorrelated for all $i\neq j$. Note that the transformation coefficients depend only on the multipole $L$. The maps constructed using these rotated harmonic moments at each $L$ constitute the principal components of the radial velocity on the light cone. The power spectrum of the principal components at each multipole $L$ is diagonal and is obtained by rotating the bin basis power spectrum  $\mathrm{\bf{c}}_L\mathrm{\bf{C}}_{L}(\mathrm{\bf{c}}_L)^{\dagger}$, where $\mathrm{\bf{c}}_L$ is the matrix defined by the coefficients $c_L^{j\beta}$. Fig.~\ref{fig:maps_pc} shows the true maps, reconstructed maps and spectra for the 2 highest signal to noise principal components from a single realization. The `unit' of the principle component power spectra and maps is signal to noise, since the noise has been normalized to unity. Clearly, the fidelity of the principle component reconstruction is far higher than for the single bins presented in Fig.~\ref{fig:maps} (although the information in the set of principle component maps is equivalent to the information in the set of bin maps), making the principle component basis desirable for a visual representation of the results. In the top panel of Fig.~\ref{fig:pc_spectra}, we show the ensemble-averaged power spectra of the first two principle components without masking. Performing the comparisons described above in the bin basis between theoretical expectations and data from the reconstructions, we again find excellent agreement, aside from sub-percent level effects at the lowest L.

\begin{figure}[ht]
  \includegraphics[scale=0.4]{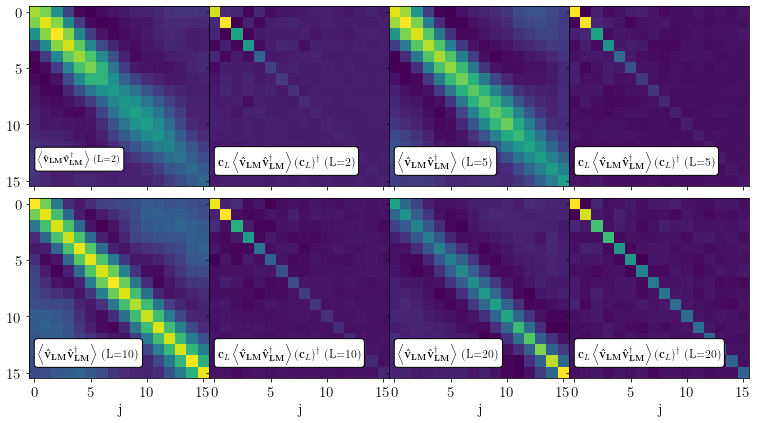}
  \caption{Pseudo principal component transformation of the reconstructed spectra $\left\langle\bf{\hat{v}}_{LM}\bf{\hat{v}}^{\dagger}_{LM}\right\rangle$ at different L multipoles, averaged over 30 simulations. }
  \label{fig:spectra_diag}
\end{figure}

The principal component analysis discussed here is subject to some complications when a mask is introduced. Statistical isotropy is broken for the masked map, which introduces $L\neq L'$ correlations within and between bins. Using the $c_L^{j\beta}$ coefficients defined for the full-sky scenario would not lead to statistically independent maps. A more rigorous procedure to find the uncorrelated combinations of the data could be done in pixel space rather than harmonic space. Such a procedure would involve the construction and diagonalization of a matrix containing the covariance between every pair of pixels at every pair of redshift bins, a task that is computationally demanding. Here, we consider a `pseudo' principal component transformation at the level of the power spectra of the reconstructions with a mask. As we discussed in the previous subsection, the power spectrum of reconstructed maps with a mask traces the underlying full-sky spectra up to a factor of $f_{sky}$ and some scale-dependent corrections due to convolution with the mask. For multipoles where the scale dependent correction is small at all redshifts, the masked $C_{L}$ is approximately proportional to the unmasked one and therefore can be diagonalized using the full-sky transformation matrix $\mathrm{\bf{c}}_L$. In the bottom panel of Fig.~\ref{fig:pc_spectra} we compare the theory signal and noise (reduced by a factor of $f_{\rm sky}$) to the reconstructed velocity for the first two principal components. Despite the complications from mode coupling with the mask, there is reasonable agreement with the theory curves. Finally, we can explicitly check that the rotation associated with the principal components $\mathrm{\bf{c}}_L$ take the reconstructed spectra $\left\langle\bf{\hat{v}}_{LM}\bf{\hat{v}}^{\dagger}_{LM}\right\rangle$ to a nearly diagonal form. This is shown in Fig.~\ref{fig:spectra_diag}. where we plot $\left\langle\bf{\hat{v}}_{LM}\bf{\hat{v}}^{\dagger}_{LM}\right\rangle$ averaged over 30 masked reconstructions. The rotation associated with the pseudo principal component basis does indeed result in a more diagonal signal covariance matrix, even in the presence of a mask.

\subsection{Reconstruction on simulated maps: Transverse velocity}\label{sec:transverse_velo_from_sims}

Next, we focus on applying the real-space estimator for the transverse velocity potential defined in Eq.~\ref{eq:ML_realspace} to our simulations. The story is a bit more complicated than the radial velocity estimator due to numerical errors. We generate a set of 30 realizations of moving lens temperature maps and LSST-like galaxy density, using 32 bins in the redshift range $0.2\leq z\leq5$ (the same binning used for radial velocity reconstruction). Maps are output at HEALPix NSIDE of 2048. As described in more detail below, we perform a reconstruction of the transverse velocity potential with and without the primary CMB in order to characterize numerical errors and confirm the estimator in some limited regimes. In all cases, we apply the mask described above with $f_{\rm sky} = 0.45$.  

\begin{figure}[ht]
  \includegraphics[scale=0.37]{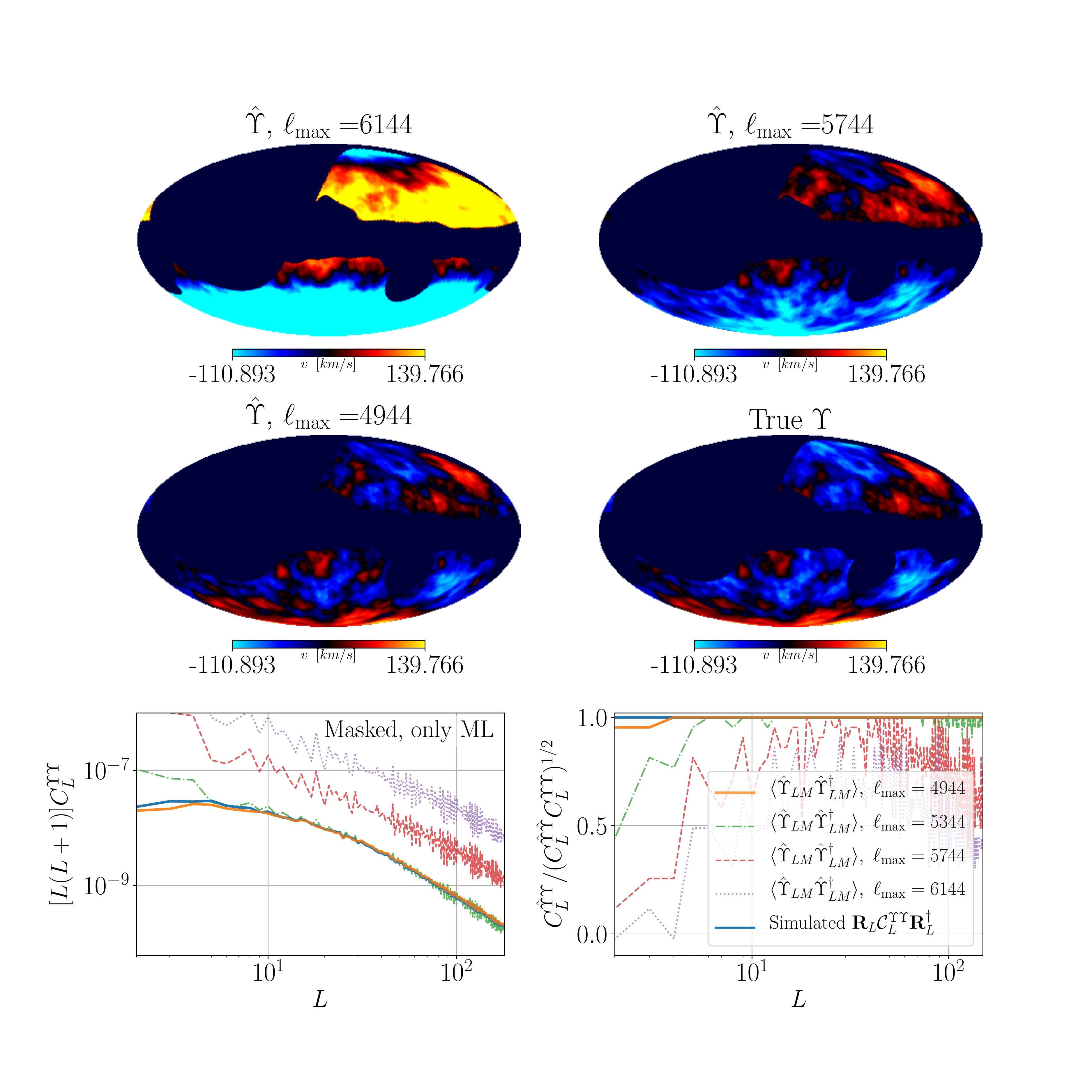}
  \caption{Demonstration of the transverse-velocity reconstruction with CMB maps including only the moving-lens signal for the 10th redshift bin where we have taken 32 bins in the redshift range $0.2\geq z\geq5$. The deterioration in the accuracy of the reconstruction due to amplified numerical errors in our estimation scheme is shown with varying high-multipole cutoffs $\ell_{\rm max}=\{6144,5744,4944\}$. The true transverse velocity field (shown in the middle-right map) is recovered for $\ell_{\rm max}\simeq5000$. Bottom-left panel compares the power-spectra of the reconstructed transverse velocity to the true transverse velocity, both averaged over 30 realisations.  The bottom-right panel shows the cross-correlation coefficient between the true and the reconstructed transverse-velocity fields.}
  \label{fig:ML_recon_CMB0_masked_new2}
\end{figure}

We begin with a reconstruction of the transverse velocity potential from temperature maps containing only the moving lens effect. The result for one realization in a redshift bin centred on $z \sim 1$ is shown in Fig.~\ref{fig:ML_recon_CMB0_masked_new2}. The result obtained when using $\ell_{\rm max} = 6144$ ($=3 \times $NSIDE) in the estimator is shown in the upper-left map in this figure. Comparing with the actual rotated transverse velocity potential (e.g. $\boldsymbol{R}_L\cdot\mathbf{\Upsilon}_{LM}$) in the lower-right map, there is no visible agreement. In the bottom left panel of this figure, we compute the power spectra of the reconstruction (purple dotted line) and actual rotated transverse velocity potential (blue solid line) averaged over all 30 realizations. It can be seen that the reconstruction (which is mostly noise) has roughly an order of magnitude more power over all scales. Computing the ensemble-averaged correlation coefficient in the bottom right panel, we see that there are some traces of the true map in the reconstruction, but nothing close to what is to be expected in this essentially noise-free example.  To compute the real-space estimator Eq.~\ref{eq:ML_realspace} it is necessary to go from map to harmonic space and back again to construct the auxiliary fields, and then one must go back to harmonic space to obtain the estimated transverse velocity potential. We attribute the observed catastrophic numerical error to information-loss incurred when performing these spherical harmonic transforms between map and harmonic space, as harmonic transforms in HEALPix are not information-preserving~\footnote{Nearly identical results are obtained when using pixel weights and/or increasing the number of iterations in the HEALPix map2alm function.}. We conjecture that this error is more severe here than for radial velocity reconstruction due to the fact that transverse velocity reconstruction relies on information from smaller angular scales (as illustrated in Fig.~\ref{fig:beams}), where we expect the spherical harmonic transform to be less accurate.  

To mitigate the numerical errors described above, we low-pass filter the maps before applying the real space estimator by cutting out all multipoles greater than some $\ell_{\rm max}$. The result is shown in Fig.~\ref{fig:ML_recon_CMB0_masked_new2} for two choices $\ell_{\rm max} = 5744, 4944$. At the level of the single realization maps and the ensemble averaged power spectrum and correlation coefficient, it can be seen that as $\ell_{\rm max}$ is decreased, the numerical error decreases. For the temperature maps containing only moving lens analyzed here, it is sufficient to take $\ell_{\rm max} \sim 5000$ to obtain a high-fidelity reconstruction. Even in this case, we still observe a residual error at the lowest multipoles. We saw something similar for the radial velocity reconstruction e.g. in Fig.~\ref{fig:cl_sims_vr}, which we believe can be attributed to same form of numerical error. 

\begin{figure}[h!]
  \includegraphics[scale=0.37]{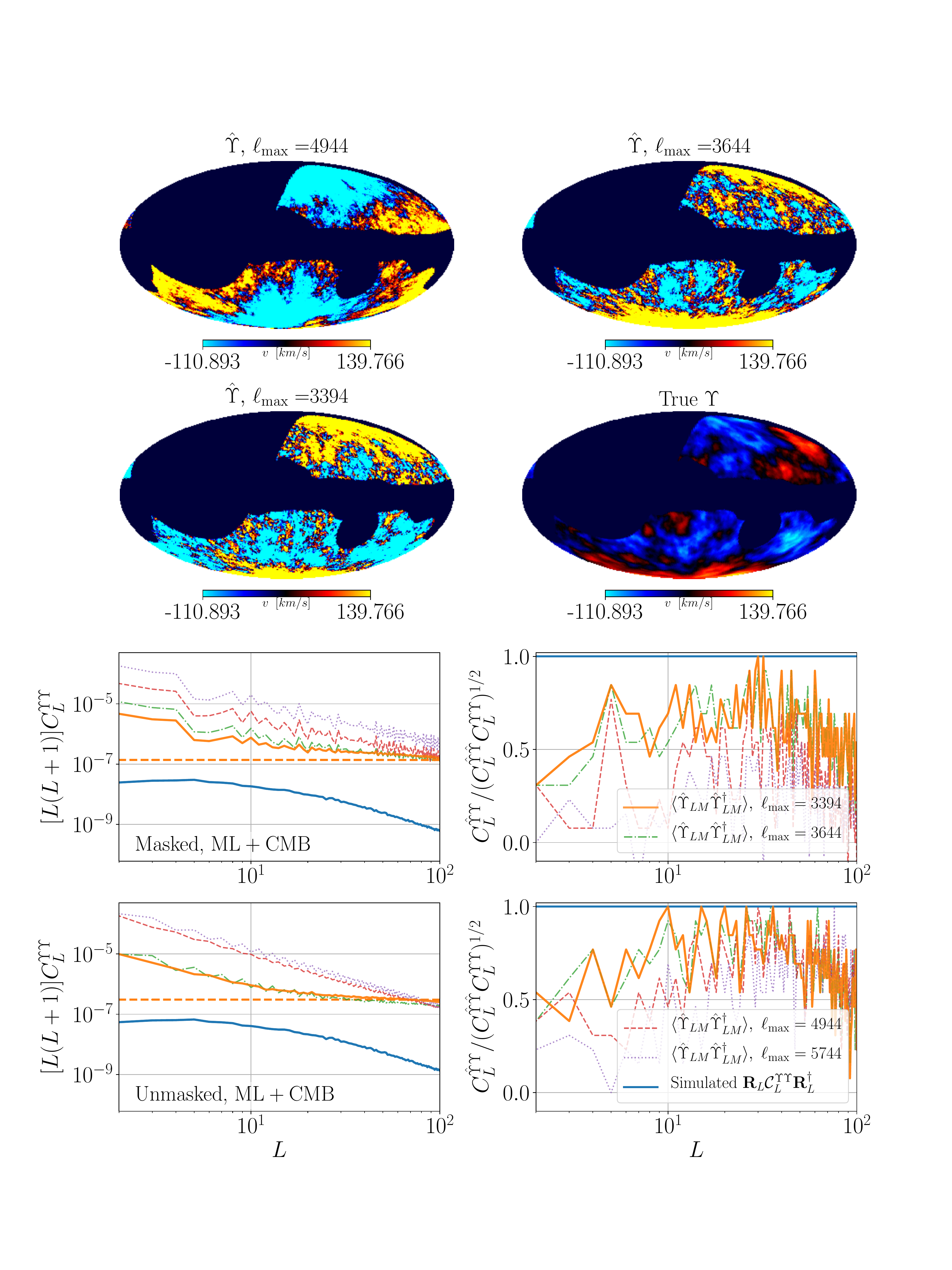}
  \caption{Similar to Figs.~\ref{fig:ML_recon_CMB0_masked_new2}. The transverse-velocity reconstruction from the moving-lens effect appears to be limited by the numerical accuracy available from HEALPix, as discussed in Sec.~\ref{sec:transverse_velo_from_sims}. We will address this issue in an upcoming study. Dashed orange lines correspond to expected reconstruction noise curves for $\ell_{\rm max}=3394$.}
  \label{fig:ML_recon_CMB1_masked_new2}
\end{figure}

The low-pass filter introduced to mitigate the numerical errors described above is not problematic when the temperature map contains only moving lens, but of course, the observed CMB contains more than just moving lens. Including the primary CMB, noise, and foreground residuals in the temperature maps is more problematic. Removing small-scale information in this case degrades the signal to noise of the reconstruction. For the case of SO x LSST, we demonstrated in Sec.~\ref{sec:transv_sorubin_forecast} that the total signal to noise for transverse velocity reconstruction was only of order $\sim 50$, with the reconstruction in each bin being noise dominated. Losing information on the transverse velocity potential from small scales is a big deal in this case. In Fig.~\ref{fig:ML_recon_CMB1_masked_new2}, we show the results of performing transverse velocity potential reconstruction including all contributions to the fiducial SO x LSST case. To reduce numerical noise, severe cuts of order $\ell_{\rm max} \sim 3000$ are necessary. Unfortunately, there is very little signal to noise in the reconstruction after so much small-scale information is removed, as demonstrated by the low correlation coefficients. In particular, comparing the expected reconstruction noise (orange dashed lines) to the power spectrum of the reconstruction, we see that there is a large amount of residual power on large angular scales. Performing the reconstruction on the full sky does not provide any significant improvement, as shown in the bottom panels of Fig~\ref{fig:ML_recon_CMB1_masked_new2}. We conclude that it will be necessary to identify alternative methods to mitigate numerical noise associated with spherical harmonic transforms before an analysis of future datasets can be undertaken. We leave this task to future work.

\section{Conclusions}\label{sec:conclusion}

This paper has outlined the formalism for velocity reconstruction in the light cone picture using CMB experiments and galaxy surveys. One of the main goals of developing this formalism has been to explore some of the challenges posed by systematics and foregrounds for velocity reconstruction. The range of effects we have explored include: properly correlated extragalactic foregrounds, large-angular scale systematics in the galaxy survey, photometric redshift errors, masking of regions contaminated by galactic emission, modelling errors in the galaxy-electron correlation function (the optical depth degeneracy), biases introduced due to additional physical effects that lead to a statistically anisotropic CMB-galaxy correlation (e.g. lensing of the primary CMB), biases introduced by coarse graining on the light cone (e.g. the `fine mode' noise), CMB instrumental noise and beam, choice of frequency channels for cleaning extragalactic foregrounds in the CMB, and the effect of performing foreground cleaning on reconstructed maps. We have developed a numerical pipeline to compute the properly correlated auto- and cross-spectra necessary to assess this range of effects. We have also developed a real-space reconstruction pipeline that we have validated using Gaussian simulations. This pipeline was used to assess the impact of systematics in real-space such as masking. The good news is that none of the systematic effects we have explored seriously degrade the fidelity of the reconstruction, indicating the promise of velocity reconstruction for extracting new cosmological information from future datasets. 

Our fiducial datasets were a LSST-like galaxy survey and an SO-like CMB experiment. We also considered the data combination of the existing unWISE galaxy catalogue with SO. These choices determine factors such as: redshift error, depth of the survey, galaxy shot noise, the level of large-angular scale systematics, frequency channels assumed for the CMB experiment, the associated level of instrumental noise and resolution, and sky coverage. For these datasets, some of the take-away points of our analysis include:
\begin{itemize}
\item The total information available in the reconstructed velocity fields (quantified by the total signal to noise) is limited mainly by the redshift error, sky coverage, and factors contributing to the Gaussian reconstruction noise (CMB instrumental noise and beam, foreground residuals, and the level of galaxy shot noise). 
\item It is essential to incorporate the `fine mode' noise associated with coarse-graining fields on the light cone into the estimator formalism for velocity reconstruction. This source of bias can be mitigated by ensuring that velocity reconstruction is performed in a sufficient number of bins along the radial direction. For radial velocity reconstruction in the fiducial SO x LSST scenario, it is necessary to use around 64 bins to mitigate this bias and include most of the signal to noise in the reconstruction; for transverse velocity reconstruction 32 bins is sufficient.
\item For radial velocity reconstruction, large-scale systematics in the galaxy survey have a significant ($\sim 10 \% $-level) effect on the total signal to noise due to the additional bin-bin correlations it introduces. There is negligible impact of this systematic on transverse velocity reconstruction.
\item The biases induced by CMB lensing and moving lens on radial velocity reconstruction are negligible; the bias introduced by CMB lensing on transverse velocity reconstruction is significant at the $\sim 10\%$ level and should be taken into account, while the bias from kSZ is negligible.
\item From the principle components of the radial and transverse velocity reconstructions (see Figs.~\ref{fig:pcs}, \ref{fig:ML_PC_coeffs}), most of the signal to noise in the reconstruction comes from large angular scales $L \alt 30$ and redshifts $z \alt 1$.
\item For SO x unWISE, it is possible to reconstruct a single principal component on the very largest angular scales. This demonstrates that even for a single broad photometric redshift bin it is possible to reconstruct the large-scale radial velocity field. For the SO x unWISE data combination, it will not be possible to reconstruct the transverse velocity field.
\item The real space estimators for velocity reconstruction are highly local, and contamination from masking is restricted to the region in close proximity to the mask. It is therefore possible to remove the mask bias by extending the mask post-reconstruction.
\item We have validated a pipeline for velocity reconstruction using Gaussian simulations. While the numerical errors for radial velocity reconstruction were small, they are significant for transverse velocity reconstruction. Resolving the observed problems with numerical errors will require improved techniques for accurate harmonic transforms. This pipeline can serve as a prototype for analysis of future datasets, which we will pursue in future work.
\item On the full sky, we demonstrated the utility of the principal components of the reconstruction. The principal component maps are signal dominated over a wide range of angular scales with a redshift distribution that is easy to visualize. To produce similar maps on the masked sky, it will be necessary to develop methods in map-space as opposed to the harmonic-space methods introduced here, which we will pursue in future work. 
\item The code used to produce the results presented in this paper, \texttt{ReCCO}, can be found at: \href{https://github.com/jcayuso/ReCCO}{\url{https://github.com/jcayuso/ReCCO}}.
\end{itemize}

A fundamental assumption of the quadratic estimator formalism explored in this paper is that the underlying fields are Gaussian. On the small-scales that contribute the most to the quadratic estimators, this assumption is far from accurate. One consequence is the presence of the `$N^{(3/2)}$ bias', explored by Ref.~\cite{Giri:2020pkk} in the box picture. We have not performed an analysis of this contribution in the light cone picture, although a similar computation based on the Halo Model could in principle be performed (albeit with complex projection integrals to contend with). This may be a necessary ingredient for using velocity reconstruction in the light cone picture to measure and constrain cosmological parameters, although Ref.~\cite{Giri:2020pkk} showed that it can be neglected for constraints on primordial non-Gaussianity. A more comprehensive assessment of the effect on a variety of cosmological parameters should be performed. Aside from this bias, both Refs.~\cite{Giri:2020pkk} and~\cite{Cayuso:2018lhv} demonstrated that velocity reconstruction essentially works as advertised even for non-linear N-body simulations. An important future analysis will be to directly compare the importance of the various systematic effects discussed in this paper between Gaussian and non-linear N-body simulated datasets. 

In the future, our framework can be extended to assess velocity reconstruction using different tracers such as the CIB~\cite{McCarthy:2019xwk} or intensity maps~\cite{Sato-Polito:2020cil}. Other extensions include velocity reconstruction using reionization kSZ and 21cm maps~\cite{Hotinli:2020csk} or reconstruction of the remote quadrupole field~\cite{Deutsch_2018,Deutsch:2017ybc}. Having a unifying framework, or at least a unifying basis, allows one to combine the cosmological information from these various probes. Examples where this may be important include constraints on modified gravity~\cite{Pan:2019dax} and various early-Universe scenarios~\cite{Cayuso:2019hen}.

We foresee a cosmological paradigm shift, in which reconstruction of the lensing potential, velocity fields and the remote quadrupole field will provide the most precise tests of fundamental physics. This use of CMB secondaries to extract cosmological information through various cross-correlations is in some sense 'free information', since these analyses rely on data from already planned CMB experiments and galaxy surveys. However, as these techniques mature, they may motivate an even stronger push towards the low-noise, high-resolution frontier with future CMB experiments. We continue to explore the possibilities in future work.

\acknowledgements
We thank S. Ferraro, A. Krolewski, M. Madhavacheril, J. Mertens, M. Munchmeyer, and E. Schaan for useful conversations and input at various stages of this project. MCJ is supported by the National Science and Engineering Research Council through a Discovery grant. This research was supported in part by Perimeter Institute for Theoretical Physics. Research at Perimeter Institute is supported by the Government of Canada through the Department of Innovation, Science and Economic Development Canada and by the Province of Ontario through the Ministry of Research, Innovation and Science. This work was completed in part at the Aspen Center for Physics, which is supported by National Science Foundation grant PHY-1607611. SCH is supported by the Horizon Fellowship from Johns Hopkins University.  SCH also acknowledges the support of a grant from the Simons Foundation at the Aspen Center for Physics, Imperial College President's Fellowship and a postdoctoral fellowship from Imperial College London. SCH would like to thank Imperial College High Performance Computing Service at Imperial College London (UK) for providing computational resources at various early stages of this project. Some of the results in this paper have been derived using the HEALPix package~\cite{2005ApJ...622..759G}.

\bibliography{references}

\appendix

\section{Beyond Limber approximation}\label{sec:beyond_limber}
\subsection{The general picture}
We review the `Beyond Limber approximation' method from \cite{Fang:2019xat}, which we use to evaluate angular power spectra that take the form Eq.~\ref{Cl_int1}:
\begin{equation*}
C_{\ell}^{F^WG^{W'}} =  \int d \chi_{1} d \chi_{2} \ W \left(\chi_{1}\right) W' \left(\chi_{2}\right) \int \frac{k^{2} d k}{(2\pi)^3} \ \mathcal{K}^{F}_{\ell}(\chi_1,k)\, \mathcal{K}^{G}_{\ell}(\chi_2,k) \ P_{FG}(\chi_1,\chi_2,k).
\end{equation*}
The method aims to separate the integral above into a piece suitable for the Limber approximation, and a piece that can be expressed as a simple Hankel transform. The separation occurs at the level of the power spectrum $P_{FG}(\chi_1,\chi_2,k)$ by defining a `non-linear' power spectrum
\begin{equation}
P^{(nlin)}_{FG}(\chi_1,\chi_2,k) = P_{FG}(\chi_1,\chi_2,k) -P^{(lin)}_{FG}(\chi_1,\chi_2,k).
\end{equation}
where $P_{FG}(\chi_1,\chi_2,k)$ is the full power spectrum we calculate using the Halo Model described below and $P^{(lin)}_{FG}(\chi_1,\chi_2,k)$ is the linear theory power spectrum. The non-linear power spectrum defined this way is negligible on large scales and starts becoming important for scales and redshifts at which non-linearity kicks in. It is argued in \cite{Fang:2019xat} that the Limber approximation of the nonlinear correction term $(P_{FG} - P^{(lin)}_{FG})$
is sufficiently accurate in realistic cases and therefore the angular power spectrum integral can be rearranged as:
\begin{eqnarray}\label{eq:cl_int2}
C_{\ell}^{F^WG^{W'}} &=& \mbox{Limber}\left[\int d \chi_{1} d \chi_{2} \ W \left(\chi_{1}\right) W' \left(\chi_{2}\right) \int \frac{k^{2} d k}{(2\pi)^3} \ \mathcal{K}^{F}_{\ell}(\chi_1,k)\, \mathcal{K}^{G}_{\ell}(\chi_2,k) \ P^{nlin}_{FG}(\chi_1,\chi_2,k)\right]\nonumber \\
&+&\int d \chi_{1} d \chi_{2} \ W \left(\chi_{1}\right) W' \left(\chi_{2}\right) \int \frac{k^{2} d k}{(2\pi)^3} \ \mathcal{K}^{F}_{\ell}(\chi_1,k)\, \mathcal{K}^{G}_{\ell}(\chi_2,k) \ P^{lin}_{FG}(\chi_1,\chi_2,k).
\end{eqnarray}
The linear power spectrum can be related to its value at redshift zero using a growth factor. Ignoring any scale dependent growth for the moment, the linear power spectrum can be expressed as:
\begin{equation}
P^{lin}_{FG}(\chi_1,\chi_2,k) = P^{lin}_{FG}(0,0,k)g_F(\chi_1)g_G(\chi_2),
\end{equation}
which allows us to separate the $\chi_1$ and $\chi_2$ dependence in the second term of Eq.~\ref{eq:cl_int2}:
\begin{eqnarray}\label{eq:cl_int3}
C_{\ell}^{F^WG^{W'}} &=& \mbox{Limber}\left[\int d \chi_{1} d \chi_{2} \ W \left(\chi_{1}\right) W' \left(\chi_{2}\right) \int \frac{k^{2} d k}{(2\pi)^3} \ \mathcal{K}^{F}_{\ell}(\chi_1,k)\, \mathcal{K}^{G}_{\ell}(\chi_2,k) \ P^{nlin}_{FG}(\chi_1,\chi_2,k)\right]\nonumber \\
&+&\int \frac{k^{2} d k}{(2\pi)^3}P^{lin}_{FG}(0,0,k)\nonumber\\
&\times&
\left[\int d \chi_{1} W \left(\chi_{1}\right)g(\chi_1)\mathcal{K}^{F}_{\ell}(\chi_1,k)\right]
\left[\int d \chi_{2} W' \left(\chi_{2}\right)g(\chi_2)\mathcal{K}^{G}_{\ell}(\chi_2,k)\right] .
\end{eqnarray}
For kernels $\mathcal{K}_{\ell}(\chi,k)$ of the form $f_1(\chi)f_2(k)f_3(\ell)j_{\ell'}(k\chi)$, where $f_1,f_2,f3$ are arbitrary functions and $j_{\ell'}(k\chi)$ is a spherical Bessel function\footnote{All the integral kernels we consider in this work can be expressed as sums of terms with this form.}, the $\chi$ space integrals between brackets can be expressed in terms of Hankel transforms, which can be calculated much faster and with more accuracy than brute force integrations of spherical Bessel functions. If the growth factors $g_F,g_G$ are scale dependent,
\begin{equation}\label{eq:lin_sep}
    P^{lin}_{FG}(\chi_1,\chi_2,k) = P^{lin}_{FG}(0,0,k)g_F(\chi_1,k)g_G(\chi_2,k)
\end{equation}
then the terms in brackets in the second line of Eq.~\ref{eq:cl_int3} cannot be expressed as Hankel transforms. The authors of \cite{Fang:2019xat} work around this problem by splitting the $\chi$ space integrations into narrow enough bins such that the evolution of the scale dependence inside each bin can be ignored. Inside each bin, the growth factor can be approximated as :
\begin{equation}
    g(\chi,k) = g(\bar{\chi},k)\frac{g(\chi,k)}{g(\bar{\chi},k)}
    \approx g(\bar{\chi},k)g^{\mbox{eff}}(\bar{\chi},\chi)
\end{equation}
where $\bar{\chi}$ is the mean $\chi$ in the bin and the approximation comes from ignoring the evolution of the k-dependence. With this, we can approximate the linear power spectrum for $\chi_1$ and $\chi_2$ inside bins with mean $\bar{\chi}_1$ and $\bar{\chi}_2$ as:
\begin{eqnarray}
    P^{lin}_{FG}(\chi_1,\chi_2,k) &=& P^{lin}_{FG}(0,0,k)g_F(\chi_1,k)g_G(\chi_2,k) \nonumber\\
    &\approx& P^{lin}_{FG}(\bar{\chi}_1,\bar{\chi}_1,k)g^{\mbox{eff}}_F(\bar{\chi_1},\chi_1)g^{\mbox{eff}}_G(\bar{\chi}_2, \chi_2)
\end{eqnarray}
and this allows us to restore the separability necessary to construct the Hankel transforms:

\begin{eqnarray}
C_{\ell}^{F^WG^{W'}} &=& \mbox{Limber}\left[\int d \chi_{1} d \chi_{2} \ W \left(\chi_{1}\right) W' \left(\chi_{2}\right) \int \frac{k^{2} d k}{(2\pi)^3} \ \mathcal{K}^{F}_{\ell}(\chi_1,k)\, \mathcal{K}^{G}_{\ell}(\chi_2,k) \ P^{nlin}_{FG}(\chi_1,\chi_2,k)\right]\nonumber \\
&+&\int \frac{k^{2} d k}{(2\pi)^3}\sum_{i}\sum_{j} P^{lin}_{FG}(\bar{\chi_i},\bar{\chi_j},k)\nonumber\\
&\times&
\left[\int d \chi_{i} W \left(\chi_{i}\right)g^{\mbox{eff}}_F(\bar{\chi_i},\chi_i)\mathcal{K}^{F}_{\ell}(\chi_i,k)\right]
\left[\int d \chi_{j} W' \left(\chi_{j}\right)g^{\mbox{eff}}_G(\bar{\chi_j},\chi_j)\mathcal{K}^{G}_{\ell}(\chi_j,k)\right] \nonumber\\
\end{eqnarray}
where the sums are over the auxiliary bins constructed to do the approximation. Since there is no limitation on how small these auxiliary bins can be, this approximation can be made as accurate as necessary.

\subsection{Our implementation}

In our implementation of the Beyond Limber method, we define the non-linear piece of the power spectrum in the following ways depending on the particular observables involved:
\begin{itemize}
    \item{For power spectra involving only dark matter, electrons or galaxies, we define the non-linear spectrum as:
    \begin{equation}
    P^{(nlin)}_{FG}(\chi_1,\chi_2,k) = P^{1h+2h}_{FG}(\chi_1,\chi_2,k) -b^F(\chi_1,k)b^G(\chi_2,k)P^{lin}_{mm}(\chi_1,\chi_2,k)
    \end{equation}
    where $P^{1h+2h}_{FG}(\chi_1,\chi_2,k)$ is the full power spectrum computed using the Halo Model containing the 1-halo term and 2-halo term (see Appendix \ref{sec:halomodel}), $P^{lin}_{mm}(\chi_1,\chi_2,k)$ is the linear dark matter power spectrum from CAMB, and $b^X(\chi,k)$ is the large scale linear bias function computed with the Halo Model, only different from 1 for galaxies (we assume electrons trace dark matter for linear modes). 
    }
    \item{For power spectra involving at least one power of the CIB or tSZ,we define the non-linear spectrum as:
    \begin{equation}
    P^{(nlin)}_{FG}(\chi_1,\chi_2,k) = P^{1h+2h}_{FG}(\chi_1,\chi_2,k) -P^{2h}_{FG}(\chi_1,\chi_2,k)
    \end{equation}
    where $P^{2h}_{FG}(\chi_1,\chi_2,k)$ is the 2-halo term computed using the Halo Model. Effectively, we are treating the 1-halo term as the non-linear piece and the 2-halo term as the linear piece. This is not entirely correct, because the 2-halo term does account for part of the non-linearities on small scales and it is not strictly separable as in Eq.~\ref{eq:lin_sep}. However, a detailed inspection reveals that the 2-halo term is separable on the scales for which the second term of Eq.~\ref{eq:cl_int3} finds most of its support, and that the 1-halo term dominates the regime for which the Limber approximation is adequate.
    }
\end{itemize}

\section{Quadratic estimators}\label{sec:quad_est_deriv}

In this appendix, we derive the unbiased and minimum variance quadratic estimator Eq.~\ref{estimator0} for a modulating field $M^{\alpha}_{\ell_1 m_1}$. The starting point is the statistically anisotropic cross-power Eq.~\ref{anisotropic2}:
\begin{equation}
\Big{\langle}\Theta_{\ell m}\;\delta^{W}_{\ell'm'}\Big{\rangle} = 
(-1)^m C^{I\delta^{W}}_{\ell}\delta_{\ell\ell'}\delta_{m-m'} +
\sum_{\ell_1 m_1} (-1)^{m_1} 
\begin{pmatrix}
\ell & \ell' & \ell_1 \\
m & m' & - m_1
\end{pmatrix}
f^{M^\alpha W}_{\ell\ell_1\ell'} \, M^{\alpha}_{\ell_1 m_1} 
\end{equation}
where the form of $f^{M^\alpha W}_{\ell\ell_1\ell'}$ depends on the observable. The quadratic estimator is of the form:
\begin{equation}
\hat{M}^{\alpha}_{LM} = A^{M^\alpha}_L \sum_{\ell m; \ell' m'} (-1)^M 
\begin{pmatrix}
\ell & \ell' & L \\
m & m' & -M
\end{pmatrix}
G^{M^\alpha W}_{\ell \ell' L}\ \Theta_{\ell m} \delta^{W}_{\ell'm'}
\end{equation}
Our goal is to find the appropriate weights $G^{M^\alpha W}_{\ell \ell' L}$ such that we minimize $\langle \hat{M}^{\alpha}_{LM} \hat{M}^{\alpha}_{LM} \rangle$ subject to the constraint $\langle \hat{M}^{\alpha}_{LM} \rangle = M^{\alpha}_{LM}$. 

First, we find the mean of the estimator:
\begin{eqnarray}
\langle \hat{M}^{\alpha}_{LM} \rangle &=& A^{M^\alpha}_L \sum_{\ell m; \ell' m'} (-1)^M 
\begin{pmatrix}
\ell & \ell' & L \\
m & m' & -M
\end{pmatrix}
G^{M^\alpha W}_{\ell \ell' L}\ \langle \Theta_{\ell m} \delta^{W}_{\ell'm'} \rangle \\
&=& A^{M^\alpha}_L \sum_{\ell' m'} (-1)^{M-m'} 
\begin{pmatrix}
\ell' & \ell' & L \\
-m' & m' & -M
\end{pmatrix}
G^{M^\alpha W}_{\ell \ell' L} C^{I\delta^{W}}_{\ell} \\
&+& A^{M^\alpha}_L \sum_{\ell m; \ell' m';L'M'} (-1)^{M+M'} 
\begin{pmatrix}
\ell & \ell' & L \\
m & m' & -M
\end{pmatrix}
\begin{pmatrix}
\ell & \ell' & L' \\
m & m' & - M'
\end{pmatrix}
G^{M^\alpha W}_{\ell \ell' L} \ f^{M^\alpha W}_{\ell L' \ell'} \, M^{\alpha}_{L'M'} 
\end{eqnarray}
We now use: 
\begin{eqnarray}
(-1)^{-m'} \begin{pmatrix}
\ell' & \ell' & L \\
-m' & m' & -M
\end{pmatrix} &=& (-1)^{-m'} \begin{pmatrix}
\ell' & \ell' & L \\
-m' & m' & 0
\end{pmatrix} \delta_{M0}
\end{eqnarray}
as well as the following properties of 3j symbols:
\begin{eqnarray}\label{eq:3j1}
\sum_{m'} ( -1)^{-m'}  \begin{pmatrix}
\ell' & \ell' & L \\
-m' & m' & 0
\end{pmatrix}
&=& (-1)^{\ell'} \sqrt{2\ell' +1 } \delta_{L0}
\end{eqnarray}
and 
\begin{equation}
\label{eq:3j2}
\sum_{m, m'} \begin{pmatrix}
\ell & \ell' & L \\
m & m' & -M
\end{pmatrix} 
\begin{pmatrix}
\ell & \ell' & L' \\
m & m' & -M'
\end{pmatrix} 
= \frac{\delta_{LL'} \delta_{MM'}}{2L+1}
\end{equation}
Substituting these relations into the estimator, we obtain:
\begin{eqnarray}
\langle \hat{M}^{\alpha}_{LM} \rangle &=& A^{M^\alpha}_L \sum_{\ell} (-1)^\ell  \sqrt{2 \ell + 1} \ G^{M^\alpha W}_{\ell \ell' L} C^{I\delta^{W}}_{\ell} \ \delta_{L0}\delta_{M0} \\
&+& M^{\alpha}_{LM}   \frac{ A^{M^\alpha}_L }{2L+1} \sum_{\ell; \ell'} G^{M^\alpha W}_{\ell \ell' L}  f^{M^\alpha W}_{\ell L \ell'}
\end{eqnarray}
Aside from the monopole, we can make the estimator unbiased so long as:
\begin{equation}\label{eq:nobias}
A_L = (2L+1) \left( \sum_{\ell; \ell'} G^{M^\alpha W}_{\ell \ell' L}  f^{M^\alpha W}_{\ell L \ell'}  \right)^{-1}
\end{equation}

We can now fix $G^{M^\alpha W}_{\ell \ell' L}$ by minimizing the variance of the estimator. We compute:
\begin{eqnarray}\label{eq:app_estimator_variance}
\langle  \hat{M}^{\alpha*}_{LM}  \hat{M}^{\alpha}_{LM} \rangle = A_L^2 \sum_{\ell_1 m_1; \ell_2 m_2} \sum_{\ell_1' m_1'; \ell_2' m_2'}  &&
\begin{pmatrix}
\ell_1 & \ell_2 & L \\
m_1 & m_2 & -M
\end{pmatrix}
\begin{pmatrix}
\ell_1' & \ell_2' & L \\
m_1' & m_2' & -M
\end{pmatrix} \\
&\times&G^{M^\alpha W}_{\ell_1 \ell_2 L} G^{M^\alpha W}_{\ell_1' \ell_2' L} \ \langle \Theta_{\ell_1 m_1}^* \delta^{W*}_{\ell_2 m_2} \Theta_{\ell_1' m_1'} \delta^{W}_{\ell_2' m_2'} \rangle 
\end{eqnarray}
The four-point function can be decomposed into a connected and disconnected piece:
\begin{eqnarray}\label{eq:est_4pt_discon_con}
\langle \Theta_{\ell_1 m_1}^* \delta^{W*}_{\ell_2 m_2} \Theta_{\ell_1' m_1'} \delta^{W}_{\ell_2' m_2'} \rangle = \langle \Theta_{\ell_1 m_1}^* \delta^{W*}_{\ell_2 m_2} \Theta_{\ell_1' m_1'} \delta^{W}_{\ell_2' m_2'} \rangle_{\rm con} + \langle \Theta_{\ell_1 m_1}^* \delta^{W*}_{\ell_2 m_2} \Theta_{\ell_1' m_1'} \delta^{W}_{\ell_2' m_2'} \rangle_{\rm discon} 
\end{eqnarray}
Here, we minimize the variance considering the disconnected contribution only:
\begin{eqnarray}
\langle \Theta_{\ell_1 m_1}^* \delta^{W*}_{\ell_2 m_2} \Theta_{\ell_1' m_1'} \delta^{W}_{\ell_2' m_2'} \rangle_{\rm discon} &=& \langle \Theta_{\ell_1 m_1}^* \delta^{W*}_{\ell_2 m_2} \rangle \langle \Theta_{\ell_1' m_1'} \delta^{W}_{\ell_2' m_2'}  \rangle + \langle \Theta_{\ell_1 m_1}^* \Theta_{\ell_1' m_1'} \rangle \langle  \delta^{W*}_{\ell_2 m_2} \delta^{W}_{\ell_2' m_2'}  \rangle \\
&+& \langle  \Theta_{\ell_1 m_1}^* \delta^{W}_{\ell_2' m_2'}  \rangle \langle  \Theta_{\ell_1' m_1'}\delta^{W*}_{\ell_2 m_2}  \rangle \\
&=& (-1)^{m_2} \delta_{\ell_1 \ell_2} \delta_{m_1-m_2} C_{\ell_1}^{\Theta \delta^{W}} (-1)^{m_1'} \delta_{\ell_1' \ell_2'} \delta_{-m_1' m_2'} C_{\ell_1'}^{\Theta \delta^{W}} \\
&+& \delta_{\ell_1 \ell_1'} \delta_{m_1 m_1'} C_{\ell_1}^{\Theta \Theta} \delta_{\ell_2 \ell_2'} \delta_{m_2 m_2'} C_{\ell_2}^{\delta^W \delta^W} \\
&+& \delta_{\ell_1 \ell_2'} \delta_{m_1 m_2'} C_{\ell_1}^{\Theta \delta^{W}} (-1)^{m_2 +m_1'}  \delta_{-m_2-m_1'} \delta_{\ell_2 \ell_1'} C_{\ell_1'}^{ \Theta \delta^{W}}
\end{eqnarray}
Contributions to the connected piece are discussed in Appendix~\ref{sec:variance_appendix}. Plugging this expression into the variance:
\begin{eqnarray}
\langle \hat{M}^{\alpha*}_{LM}  \hat{M}^{\alpha}_{LM} \rangle &=&  A_L^2 \sum \left[  
\begin{pmatrix}
\ell_1 & \ell_2 & L \\
m_1 & m_2 & -M
\end{pmatrix}^2
(G^{M^\alpha W}_{\ell_1 \ell_2 L})^2 C_{\ell_1}^{\Theta \Theta} C_{\ell_2}^{\delta^W \delta^W} \right. \\ 
&+& \left. (-1)^{m_2+m_1'} 
\begin{pmatrix}
\ell_2 & \ell_2 & L \\
-m_2 & m_2 & -M
\end{pmatrix}
\begin{pmatrix}
\ell_1' & \ell_1' & L \\
m_1' & -m_1' & -M
\end{pmatrix}
G^{M^\alpha W}_{\ell_2 \ell_2 L} G^{M^\alpha W}_{\ell_1' \ell_1' L} C_{\ell_2}^{\Theta \delta^{W}} C_{\ell_1'}^{\Theta \delta^{W}} \right. \\
&+& \left. 
\begin{pmatrix}
\ell_1 & \ell_1' & L \\
m_1 & m_1' & -M
\end{pmatrix}
\begin{pmatrix}
\ell_1' & \ell_1 & L \\
m_1' & m_1 & -M
\end{pmatrix}
G^{M^\alpha W}_{\ell_1 \ell_1' L} G^{M^\alpha W}_{\ell_1' \ell_1 L} C_{\ell_1}^{\Theta \delta^{W}} C_{\ell_1'}^{\Theta \delta^{W}}
 \right]
\end{eqnarray}
We now perform the sums over $m$. Using Eq.~\ref{eq:3j1}, the second term in parentheses contributes only to the monopole. We neglect this term in the following. To evaluate the third term, we use
\begin{equation}
\begin{pmatrix}
\ell_1' & \ell_1 & L \\
m_1' & m_1 & -M
\end{pmatrix} = (-1)^{\ell_1+\ell_1'+L} 
\begin{pmatrix}
\ell_1 & \ell_1' & L \\
m_1 & m_1' & -M
\end{pmatrix}
\end{equation}
Changing dummy indices, the variance is
\begin{eqnarray}
\langle \hat{M}^{\alpha*}_{LM}  \hat{M}^{\alpha}_{LM} \rangle &=& A_L^2 \sum_{\ell_1 m_1; \ell_2 m_2} \left[  
\begin{pmatrix}
\ell_1 & \ell_2 & L \\
m_1 & m_2 & -M
\end{pmatrix}^2
(G^{M^\alpha W}_{\ell_1 \ell_2 L})^2 C_{\ell_1}^{\Theta \Theta} C_{\ell_2}^{\delta^W \delta^W} \right. \\ 
&+& \left.  (-1)^{\ell_1+\ell_1'+L} 
\begin{pmatrix}
\ell_1 & \ell_2 & L \\
m_1 & m_2 & -M
\end{pmatrix}^2
G^{M^\alpha W}_{\ell_1 \ell_2 L} G^{M^\alpha W}_{\ell_2 \ell_1 L} C_{\ell_1}^{\Theta \delta^{W}} C_{\ell_2}^{\Theta \delta^{W}}
 \right]
\end{eqnarray}
Using Eq.~\ref{eq:3j2}, we can perform the sums over $m_1, m_2$ to obtain:
\begin{equation}
\langle \hat{M}^{\alpha*}_{LM}  \hat{M}^{\alpha}_{LM} \rangle = \frac{ 1}{2L+1} \sum_{\ell_1; \ell_2} A_L G^{M^\alpha W}_{\ell_1 \ell_2 L} \left[  
 A_L G^{M^\alpha W}_{\ell_1 \ell_2 L} C_{\ell_1}^{\Theta \Theta} C_{\ell_2}^{\delta^W \delta^W}
+ (-1)^{\ell_1+\ell_2+L} 
 A_L G^{M^\alpha W}_{\ell_2 \ell_1 L} C_{\ell_1}^{\Theta \delta^{W}} C_{\ell_2}^{\Theta \delta^{W}}
 \right]
\end{equation}

To minimize the variance, we can use the Lagrange Multiplier method. First, let's define 
\begin{equation}
F_{\ell_1 \ell_2 L} \equiv A_L G^{M^\alpha W}_{\ell_1 \ell_2 L}
\end{equation}
The variance can therefore be written as:
\begin{equation}
\langle \hat{M}^{\alpha*}_{LM}  \hat{M}^{\alpha}_{LM} \rangle = \frac{1}{2L+1} \sum_{\ell_1; \ell_2} F_{\ell_1 \ell_2 L} \left[  
 F_{\ell_1 \ell_2 L} C_{\ell_1}^{\Theta \Theta} C_{\ell_2}^{\delta^W \delta^W}
+ (-1)^{\ell_1+\ell_2+L} 
 F_{\ell_2 \ell_1 L} C_{\ell_1}^{\Theta \delta^{W}} C_{\ell_2}^{\Theta \delta^{W}}
 \right]
\end{equation}
We want to minimize the variance subject to the constraint that the estimator is unbiased, which is enforced by Eq.~\ref{eq:nobias}. This condition translates to:
\begin{equation}\label{eq:nobbias2}
\frac{1}{2L+1} \sum_{\ell_1;\ell_2} F_{\ell_1 \ell_2 L} f^{M^\alpha W}_{\ell_1 L \ell_2} - 1 = 0
\end{equation}
We therefore want to evaluate:
\begin{eqnarray}
0 = \frac{\delta}{\delta F_{\ell_1 \ell_2 L}} && \left( \frac{1}{2L+1} \sum_{\ell_1; \ell_2} F_{\ell_1 \ell_2 L} \left[  
 F_{\ell_1 \ell_2 L} C_{\ell_1}^{\Theta \Theta} C_{\ell_2}^{\delta^W \delta^W}
+ (-1)^{\ell_1+\ell_2+L} 
 F_{\ell_2 \ell_1 L} C_{\ell_1}^{\Theta \delta^{W}} C_{\ell_2}^{\Theta \delta^{W}}
 \right] \right. \\
 &+& \left. \lambda \left[ \frac{1}{2L+1} \sum_{\ell_1;\ell_2} F_{\ell_1 \ell_2 L} f^{M^\alpha W}_{\ell_1 L \ell_2} - 1 \right]  \right)
\end{eqnarray}
where $\lambda$ is the Lagrange multiplier. Evaluating the derivative yields
\begin{equation}\label{eq:minimize}
F_{\ell_1 \ell_2 L} C_{\ell_1}^{\Theta \Theta} C_{\ell_2}^{\delta^W \delta^W}
+ (-1)^{\ell_1+\ell_2+L} 
 F_{\ell_2 \ell_1 L} C_{\ell_1}^{\Theta \delta^{W}} C_{\ell_2}^{\Theta \delta^{W}} + \lambda f^{M^\alpha W}_{\ell_1 L \ell_2} = 0
\end{equation}
Multiplying by $C_{\ell_2}^{\Theta \Theta}C_{\ell_1}^{\delta^W \delta^W}$ and subtracting $(-1)^{\ell_1+\ell_2+L} C_{\ell_1}^{\Theta \delta^{W}} C_{\ell_2}^{\Theta \delta^{W}}$ times Eq.~\ref{eq:minimize} with permuted indices $\ell_1 \leftrightarrow \ell_2$ we obtain:
\begin{eqnarray}
F_{\ell_1 \ell_2 L}  + \lambda \frac{ C_{\ell_2}^{\Theta \Theta}C_{\ell_1}^{\delta^W \delta^W} f^{M^\alpha W}_{\ell_1 L \ell_2} - (-1)^{\ell_1+\ell_2+L} C_{\ell_1}^{\Theta \delta^{W}} C_{\ell_2}^{\Theta \delta^{W}} f^{M^\alpha W}_{\ell_2 L \ell_1} }{ C_{\ell_1}^{\Theta \Theta} C_{\ell_2}^{\Theta \Theta} C_{\ell_1}^{\delta^W \delta^W} C_{\ell_2}^{\delta^W \delta^W} - (C_{\ell_1}^{\Theta \delta^{W}})^2 (C_{\ell_2}^{\Theta \delta^{W}})^2 } = 0
\end{eqnarray}
For now, let's define a new function:
\begin{equation}
h_{\ell_1 \ell_2 L} \equiv \frac{ C_{\ell_2}^{\Theta \Theta}C_{\ell_1}^{\delta^W \delta^W} f^{M^\alpha W}_{\ell_1 L \ell_2} - (-1)^{\ell_1+\ell_2+L} C_{\ell_1}^{\Theta \delta^{W}} C_{\ell_2}^{\Theta \delta^{W}} f^{M^\alpha W}_{\ell_2 L \ell_1} }{ C_{\ell_1}^{\Theta \Theta} C_{\ell_2}^{\Theta \Theta} C_{\ell_1}^{\delta^W \delta^W} C_{\ell_2}^{\delta^W \delta^W} - (C_{\ell_1}^{\Theta \delta^{W}})^2 (C_{\ell_2}^{\Theta \delta^{W}})^2 }
\end{equation}
so that:
\begin{equation}\label{eq:fstar}
F_{\ell_1 \ell_2 L} = - \lambda h_{\ell_1 \ell_2 L}.
\end{equation}
Multiplying by $f^{M^\alpha W}_{\ell_1 L \ell_2}$ and using the no bias condition Eq.~\ref{eq:nobbias2}, we can solve for the Lagrange multiplier:
\begin{equation}
\lambda = -(2L+1) \left[  \sum_{\ell_1 \ell_2} h_{\ell_1 \ell_2 L}  f^{M^\alpha W}_{\ell_1 L \ell_2} \right]^{-1}
\end{equation}
Substituting this into Eq.~\ref{eq:fstar}, we have
\begin{equation}
F_{\ell_1 \ell_2 L}  = (2L+1) h_{\ell_1 \ell_2 L} \left[  \sum_{\ell \ell'} h_{\ell \ell' L} f^{M^\alpha W}_{\ell L \ell'} \right]^{-1} = (2L+1) G^{M^\alpha W}_{\ell_1 \ell_2 L} \left[  \sum_{\ell \ell'} G^{M^\alpha W}_{\ell \ell' L} f^{M^\alpha W}_{\ell L \ell'} \right]^{-1}
\end{equation}
and so we can identify $G^{M^\alpha W}_{\ell_1 \ell_2 L} = h_{\ell_1 \ell_2 L}$ as the choice that minimizes the variance:
\begin{equation}
G^{M^\alpha W}_{\ell_1 \ell_2 L} \equiv \frac{ C_{\ell_2}^{\Theta \Theta}C_{\ell_1}^{\delta^W \delta^W} f^{M^\alpha W}_{\ell_1 L \ell_2} - (-1)^{\ell_1+\ell_2+L} C_{\ell_1}^{\Theta \delta^{W}} C_{\ell_2}^{\Theta \delta^{W}} f^{M^\alpha W}_{\ell_2 L \ell_1} }{ C_{\ell_1}^{\Theta \Theta} C_{\ell_2}^{\Theta \Theta} C_{\ell_1}^{\delta^W \delta^W} C_{\ell_2}^{\delta^W \delta^W} - (C_{\ell_1}^{\Theta \delta^{W}})^2 (C_{\ell_2}^{\Theta \delta^{W}})^2 }
\end{equation}

\section{Additional contributions to the estimator mean and variance}\label{sec:variance_appendix}

In this appendix we discuss the contributions to the mean and variance of the quadratic estimator for the radial and transverse velocity fields that arise in the presence of additional non-Gaussian contributions to correlation functions between the CMB temperature and density field. The various contributions to the CMB temperature listed in Eq.~\ref{eq:CMB_contributions} generate non-trivial 3- and 4-point functions between the measured CMB temperature and density tracer. There are two distinct types of non-Gaussian contributions that we must consider. The first type is what our quadratic estimators are based on: the non-Gaussianity associated with the fact that CMB secondaries are line-of-sight integrals over products of fields. The second type is the intrinsic non-Gaussianity of the density and velocity fields due to gravitational collapse (or primordial non-Gaussianity, though we expect this to be negligibly small). A complete assessment of the magnitude of the many contributions to the mean and variance of the quadratic estimators is beyond the scope of the present paper, and will be evaluated in future work. Here, we only attempt to enumerate the contributions that must be considered, and in some cases, estimate their magnitude.

\subsection{Estimator mean}

The mean of the estimators $\langle \hat{M}^{\alpha}_{LM} \rangle$ depend on the two-point function $\langle \Theta_{\ell m} \delta^{W}_{\ell'm'} \rangle$. Quadratic estimators for the radial and transverse velocity fields are based on non-Gaussian contributions to the correlation functions $\langle \Theta^{kSZ}_{\ell m} \delta^{W}_{\ell'm'} \rangle$ and $\langle \Theta^{ML}_{\ell m} \delta^{W}_{\ell'm'} \rangle$. Because the kSZ and moving lens temperature anisotropies depend on the product of density (contrast or gradient) and velocity, these are in fact three-point functions. Above, we considered the squeezed limit of these correlators, where the velocity mode is of much larger wavelength than the density modes. Additionally, we treated the velocity and density fields as Gaussian. When the velocity mode is of comparable wavelength to the density modes, there will be a contribution to the three-point function due to gravitational collapse. We expect this to be important at high $L$, beyond the regime of interest for velocity reconstruction. Likewise, the contribution from lensing $\langle \Theta^L_{\ell m} \delta^{W}_{\ell' m'} \rangle$ will receive contributions from non-linearities on small scales, but since the leading order bias from lensing is small, we expect these additional contributions to be completely negligible on scales of interest. In Sec.~\ref{sec:galaxy_sys} we also considered large angular scale systematics that modulate the observed density field. Similar systematics in a CMB experiment will lead to similar effects although we expect their magnitude to be far smaller. 

Another contribution to the mean of the estimators, which was not considered above, arises from non-linear terms in $\langle \Theta^{XG}_{\ell m} \delta^{W}_{\ell'm'} \rangle$. On scales $\ell, \ell' \gg 1$ where the estimator receives most of its weight, we must include non-linear contributions to the galaxy density field as well as the extragalactic foregrounds (here, the CIB and tSZ). At second order in perturbation theory, schematically we must consider correlators of the form $\langle t \delta \delta \rangle$ where $t$ is the large-scale tidal field (see e.g. Ref.~\cite{Schmidt_2014,Zhu_2016}). It is difficult to imagine this term being larger than the bias induced by calibration error, which takes a similar form, and which is likely far larger in magnitude than the large-scale tidal field. Related to the tidal field, systematics associated with the intrinsic aligment of galaxies lead to a large-scale statistical anisotropy in the galaxy number counts~\cite{0903.4929}; again, it is difficult to imagine that the amplitude of this effect is large enough to cause a significant bias. We defer a detailed estimate of these and other effects to future work.

\subsection{Estimator variance}

Above, we considered only the disconnected contributions to the estimator variance $\langle \hat{M}^{\alpha}_{LM} \hat{M}^{\beta}_{LM} \rangle$. There are a number of additional contributions to the variance, arising from the non-Gaussian nature of the kSZ effect as well as other non-Gaussian contributions to the CMB temperature and galaxy survey. Concentrating on non-kSZ, non-Gaussian contributions to the estimator variance Eq.~\ref{eq:app_estimator_variance}, we conjecture that the most important terms arise from extragalactic foregrounds and CMB lensing: 
\begin{equation}\label{eq:4pt_xgal_L}
\langle \Theta^{XG*}_{\ell_1 m_1} \delta^{W*}_{\ell_2 m_2} \Theta^{XG}_{\ell_1' m_1'} \delta^{W}_{\ell_2' m_2'} \rangle_{\rm con},  \ \ \langle \Theta^{L*}_{\ell_1 m_1} \delta^{W*}_{\ell_2 m_2} \Theta^{L}_{\ell_1' m_1'} \delta^{W}_{\ell_2' m_2'} \rangle_{\rm con}, \ \ 
\end{equation}
Note that the relevant shape of the four point function for the estimator variance is the `collapsed' configuration where $\ell_1 \sim \ell_2$ and $\ell_1' \sim \ell_2'$, since the relevant scales are $L \ll \ell_1, \ell_2, \ell_1', \ell_2'$. The terms in Eq.~\ref{eq:4pt_xgal_L} should be calculable analytically within the Halo Model since the collapsed four-point function typically has a simple form~\cite{Scoccimarro:1999kp}. For example, similar computations have been performed in the context of the CIB have been performed~\cite{Schaan:2018yeh}. Roughly speaking, we expect the disconnected four-point function to dominate the connected four-point function by a power of the matter power spectrum. Therefore, including the connected four-point function will most likely not make a large contribution to the estimator variance. We leave a detailed computation to future work.

Because the kSZ temperature anisotropies arise due to the product of the optical depth and radial velocity, evaluating kSZ contributions to the estimator variance involves computing a six-point function:
\begin{equation}\label{eq:4point_to_6point}
\langle \Theta^{kSZ*}_{\ell_1 m_1} \delta^{W*}_{\ell_2 m_2} \Theta^{kSZ}_{\ell_1' m_1'} \delta^{W}_{\ell_2' m_2'} \rangle =  \sum_{\bar{\ell}_1 \bar{m}_1; \bar{\ell}_2 \bar{m}_2} \sum_{\tilde{\ell}_1 \tilde{m}_1; \tilde{\ell}_2 \tilde{m}_2} R^{\bar{\ell}_1 \bar{\ell}_2 \ell_1}_{\bar{m}_1 \bar{m}_2 -m_1} R^{\tilde{\ell}_1 \tilde{\ell}_2 \ell_1'}_{\tilde{m}_1 \tilde{m}_2 -m_1'} \sum_{ss'} \langle v_{\bar{\ell}_1 \bar{m}_1}^{s*} \dot{\tau}_{\bar{\ell}_2 \bar{m}_2 }^{s*} \delta_{\ell_2 m_2}^{W*} v_{\tilde{\ell}_1 \tilde{m}_1}^{s'} \dot{\tau}_{\tilde{\ell}_2 \tilde{m}_2 }^{s'} \delta^{W}_{\ell_2' m_2'} \rangle
\end{equation}
where
\begin{equation}
R^{\ell_1 \ell_2 \ell}_{m_1 m_2 -m} = (-1)^m \sqrt{\frac{(2\ell+1)(2\ell_1+1)(2\ell_2+1)}{4\pi}} 
\begin{pmatrix}
\ell_1 & \ell_2 & \ell \\
0 & 0 & 0
\end{pmatrix}
\begin{pmatrix}
\ell_1 & \ell_2 & \ell \\
m_1 & m_2 & -m
\end{pmatrix}
\end{equation}
To compute the six point function we must consider both connected and disconnected components. There are a total of 15 terms in the disconnected six point function. We can use the fact that the four-point function Eq.~\ref{eq:4point_to_6point} takes the collapsed configuration, together with the property that the velocity power spectrum falls rapidly with $\ell$ to argue that the relevant scales are $\bar{\ell}_1 \ll \ell_1$, $\tilde{\ell}_1 \ll \ell_1'$, $\ell_1 \sim \ell_2$, and $\ell_1' \sim \ell_2'$. From the 3j symbols in the coupling functions $R^{\ell_1 \ell_2 \ell}_{m_1 m_2 -m} $, this in turn implies that $\bar{\ell_2} \sim \ell_2$ and $\tilde{\ell}_1 \sim \ell_2'$. Therefore, correlators involving the velocity (which is relevant at low-$\ell$) and either $\dot{\tau}$ or $\delta^W$ (which are relevant at high-$\ell$) will not make a significant contribution to the disconnected six-point function. We can therefore make the approximation:
\begin{eqnarray}
\langle v_{\bar{\ell}_1 \bar{m}_1}^{s*} \dot{\tau}_{\bar{\ell}_2 \bar{m}_2 }^{s*} \delta_{\ell_2 m_2}^{W*} v_{\tilde{\ell}_1 \tilde{m}_1}^{s'} \dot{\tau}_{\tilde{\ell}_2 \tilde{m}_2 }^{s'} \delta^{W}_{\ell_2' m_2'} \rangle_{\rm discon} &\simeq& \langle v_{\bar{\ell}_1 \bar{m}_1}^{s*} v_{\tilde{\ell}_1 \tilde{m}_1}^{s'} \rangle \left[ \langle \dot{\tau}_{\bar{\ell}_2 \bar{m}_2 }^{s*} \delta_{\ell_2 m_2}^{W*} \rangle \langle  \dot{\tau}_{\tilde{\ell}_2 \tilde{m}_2 }^{s'} \delta^{W}_{\ell_2' m_2'} \rangle \right. \\
&+& \langle \dot{\tau}_{\bar{\ell}_2 \bar{m}_2 }^{s*} \dot{\tau}_{\tilde{\ell}_2 \tilde{m}_2 }^{s'} \rangle \langle\delta_{\ell_2 m_2}^{W*} \delta^{W}_{\ell_2' m_2'} \rangle +  \left. \langle \dot{\tau}_{\bar{\ell}_2 \bar{m}_2 }^{s*} \delta^{W}_{\ell_2' m_2'} \rangle \langle  \dot{\tau}_{\tilde{\ell}_2 \tilde{m}_2 }^{s'}  \delta_{\ell_2 m_2}^{W*}  \rangle  \right]
\end{eqnarray}
As we now show, the first term gives rise to the signal covariance, the second term reproduces the Gaussian estimator variance, and the third term yields the `$N^{(1)}$ bias' from Ref.~\cite{Giri:2020pkk}.

Substituting the first term into the estimator variance Eq.~\ref{eq:app_estimator_variance}, we obtain:
\begin{eqnarray}
\langle \hat{M}^{\alpha}_{LM} \hat{M}^{\beta}_{LM} \rangle_{kSZ, 1} &=& \sum_{s, s' =1}^\infty (C^{vv})_L^{ss'} \\
&&\left(\sum_{\ell_1 \ell_2} \frac{A^{M^\alpha}_L}{2L+1} G^{M^\alpha W}_{\ell_1 \ell_2 L} \sqrt{ \frac{(2 \ell_1+1) (2 \ell_2+1) (2L+1)}{4\pi}}  \begin{pmatrix}
\ell_1 & \ell_2 & L \nonumber \\
0 & 0 & 0
\end{pmatrix} C_{\ell_2}^{\dot{\tau}^s \delta^W} \right)  \\ 
&&\left(\sum_{\ell_1 \ell_2} \frac{A^{M^\beta}_L}{2L+1} G^{M^\beta W}_{\ell_1 \ell_2 L} \sqrt{ \frac{(2 \ell_1+1) (2 \ell_2+1) (2L+1)}{4\pi}}  \begin{pmatrix}
\ell_1 & \ell_2 & L \nonumber \\
0 & 0 & 0
\end{pmatrix} C_{\ell_2}^{\dot{\tau}^{s'} \delta^W} \right) \\
&=& \sum_{s, s' =1}^\infty (C^{vv})_L^{s s'}  \left(\sum_{\ell_1 \ell_2} \frac{A^{M^\alpha}_L}{2L+1} G^{M^\alpha W}_{\ell_1 \ell_2 L} f_{\ell_1 L \ell_2}^{v^s W}  \right) \left(\sum_{\ell_1 \ell_2} \frac{A^{M^\beta}_L}{2L+1} G^{M^\beta W}_{\ell_1 \ell_2 L} f_{\ell_1 L \ell_2}^{v^{s'} W}  \right) \\
&=&\sum_{s, s' =1}^\infty (C^{vv})_L^{ss'} R_L^{v^s M^\alpha} R_L^{v^{s'} M^\beta}
\end{eqnarray}
This is the signal covariance rotated into the basis defined by the estimators.

 Moving to the second term:
 \begin{eqnarray}
 \langle \hat{M}^{\alpha}_{LM} \hat{M}^{\beta}_{LM} \rangle_{kSZ, 2} &=& A^{M^\alpha}_L A^{M^\beta}_L   \sum_{\ell_1 \ell_2} \frac{ (-1)^{\ell_1 +\ell_2 + L}}{2L+1} G^{M^\alpha W}_{\ell_1 \ell_2 L} G^{M^\beta W}_{\ell_1 \ell_2 L} C_{\ell_2}^{\delta^W \delta^W} \nonumber \\ 
 &&   \sum_{\bar{\ell}_1 \bar{\ell}_2} \frac{(2 \bar{\ell}_1+1) (2 \bar{\ell}_2+1)}{4\pi }
\begin{pmatrix}
\bar{\ell}_1 & \bar{\ell}_2 & \ell_1  \\
0 & 0 & 0
\end{pmatrix}^2
  \sum_{s, s' =1}^\infty 
  (C^{vv})_{\bar{\ell}_1}^{ss'}   (C^{\dot{\tau} \dot{\tau}})_{\bar{\ell}_2}^{ss'}  \\
 &=& A^{M^\alpha}_L A^{M^\beta}_L   \sum_{\ell_1 \ell_2} \frac{ (-1)^{\ell_1 +\ell_2 + L}}{2L+1} G^{M^\alpha W}_{\ell_1 \ell_2 L} G^{M^\beta W}_{\ell_1 \ell_2 L} C_{\ell_2}^{\delta^W \delta^W} C_{\ell_1}^{\rm kSZ}
 \end{eqnarray}
This term combines with the non-kSZ disconnected components of the temperature galaxy four-point function to yield the estimator noise.

The third term is somewhat more complicated, 
\begin{eqnarray}
 \langle \hat{M}^{\alpha}_{LM} \hat{M}^{\beta}_{LM} \rangle_{kSZ, 3} &=&  A^{M^\alpha}_L A^{M^\beta}_{L} \sum_{\ell_1; \ell_2} \sum_{\ell_1'; \ell_2'}  \sum_{\bar{\ell}_1} G^{M^\alpha W}_{\ell_1 \ell_2 L} G^{M^\beta W}_{\ell_1' \ell_2' L'}   \sum_{s, s' =1}^\infty   (C^{vv})_{\bar{\ell}_1}^{s s'} C_{\ell_2'}^{\dot{\tau}^s \delta^W}  C_{\ell_2}^{\dot{\tau}^{s'} \delta^W} \frac{2\bar{\ell}_1+1}{4\pi}  \nonumber\\
&&  \sqrt{(2\ell_1+1) (2\ell_2+1) (2\ell_1'+1) (2\ell_2'+1)} 
\begin{pmatrix}
\bar{\ell}_1 & \ell_2' & \ell_1 \\
0 & 0 & 0
\end{pmatrix}
\begin{pmatrix}
\bar{\ell}_1 & \ell_2 & \ell_1' \\
0 & 0 & 0
\end{pmatrix}
\frac{(-1)^{L+\bar{\ell}_1}}{2L+1}
\begin{Bmatrix}
\ell_1' & \ell_2' & L \\
\ell_1 & \ell_2 & \bar{\ell}_1
\end{Bmatrix}\nonumber\\
\end{eqnarray}
When $L$ is much smaller than the other factors in the 6j symbol, we can simplify using:
\begin{equation}
\begin{Bmatrix}
\ell_1' & \ell_2' & 0 \\
\ell_1 & \ell_2 & \bar{\ell}_1
\end{Bmatrix} = \frac{\delta_{\ell_1, \ell_2} \delta_{\ell_1', \ell_2'}}{\sqrt{\left(2 \ell_1'+1\right)\left(2 \ell_2+1\right)}}(-1)^{\ell_1'+\ell_2+\bar{\ell}_1}\left\{\ell_1',\ell_2,\bar{\ell}_1\right\}
\end{equation}
This gives 
\begin{eqnarray}
  \langle \hat{M}^{\alpha}_{LM} \hat{M}^{\beta}_{LM} \rangle_{kSZ, 3} &=& A^{M^\alpha}_L A^{M^\beta}_{L}  \sum_{\ell_1'; \ell_2}  \sum_{\bar{\ell}_1} G^{M^\alpha W}_{\ell_2 \ell_2 L} G^{M^\beta W}_{\ell_1' \ell_1' L}  \sum_{s, s' =1}^\infty  \frac{2\bar{\ell}_1+1}{4\pi}  (C^{vv})_{\bar{\ell}_1}^{ss'} C_{\ell_1'}^{\dot{\tau}^s \delta^W}  C_{\ell_2}^{\dot{\tau}^{s'} \delta^W} 
   \nonumber\\
&&  \sqrt{(2\ell_2+1)(2\ell_1'+1)} 
\begin{pmatrix}
\bar{\ell}_1 & \ell_1' & \ell_2 \\
0 & 0 & 0
\end{pmatrix}^2
\end{eqnarray}
This is the $N^{(1)}$ bias first computed in \cite{Giri:2020pkk}. Evaluating it, we find, in agreement with \cite{Giri:2020pkk}, that this term is negligible compared to the Gaussian estimator noise. 

Another contribution to the estimator variance arises due to the connected six-point function, the `$N^{(3/2)}$ bias', which was found in Ref.~\cite{Giri:2020pkk} to be even larger than the Gaussian estimator noise in the high signal-to-noise regime. A full computation of this term is beyond the scope of this paper, but will be necessary for a complete analysis in the future.

\section{Halo Model}\label{sec:halomodel}

In this Appendix, we describe the assumptions made in our Halo Model description of various tracers of large scale structure including: dark matter density, galaxy number counts, electron density, cosmic infrared background (CIB), and the thermal Sunyaev Zel'dovich (tSZ) effect. For a general review of the Halo Model of large scale structure, see e.g. Ref.~\cite{Cooray:2002dia}. The final product of the numerical computations is a set of auto- and cross-power spectra at a set of redshifts (dark matter density, galaxy number counts, and electron density) and/or frequencies (CIB and tSZ). These auto- and cross-power spectra are then converted to angular spectra as a function of redshift bin and/or frequency using the techniques described in Sec.~\ref{sec:isotropic_correlations}.

 \subsection{Halo mass function, halo bias, and the matter power spectrum}\label{sec:halomass_bias_etc}
 
 \subsubsection{Halo mass function}
 Within the Halo Model, all matter is distributed in discrete halos of different sizes. The halo mass function $\frac{dN}{dM}$ describes the distribution of the halos: the number density of halos $n^h(z)$ between masses $M_1$ and $M_2$ at $z$ is given by
 \begin{equation}
 n^h(z) = \int _{M_1}^{M_2}\frac{dN}{dM}(M,z)dM.
 \end{equation}
 In our Halo Model, we use the halo mass function of~\cite{2010ApJ...724..878T}, which parametrizes the halo multiplicity function $f(\upsilon)$ as
 \begin{equation}\label{eq:f_upsilon}
f(\upsilon)= \alpha (1+(\beta\upsilon)^{-2\phi})\upsilon^{2\eta}e^{-\eta\upsilon^2/2}.
\end{equation}
 $f(\upsilon)$ is related to $\frac{dN}{dM}$ as
 (see Eq.~2 of \cite{2008ApJ...688..709T})
\begin{equation}
\frac{dN}{dM} = \upsilon f(\upsilon)\frac{\rho_m}{M}\frac{d\ln \sigma^{-1}}{dM}
\end{equation}
where $\rho_m$ is the present day cosmological matter density and the peak height $\upsilon$ is
\begin{equation}
\upsilon\equiv \frac{\delta_c}{\sigma}
\end{equation}
with $\delta_c = 1.686$ the critical density required for collapse. $\sigma$ is the linear matter variance smoothed with a top-hat function over the radius of the halo $R=\left(\frac{3M}{4\pi \rho_m}\right)^{1/3}$
\begin{equation}
\sigma^2(R,z)=\frac{1}{2\pi^2}\int P(k,z) \hat W(k,R)k^2dk
\end{equation}
where $\hat W(k,R)$ is the Fourier transform (in $k$) of a top-hat function with radial extent $R$. 

The values of the parameters $\{\beta,\gamma,\phi,\eta\}$ are listed in Table 4 of~\cite{2010ApJ...724..878T} with a mild redshift dependence given in Eqs.~9-12 of \cite{2010ApJ...724..878T}. The value of $\alpha$ results from applying the $z$-dependent normalization condition to be discussed below in Sec.~\ref{sec:dmpowerspectrum}.

 \subsubsection{Halo bias}
 
 Halos are biased with respect to the underlying dark matter power spectrum; in particular, the power spectrum of halos of masses $M$ at redshift $z$ $P_{hh}$ can be written (on large scales, where the bias is scale-independent) as
 \begin{equation}
 P_{hh}(k,M,z) = b_h(M,z) P_{\mathrm{lin}}(k,z),
\end{equation}
where $P_{\mathrm{lin}}(k,z)$ is the linear dark matter power spectrum and $b_h(M,z)$ is the halo bias. In our Halo Model, we use the halo bias of~\cite{2010ApJ...724..878T}, which is parametrized as
\begin{equation}
b(\upsilon) =1-A\frac{\upsilon^a}{\upsilon^a+\delta_c^a}+B\upsilon^b+C\upsilon^c .
\end{equation}
The values of the parameters $\{A,a,B,b,C,c\}$ are listed in Table 2 of \cite{2010ApJ...724..878T}.

\subsubsection{Halo density profile}

The halo densty profile  $\rho(\vec r, M,z)$ gives the density at a displacement $\vec r$ from the centre of a halo and thus governs the distribution of dark matter within a halo. For spherically symmetric halos, $\rho(\vec r)=\rho(r)$. We take $\rho(r)$ to be Navarro--Frenk--White (NFW)~\cite{1996ApJ...462..563N}, ie
\begin{equation}\label{eq:NFW}
\rho(r) = \frac{\rho_S}{\frac{r}{r_S}\left(1+\frac{r}{r_S}\right)^2}
\end{equation}
with $r_S$ the scale radius, a parameter which is related to the halo radius $r_M$ by the concentration parameter $c=\frac{r_M}{r_S}$;  the scale density $\rho_S$ defines the density of the halo, and can be eliminated in favour of the virial radius and mass by using the definition of mass $M=\int_0^{r_M}4 \pi r^3 \rho(r) dr$. We use the halo concentration parametrization found in~\cite{2008MNRAS.390L..64D}, which parametrizes the concentration of halos as
\begin{equation}
c=A\left(\frac{M}{M_{\mathrm{pivot}}}\right)^{B}(1+z)^{C}.
\end{equation}
The values of $A,B,C$ depend on the definition of the halo mass one is using 
and can be found in Table 1 of~\cite{2008MNRAS.390L..64D} (in particular we take the Sample-F redshift 0-2 row). Note that Ref.~\cite{2008MNRAS.390L..64D} provides different values for the parameters depending on the definition of the halo mass considered; we take $M$ to be the mass within the radius $R_{200m}$ for which the mean density of the halo is 200 times the mean matter density (labeled $M_{\rm{mean}}$ in Ref.~\cite{2008MNRAS.390L..64D}).

In power spectra, the normalized Fourier transform of $\rho(r)$
\begin{equation}
u(k,M,z) \equiv \frac{\int_0^{R} dr 4\pi r^2 \frac{\sin\left( kr\right)}{kr}\rho(r,M,z)}{\int_0^{R} dr 4\pi r^2 \rho(r,M,z)}
\end{equation}
is used. The halo radius $R$ at which we cut off the integral is $R_{200m}$.

\subsubsection{Dark matter power spectrum}\label{sec:dmpowerspectrum}

Within the Halo Model,  power spectra are split into a term sourced by correlations in different halos (inter-halo correlations), and correlations within a single halo (intra-halo correlations). These terms are known as the 2-halo and 1-halo power spectra respectively, so we have
\begin{equation}
P_{mm}(k,z) = P_{mm}^{2h}(k,z) +P_{mm}^{1h}(k,z) 
\end{equation}
where $P_{mm}^{2h}$ and $P_{mm}^{1h}$ denote the 2-halo and 1-halo dark matter power spectra respectively, and $P_{mm}$ is the total dark matter power spectrum.
 Each term  is an integral over all halo masses:
\begin{align}
P_{mm}^{2h}(k,z) = &\left(\int dM\frac{dN}{dM} b_h (M,z)\frac{M}{\rho_m}u(k,M,z) \right) ^2 P_{\mathrm{lin}}(k,z);\\
P_{mm}^{1h}(k,z) = &\int dM\frac{dN}{dM}\left(\frac{M}{\rho_m}u(k,M,z) \right)^2.
\end{align}
 On large scales, it is a requirement that $P_{mm}^{2h}(k,z)=P_{\mathrm{lin}}(k,z)$; this is a consistency condition that ensures that all dark matter resides in halos, and that it is unbiased with respect to itself. This results in the following normalization condition:
 \begin{equation}
 \int b(\nu) f(\nu) d\nu = 1.
 \end{equation}
This consistency condition results in a $z$-dependent constraint on the normalization of the halo mass function: it fixes the value of the parameter $\alpha$ in Eq.~\ref{eq:f_upsilon}.

\subsection{Large-scale structure tracers}\label{sec:halomodel_lss_tracers}

The large-scale structure tracers we are interested in are the galaxy density $g$, electron density $e$, the CIB flux density $I_\nu$ at frequency $\nu$, and the tSZ temperature anisotropy $\Theta^{{\rm{tSZ}},\nu}$ at frequency $\nu$. Below, we summarize for each tracer the essential details necessary for constructing auto-power and cross-power spectra in the Halo Model.

\subsubsection{Galaxy density}

Galaxies are distributed in halos according to a \textit{halo occupation distribution} (HOD). In our HOD, we assign one `central'' galaxy to the centre of halos in a mass-dependent way, and additional `satellite'' galaxies which are distributed throughout the halo according to the dark matter distribution. Thus, the number of galaxies in a halo of mass $M$ at redshift $z$ is
\begin{equation}
N^{\rm{gal}}(M,z) = N^{\mathrm{cen}}(M,z) + N^{\mathrm{sat}}(M,z) 
\end{equation}
where $N^{\mathrm{cen}}$ denotes the number of central galaxies (always 0 or 1) and $N^{\mathrm{sat}}$ the number of satellite galaxies. We use the same HOD as Ref.~\cite{Smith:2018bpn}. The mean number density of galaxies at $z$ is then
\begin{equation}
\bar n^{g}(z) = \int dM\frac{dN}{dM}\left(N^{\mathrm{cen}}(M,z) + N^{\mathrm{sat}}(M,z) \right).
\end{equation}

\subsubsection{Electron density}
Electrons are distributed inside dark matter halos according to a radial density profile $\rho_{e}(r)$. As a fiducial model, we choose the `AGN' gas profiles from \cite{Battaglia:2016xbi}. The Fourier space density profile for electrons is then:

\begin{equation}
u_e(k,M,z)=\frac{1}{M_{AGN}}\int _0 ^{R}4 \pi r^2 \rho_{AGN}(r) \frac{\sin\left( kr \right) }{kr}dr\label{electronprofiles}
\end{equation}
where $M_{AGN}$ is the AGN `mass'' $M_{AGN}=\int_0^{R}4 \pi r^2 \rho_{AGN}(r)dr$, with $R$ the cutoff radius at which we cut off the NFW profile Eq.~\ref{eq:NFW} ($R_{200m}$ for us). 

 \subsubsection{CIB flux density}
 
 The CIB flux density at frequency $\nu$ $I_\nu$ is given by an integral over the CIB emissivity density $j_\nu(\hat n,\chi)$:
 \begin{equation}
 I_\nu(\hat n) = \int d\chi a(\chi) j_\nu(\hat n,\chi).
 \end{equation}
 This can be written as an integral over galaxies with different luminosity densities: the mean emissivity density is
 \begin{equation}
 \bar j_\nu( \chi) =\int dL_{(1+z)\nu} \frac{dN}{dL_{(1+z)\nu}} \frac{L_{(1+z)\nu}}{4\pi}
 \end{equation}
 where $L_{(1+z)\nu}$ is the luminosity density and $ \frac{dN}{dL_{(1+z)\nu}}$ is the halo luminosity function defined in analogy with the halo mass function; the factor of $(1+z)$ in the frequency accounts for the fact that the photons that we receive have been redshifted. Neglecting scatter between $M$ and $L_{\nu}$, this can be written as an integral over the halo mass function
 \begin{equation}
  \bar j_\nu( \chi) =\int dM \frac{dN}{dM}\frac{L_{(1+z)\nu}}{4\pi}.
 \end{equation}
As all luminosity is sourced by galaxies, $L_\nu$ can be separated into that sourced by the central galaxies and the satellite galaxies:
\begin{equation}
L_\nu(M,z) = L^{\mathrm{cen}}_\nu(M,z) +L^{\mathrm{sat}}_\nu(M,z).
\end{equation}
 
 Point sources can be identified and removed form CIB maps by imposing a flux cut and removing sources above this; in our CIB calculations we impose a flux cut of 400mJy at all frequencies.

 \subsubsection{tSZ temperature}
 The tSZ temperature anisotropy at frequency $\nu$ is given by
 \begin{equation}
 \frac{\Delta T^{{\rm{tSZ}}}}{T}(\hat n, \chi) = g_\nu y(\hat n,\chi)
 \end{equation}
where $y(\hat n, x)$ is the Compton $y$-parameter and $g_\nu$ is the spectral function of the tSZ
 \begin{equation}
 g_\nu = x\coth\frac{x}{2}-4
 \end{equation}
 with the dimensionless variable $x$ given by $x\equiv\frac{h\nu}{k_B T_{\rm{CMB}}}$ (where $h$ is Planck's constant; $k_B$ is the Boltzmann constant; and $T_{\rm{CMB}}$ is the temperature of the black-body CMB). The Compton $y$-parameter is a line-of-sight integral over electron pressure
 \begin{equation}
 y(\hat n, \chi) = \frac{\sigma_T}{m_e c^2}\int d\chi a(\chi)P_e(\hat n,\chi) 
 \end{equation}
 where $P_e(\hat n,\chi)$ is the electron pressure at $(\hat n,\chi)$, and where $\sigma_T$ is the Thompson scattering cross section; $m_e$ is the electron mass; and $c$ is the speed of light. 
 
 To calculate the power spectrum of $y$, we need the three-dimensional Fourier transform of $P_e(r)$; for spherically symmetric halos this allows us to define the profile
  \begin{equation}
 y(k,M,z) \equiv  \frac{4\pi\sigma_T a}{m_e c ^2} \int dr r^2 \frac{\sin \left(k r\right)}{kr} P_e(r).\label{ykmzdef}
 \end{equation}
 We use the pressure profiles of~\cite{Battaglia:2016xbi} in our model.

\subsection{2-halo power spectra}\label{sec:2halo}
The 2-halo power spectra are all of the form
\begin{equation}
P^{2h}_{XY}(k,z)=D_X(k,z)D_Y(k,z) P_{\mathrm {lin}}(k,z)
\end{equation}
where $P_{\mathrm {lin}}(k,z)$ is the linear dark matter power spectrum and $D_X(k,z)$ takes the form 
\begin{equation}\label{eq:Dx}
D_X(k,z)=\int dM \frac{dN}{dM}b_h(M,z) A_X(M,k,z)
\end{equation}
with $A_X(M,k,z)$ the Fourier profile of the observable $X$.

 The profile $A_X(M,k,z)$ in Eq.~\ref{eq:Dx} is specific to the observable $X$. For matter, we saw in Sec.~\ref{sec:halomass_bias_etc} that
\begin{equation}
A_m(M,k,z)= \left( \frac{M}{\rho_m}\right) u(k,M,z)
\end{equation}
where $u(k,M,z)$ is the normalized Fourier-transformed dark matter halo density profile. 
For the other observables we have
\begin{align}
A_e (M,k,z)=&\left(\frac{M}{\rho_m}\right) u_e (k,M,z);\label{tau_prof}\\
A_g(M,k,z)=&\frac{N^{\mathrm{cen}}(M,z)+N^{\mathrm{sat}}(M,z) u(k,m,z)}{\bar n_g(z)};\\
A_{j_\nu}(M,k,z)=&\frac{1}{4\pi}\left( L_{(1+z)\nu}^{\mathrm{cen}}(M,z)+L_{(1+z)\nu}^{\mathrm{sat}}(M,z)u(k,M,z)\right);\label{cib_profile}\\
A_y(M,k,z) =&y(k,M,z)\label{y_prof},
\end{align}
with $m$ referring to dark matter; $e$ to the electron density profile; $g$ to the galaxy density; $j_\nu$ to the CIB luminosity density at frequency $\nu$; and $y$ to the Compton $y$ parameter. $y(k,M,z)$ is defined in Eq.~\ref{ykmzdef}.
Note that central galaxies are always taken to be at the centre of the halo which is why they are not multiplied by a $k$-dependent factor\footnote{One could also replace $N^{\mathrm{cen}} \rightarrow N^{\mathrm{cen}}u_c(k)$ to take into account central galaxies that are mis-centered. Here, we take $u_c(k)$ to be $1$, as in Appendix B of \cite{Smith:2018bpn}. One could similarly take into account mis-centering in $ L_{(1+z)\nu}^{\mathrm{cen}}(M,z)$.}. 
The satellite galaxies (and luminosity) are weighted by the dark matter profile $u(k,M,z)$; this is because that the galaxy distribution is modelled as following the dark matter distribution in the halo. 

\subsection{1-halo power spectra}

For the 1-halo power spectra, we distinguish between `discrete'' observables (galaxies and CIB) and `continuous' observables (everything else). For dark matter, electrons, and Compton y, we have
\begin{align}
P^{1h}_{mm}=&\int dM \frac{dN}{dM}\left(\frac{M}{\rho_m}u(k,M,z)\right)^2\\
P^{1h}_{ e e }=&\int dM \frac{dN}{dM}\left(\frac{M}{\rho_m}u_e (k,M,z)\right)^2\\
P^{1h}_{yy}=&\int dM  \frac{dN}{dM}\left(\frac{4\pi\sigma_Ta}{m_ec^2}y(k,M,z)\right)^2,
\end{align}
These power spectra are all of the form 
\begin{equation}
P^{1h}_{XX}= \int dM  \frac{dN}{dM}A_X(M,k,z)^2,
\end{equation}
and their cross spectra are 
\begin{equation}
P^{1h}_{XY}= \int dM  \frac{dN}{dM}A_X(M,k,z)A_Y(M,k,z).
\end{equation}
For galaxies we have~\cite{Smith:2018bpn} (in the `maximally correlated'' model)
\begin{equation}
P^{1h}_{gg}=\int dM \frac{dN}{dM}\frac{2 N^{\mathrm {sat}}(M,z) u(k,M,z)+ ( N^{\mathrm {sat}}(M,z)^2 / N^{\mathrm {cen}}) u(k,M,z)^2}{\bar n_g(z)^2}.
\end{equation}
The 1-halo power spectrum for CIB at frequencies $\nu$ and $\nu'$ is
\begin{equation}
P^{1h}_{\nu\nu^\prime}=\int dM \frac{dN}{dM}\frac{1}{\left(4\pi\right)^2}\left( L_{\nu(1+z)}^{\cen}L_{\nu^\prime(1+z)}^{\sat}u(k,M,z)+L_{\nu^\prime(1+z)}^{\cen}L_{\nu(1+z)}^{\sat}u(k,M,z)+L_{\nu(1+z)}^{\sat}L_{\nu^\prime(1+z)}^{\sat}u(k,M,z)^2\right)
\end{equation}
Within this paradigm, the cross power spectrum between the CIB at frequency $\nu$ and galaxies is
\begin{eqnarray}
P^{1h}_{g\nu}&=&\int dM \frac{dN}{dM}\frac{1}{4\pi \bar n_g(z)}\left( L_{\nu(1+z)}^{\cen} N^{\sat}(M,z) u(k,M,z) \right. \\
&+& \left. L_{\nu(1+z)}^{\sat} N^{\cen}(M,z) u(k,M,z)+L_{\nu(1+z)}^{\sat} N^{\sat}(M,z) u(k,M,z)^2\right). \nonumber
\end{eqnarray}
The cross power-spectra of the `continuous'' and the `discrete'' observables is:
\begin{equation}
P^{1h}_{XY} = \int dM \frac{dN}{dM}A_X(M,k,z)A_Y(M,k,z).
\end{equation}

\subsection{Poissonian noise}

In all galaxy-galaxy, CIB-CIB, and galaxy-CIB power spectra, we must also include the scale-independent Poissonian noise (or shot noise). 
\begin{equation}
C_\ell^{gg,\mathrm{shot}} (z)=\frac{1}{\bar n (z)}
\end{equation}
where $\bar n (z)$ is the total galaxy number density in the map in a redshift bin. For the CIB, the shot noise is
\begin{equation}
C_\ell^{\nu\nu,\mathrm{shot}}=\int dS_\nu \frac{dN}{dS_\nu}S_\nu^2
\end{equation}
where $S_\nu$ represents flux  measured at frequency $\nu$.

\end{document}